\documentclass[twocolumn,a4paper,reprint,amssymb,aps,prd,groupedaddress,superscriptaddress,nofootinbib]{revtex4-2}
\pdfoutput=1
\usepackage{graphicx}
\usepackage{epsf}
\usepackage{bm}
\usepackage{amsmath}
\usepackage{amsfonts}
\usepackage{amssymb}
\usepackage{epstopdf}
\usepackage{natbib}
\usepackage{color}
\usepackage[dvipsnames]{xcolor}
\usepackage{physics}
\usepackage{bm}
\usepackage[colorlinks = true,
            linkcolor = blue,
            urlcolor = blue,
            citecolor = blue,
            anchorcolor = blue]{hyperref}
\usepackage{lipsum}
\providecommand{\U}[1]{\protect\rule{.1in}{.1in}}

\usepackage[capitalize]{cleveref}
\usepackage{soul}
\usepackage{braket}
\usepackage{enumitem}

\begin{document}

\title{Transition dynamics in the $\Lambda_{\rm s}$CDM model: Implications for bound cosmic structures}
\author{Evangelos A. Paraskevas}
\email{e.paraskevas@uoi.gr}
\affiliation{Department of Physics, University of Ioannina, GR-45110, Ioannina, Greece}
\author{Arman \c{C}am}
\email{cam21@itu.edu.tr}
\affiliation{Department of Physics, Istanbul Technical University, Maslak 34469 Istanbul, Turkey}
\author{Leandros Perivolaropoulos}
\email{leandros@uoi.gr}
\affiliation{Department of Physics, University of Ioannina, GR-45110, Ioannina, Greece}
\author{\"{O}zg\"{u}r Akarsu}
\email{akarsuo@itu.edu.tr}
\affiliation{Department of Physics, Istanbul Technical University, Maslak 34469 Istanbul, Turkey}


\begin{abstract}
We explore the predictions of $\Lambda_{\rm s}$CDM, a novel framework suggesting a rapid anti-de Sitter (AdS) to de Sitter (dS) vacua transition in the late Universe, on bound cosmic structures. In its simplest version, the cosmological constant, $\Lambda_{\rm s}$, abruptly switches sign from negative to positive, attaining its present-day value at a redshift of ${z_\dagger\sim 2}$. The $\Lambda_{\rm s}$CDM model emerges as a promising solution to major cosmological tensions, particularly the $H_0$ and $S_8$ tensions, as well as other less definite tensions. A key aspect of our investigation is examining the impact of the abrupt $\Lambda_{\rm s}$CDM model on the formation and evolution of bound cosmic structures. We identify three primary influences: (i) the negative cosmological constant (AdS) phase for $z > z_\dagger$, (ii) the abrupt transition marked by a type II (sudden) singularity, leading to an abrupt increase in the universe's expansion rate at $z=z_\dagger$, and (iii) an increased expansion rate in the late universe under a positive cosmological constant for $z < z_\dagger$, compared to $\Lambda$CDM. Utilizing the spherical collapse model, we investigate the non-linear evolution of bound cosmic structures within the $\Lambda_{\rm s}$CDM framework. We find that the virialization process of cosmic structures, and consequently their matter overdensity, varies depending on whether the AdS-dS transition precedes or follows the turnaround. Specifically, structures virialize with either increased or reduced matter overdensity compared to the Planck-$\Lambda$CDM model, contingent on the timing of the transition. Additionally, our results demonstrate that the sudden singularity does not result in the dissociation of bound systems. Despite its profound nature, the singularity exerts only relatively weak effects on such systems, thereby reinforcing the model's viability in this context.
\end{abstract}

\maketitle

\section{\label{sec:introduction}Introduction}

The current standard model of cosmology, the Lambda-Cold Dark Matter ($\Lambda$CDM) model~\cite{DODELSON20211}, has been remarkably successful in explaining a broad spectrum of cosmological observations~\cite{SupernovaSearchTeam:1998fmf,SupernovaCosmologyProject:1998vns,Planck:2018vyg,ACT:2020gnv,eBOSS:2020yzd,KiDS:2020suj,DES:2021wwk}. However, setting aside the notoriously challenging theoretical issues associated with the cosmological constant $\Lambda$~\cite{Weinberg:1988cp,Carroll:1991mt,Sahni:1999gb,Peebles:2002gy,Padmanabhan:2002ji,Bull:2015stt}, the era of high-precision cosmology has seen the emergence of multiple discrepancies at various levels of statistical significance. Notably, the so-called $H_0$ and $S_8$ tensions, observed when analyzing different datasets within the $\Lambda$CDM model framework, suggest that the model might be incomplete~\cite{Bull:2015stt,DiValentino:2020zio,DiValentino:2021izs,Peebles:2022akh,Perivolaropoulos:2021jda,Abdalla:2022yfr,Vagnozzi:2023nrq,Hu:2023jqc,Akarsu:2024qiq}.

The most statistically significant disagreements lie in the values of the Hubble constant, $H_0$, and the weighted amplitude of matter fluctuations, $S_8$. Estimations of $H_0$ show a discrepancy that reaches a significance of $5\sigma$, as seen between the Planck Cosmic Microwave Background (CMB) estimate under the $\Lambda$CDM assumption~\cite{Planck:2018vyg}, and the local distance ladder measurements by the SH0ES team~\cite{Riess:2021jrx}. Additionally, the $S_8$ tension within the $\Lambda$CDM framework becomes apparent in the differing results obtained from Planck CMB data and KiDS-1000 cosmic shear measurements~\cite{KiDS:2020suj}, with discrepancies reaching a $3\sigma$ level. In particular, estimates from the Planck-$\Lambda$CDM suggest $H_{0} = (67.4\pm0.5)\,{\rm km\, s^{-1}\, Mpc^{-1}}$\cite{Planck:2018vyg}, while the local $H_0$ measurement by the SH0ES team, using luminosities of Cepheid calibrators and Supernovae Type Ia (SnIa), indicates $H_{0} = (73.04\pm1.04)\,{\rm km\, s^{-1}\, Mpc^{-1}}$~\cite{Riess:2021jrx}. The tension in $S_8$ is highlighted when comparing constraints on $S_8$ from high redshift observations, like the Planck data (TT,TE,EE+lowE), reporting $S_8 = 0.834 \pm 0.016$~\cite{Planck:2018vyg}, against those from lower-redshift observations, such as weak gravitational lensing ($S_8=0.759^{+0.024}_{-0.021}$~\cite{KiDS:2020suj} and $S_8=0.772 \pm 0.022$~\cite{Burger:2023qef}) and galaxy clustering ($S_8=0.736 \pm 0.051$~\cite{Chen:2021wdi}), implying the Planck data predict a stronger growth of cosmological perturbations than what dynamical probe observations infer. A consistency test of the Planck-$\Lambda$CDM suggests that $S_8$ determinations from $f\sigma_8$ constraints increase with effective redshift, showing a $\sim3\sigma$ tension with the Planck-$\Lambda$CDM predictions at lower redshifts, but aligning within 1$\sigma$ at higher redshifts, hinting that the $S_8$ tension is physical in origin, and potentially indicating a breakdown in the standard $\Lambda$CDM model~\cite{Adil:2023jtu}.

In addressing the $H_0$ tension, a variety of extensions to the $\Lambda$CDM model have been proposed, which can be broadly categorized as follows:
\begin{itemize}[nosep,wide]
\item \textit{Early Time Models:} These models introduce new physics before recombination ($z\gtrsim 1100$) to reduce the sound horizon scale, thereby increasing the $H_0$ value. Examples include: Early Dark Energy (EDE)~\cite{Karwal:2016vyq,Poulin:2018cxd,Poulin:2018dzj,Agrawal:2019lmo,Kamionkowski:2022pkx,Odintsov:2023cli}, New Early Dark Energy (New EDE)~\cite{Niedermann:2019olb,Cruz:2023lmn,Niedermann:2023ssr}, Anti de-Sitter Early Dark Energy (AdS-EDE)~\cite{Ye:2020btb,Ye:2020oix,Ye:2021iwa}, and modified gravity~\cite{Rossi:2019lgt,Braglia:2020iik,Adi:2020qqf,Braglia:2020auw,Ballardini:2020iws,FrancoAbellan:2023gec,Petronikolou:2023cwu}. Also notable is the approach that suggests modification at the inflationary epoch, with oscillations in the inflaton potential~\cite{Hazra:2022rdl}.
\item \textit{Intermediate/Late Time Models:} These models introduce new physics at intermediate to late times ($0.1 \lesssim z \lesssim 3.0$). Their goal is to adjust the expansion rate history $H(z)$ to align $H_0$ predictions with local measurements, while remaining consistent with CMB and late-time observational data. Examples include: The $\Lambda_{\rm s}$CDM model~\cite{Akarsu:2019hmw,Akarsu:2021fol,Akarsu:2022typ,Akarsu:2023mfb} (which posits a rapidly sign-switching cosmological constant $\Lambda_{\rm s}$, from Anti de-Sitter to de-Sitter (AdS to dS), in the late universe as conjectured from the findings in graduated dark energy (gDE)~\cite{Akarsu:2019hmw}), Phantom Crossing Dark Energy~\cite{DiValentino:2020naf,Alestas:2020mvb,Alestas:2020zol,Gangopadhyay:2022bsh,Basilakos:2023kvk,Adil:2023exv,Gangopadhyay:2023nli}, Omnipotent Dark Energy model~\cite{DiValentino:2020naf,Adil:2023exv}, dynamical DE on top of an AdS background~\cite{Visinelli:2019qqu,Dutta:2018vmq,Sen:2021wld,Adil:2023exv}, and (non-minimally) Interacting Dark Energy (IDE)~\cite{Kumar:2017dnp,DiValentino:2017iww,Yang:2018uae,Pan:2019gop,Kumar:2019wfs,DiValentino:2019jae,DiValentino:2019ffd,Lucca:2020zjb,Gomez-Valent:2020mqn,Kumar:2021eev,Nunes:2022bhn,Bernui:2023byc}.\footnote{Dark energy densities that reach negative values, consistent with a negative (AdS-like) cosmological constant, especially for $z\gtrsim1.5-2$, are also observed in model-independent/non-parametric observational reconstructions~\cite{Aubourg:2014yra,Sahni:2014ooa,Poulin:2018zxs,Wang:2018fng,Bonilla:2020wbn,Escamilla:2023shf,Escamilla:2021uoj,Malekjani:2023dky,Akarsu:2022lhx,Calderon:2020hoc}. Further more, a recent model-independent reconstruction of the IDE kernel, employing  Gaussian process methods as suggested in~\cite{Escamilla:2023shf}, reveals that DE assumes negative densities for $z\gtrsim2$, which suggests that IDE models do not preclude the possibility of negative DE densities at high redshifts.}
\item \textit{Ultra Late Time Models:} These models implement changes in either fundamental physics or stellar physics during the recent past ($z \lesssim 0.01$)~\cite{Marra:2021fvf,Alestas:2020zol,Alestas:2021nmi,Alestas:2021luu,Perivolaropoulos:2021bds}.
\end{itemize}

While our list includes some key examples from the numerous attempts to resolve the $H_0$ tension through new physics, it is by no means exhaustive. For a comprehensive overview and detailed classification of various approaches, please refer to Refs.~\cite{DiValentino:2021izs,Perivolaropoulos:2021jda,Abdalla:2022yfr}. It is fair to say that, as of now, there is no widely accepted model that is both observationally and theoretically fully satisfactory. Moreover, addressing the $H_0$ tension while ensuring compatibility with all available data and without exacerbating other, less significant discrepancies such as the $S_8$ tension, remains a challenging task.
Currently, only a few models are known to propose simultaneous solutions to both the $H_0$ and $S_8$ tensions. Among these, without claim to be exhaustive, are the $\Lambda_{\rm s}$CDM model~\cite{Akarsu:2019hmw,Akarsu:2021fol,Akarsu:2022typ,Akarsu:2023mfb}, New EDE~\cite{Cruz:2023lmn,Niedermann:2023ssr}, inflation with oscillations in the inflaton potential~\cite{Hazra:2022rdl}, some IDE models~\cite{Kumar:2019wfs,DiValentino:2019ffd,Bernui:2023byc}, sterile neutrino with a non-zero masses+dynamical DE~\cite{Pan:2023frx},  dark matter (DM) with varying equation of stat (EoS) parameter~\cite{Naidoo:2022rda}, and AdS-EDE with ultralight axion~\cite{Ye:2021iwa}. However, even with an optimistic view, it is difficult to claim that these models currently present a completed theoretical framework. Among them, the $\Lambda_{\rm s}$CDM model stands out for its simplicity, introducing only one extra free parameter compared to the standard $\Lambda$CDM model: the $z_\dagger$, which signifies the redshift of the rapid AdS-dS transition. The remainder of this paper will focus on the $\Lambda_{\rm s}$CDM model.

Another aspect drawing our attention to the $\Lambda_{\rm s}$CDM model is its potential relevance to recent findings from the James Webb Space Telescope (JWST). As initially noted in~\cite{Akarsu:2022typ} (refer to Section~IV.~C of the reference), the model's incorporation of a negative (AdS) cosmological constant for $z\gtrsim2$ could lead to enhanced structure formation at these higher redshifts. This possibility aligns with observations from JWST's deep space probes ($z\gtrsim5$), which suggest that structure formation is more intense at these higher redshifts than predicted by the standard $\Lambda$CDM framework~\cite{Boylan-Kolchin:2022kae,2023Natur.616..266L}. Specifically, JWST observations have revealed that the early formation of luminous galaxies~\cite{2023Natur.616..266L, Menci:2022wia,Biagetti:2022ode,Forconi:2023izg,Gupta:2023mgg,Glazebrook:2023vkx,Adil:2023ara,Hirano:2023auh,Parashari:2023cui} exhibits more intense growth\footnote{However, some studies such as~\cite{Yung:2023bng,McCaffrey:2023qem,Wang:2023xmm} argue that the JWST data are not robust enough to conclusively assert a tension with the $\Lambda$CDM model.}. Moreover, galaxies within the redshift range of $7 \lesssim z \lesssim 10$ display an unusually high star formation rate~\cite{Wang:2023gla,Wang:2022jvx,Boylan-Kolchin:2022kae,Wang:2023xmm}, and the observed number density of ultraviolet (UV) bright galaxies at redshift $z\sim 15$ exceeds expectations, posing a challenge to the galaxy formation models based on the $\Lambda$CDM model~\cite{Gupta:2023mgg,Melia:2023dsy}. Recent works have shown~\citep{Paraskevas:2023itu,Adil:2023ara,Menci:2024rbq} that the presence of a negative cosmological constant (or more broadly, negative energy densities contributing to the Friedmann equation) at relevant redshifts can accommodate the anomalous findings on cosmological structures observed by the JWST's deep space probes.

The $\Lambda_{\rm s}$CDM model exhibits features that can address not only the $H_0$ and $S_8$ tensions but also, potentially, the anomalous findings from the JWST. All these features are controlled by a single extra free parameter, $z_\dagger\sim1.8$~\cite{Akarsu:2022typ,Akarsu:2023mfb}, the redshift at which the cosmological constant switches its sign from negative to positive and achieves its present-day value. The mechanism by which the $\Lambda_{\rm s}$CDM model achieves these outcomes is straightforward~\cite{Akarsu:2021fol,Akarsu:2022typ,Akarsu:2023mfb}: The presence of a negative cosmological constant for $z>z_\dagger$ implies that $H(z)$ is smaller than in $\Lambda$CDM at these higher redshifts. Consequently, the smaller $H(z)$, offering less resistance against structure growth, leads to faster structure growth for $z>z_\dagger$, aligning with findings from JWST. On the other hand, due to the fact that the comoving angular diameter distance to the last scattering surface, $d_A(z_*)=c\int_0^{z_*}H^{-1}{\rm d}{z}$ (where $z_*\approx1090$), is strictly determined by the CMB power spectra, any reduction in $H(z)$ for $z>z_{\dagger}$, compared to $\Lambda$CDM, must be compensated by an increase in $H(z)$ for $z<z_\dagger$. This not only explains the higher $H_0$ values predicted by $\Lambda_{\rm s}$CDM but also the suppression of structure growth for $z<z_\dagger$, in comparison to $\Lambda$CDM, as the larger $H(z)$ for $z<z_\dagger$ implies greater resistance to structure growth. Essentially, $\Lambda_{\rm s}$CDM predicts higher values of both $H_0$ and $\sigma_8$, compared to $\Lambda$CDM. However, the decreased value of $\Omega_{\rm m0}$, due to the increased $H_0$, outweighs the increased $\sigma_8$, resulting in a decreased $S_8=\sigma_8\sqrt{\Omega_{\rm m0}/0.3}$. Thus, it is conceivable that the $\Lambda_{\rm s}$CDM model can account for intense structure growth at higher redshifts while simultaneously accommodating weaker growth at redshifts $z\lesssim2$~\cite{Kazantzidis:2019nuh,Benisty:2020kdt,Nunes:2021ipq,SPT:2018njh,Burger:2023qef}.

The $\Lambda_{\rm s}$CDM~\cite{Akarsu:2019hmw,Akarsu:2021fol,Akarsu:2022typ} model emerges as one of the promising models for addressing major cosmological tensions and stands as the most economical model among those in the literature with this capability. While the abrupt/rapid nature of the $\Lambda_{\rm s}$, along with its shift from negative to positive values, presents challenges in identifying a concrete physical mechanism, the phenomenological success of $\Lambda_{\rm s}$CDM, despite its simplicity, strongly encourages the search for possible underlying physical mechanisms. Furthermore, it could have profound implications in theoretical physics, given that $\Lambda<0$ is a theoretical sweet spot; AdS vacuum is welcome due to the AdS/CFT correspondence~\cite{Maldacena:1997re} and is preferred by string theory and string theory motivated supergravities~\cite{Bousso:2000xa}. Thereby, it would be natural to associate $\Lambda_{\rm s}$ to a possible AdS-dS (phase) transition that is derived in such fundamental theories, and the theories that find motivation from them. Recently, it was shown in~\cite{Anchordoqui:2023woo} that although the AdS swampland conjecture suggests that $\Lambda_{\rm s}$ in the late universe seems unlikely---due to the AdS vacua being an infinite distance apart from dS vacua in moduli space---it can still be realized through the Casimir forces of fields inhabiting the bulk. Another study~\cite{Alexandre:2023nmh} demonstrated that in various formulations of general relativity, it is possible to obtain a sign-switching cosmological constant through an overall sign change of the metric. In a more recent study~\cite{Akarsu:2024qsi}, it was demonstrated that within a type-II minimally modified gravity framework, known as VCDM~\cite{DeFelice:2020eju,DeFelice:2020cpt}, an auxiliary scalar field with a linear potential\footnote{Refer to Refs.~\cite{Garriga:2003hj,Garriga:2003nm,Perivolaropoulos:2004yr} for discussion on the utilization of a scalar field with a linear potential in the context of DE modeling within the GR framework.} can induce an effective cosmological constant, enabling the realization of an abrupt $\Lambda_{\rm s}$CDM model through a piece-wise linear potential with two segments and facilitating smooth $\Lambda_{\rm s}$CDM models by smoothing out this potential. This novel theoretical framework, endowed with a specific Lagrangian from the VCDM theory, elevates $\Lambda_{\rm s}$CDM to a theoretically complete physical cosmology, offering a fully predictive description of our universe. These theoretical developments, emerging shortly after the introduction of the $\Lambda_{\rm s}$CDM model, suggest that this model could potentially serve as an alternative to the standard $\Lambda$CDM model in the near future.

The simplest version of the $\Lambda_{\rm s}$CDM model is constructed phenomenologically by replacing the usual $\Lambda$ in the standard $\Lambda$CDM model with $\Lambda_{\rm s}\equiv\Lambda_{\rm s0}{\rm sgn}(z_\dagger-z)$, where a cosmological constant undergoes an abrupt sign-switch in the past, occurring at redshift $z_\dagger$, and maintains a constant positive value thereafter~\cite{Akarsu:2021fol,Akarsu:2022typ,Akarsu:2023mfb}. In this paper, we occasionally refer to this model as the abrupt $\Lambda_{\rm s}$CDM. It is characterized by the following Friedmann equation:
\begin{equation}
     \label{eq:lscdm_friedmann}
    \frac{H^2(z)}{H_0^2} = \Omega_{\rm m0}(1+z)^{3}+ \Omega_{\Lambda_{\rm s}0}\text{sgn}\qty(z_{\dagger}-z) + \Omega_{k0}(1+z)^2\,,
\end{equation}
where the transition is incorporated using the signum function (sgn), and $\Lambda_{\rm s0}>0$ represents the present-day value of $\Lambda_{\rm s}$ (or for redshifts $z<z_\dagger$).\footnote{Note that this abrupt behavior of $\Lambda_{\rm s}$, as described and considered in this work, represents as an idealized depiction of a rapid transition phenomenon, akin to a phase transition, from AdS vacuum to dS vacuum, or a dark energy (DE) model, such as gDE~\cite{Akarsu:2019hmw}, capable of mimicking this behavior~\cite{Akarsu:2022typ}. Additionally, it is important to emphasize that $\Lambda_{\rm s}$, whether exhibiting abrupt changes or not, does not violate the principle of energy conservation. For further details on this aspect, see Refs.~\cite{Ozulker:2022slu,Akarsu:2022typ}.} Here, $H(z)$ represents the Hubble parameter, while $\Omega_{\rm m0}$, $\Omega_{\Lambda_{\rm s}0}$, and $\Omega_{k0}$ denote the present-day density parameters for the pressureless fluid (baryons+CDM), sign-switching cosmological constant, and spatial curvature, respectively. In the abrupt $\Lambda_{\rm s}$CDM model, the Hubble parameter exhibits a discontinuity in the past, specifically, at redshift $z = z_{\dagger}$, indicative of a type II (sudden) singularity at this exact redshift~\cite{Barrow:2004xh}. \footnote{For a general overview of cosmological singularities, readers are referred to Refs.~\cite{Barrow:2004xh,Nojiri:2005sx,Fernandez-Jambrina:2006tkb,Balcerzak:2023ynk,deHaro:2023lbq}.} \footnote{This singularity is absent in smooth $\Lambda_{\rm s}$CDM models featuring a rapidly yet smoothly occurring sign-switch, as realized under the VCDM framework in Ref.~\cite{Akarsu:2024qsi}. However, this work delves into the effects of $\Lambda_{\rm s}$ in its most extreme form, the abrupt $\Lambda_{\rm s}$CDM model, within the GR framework.} Although this singularity is mild, being weak enough not to compromise the cosmological model's viability (see Appendix~\ref{app:sudden_cosmo_sing} for further discussion and Refs.~\cite{Balcerzak:2023ynk,Ozulker:2022slu,Trivedi:2023zlf,Paraskevas:2023aae} for additional reading), it nonetheless imparts a velocity kick to particles. This, in turn, delays the growth of overdensities following the sign-switch in the cosmological constant. Thus, compared to the standard model, the $\Lambda_{\rm s}$CDM model enhances early-universe structure growth, particularly for $z > z_\dagger$, driven by an initially negative cosmological constant. Conversely, in the late universe ($z < z_{\dagger}$), when $\Lambda_{\rm s}$ becomes positive, it predicts weaker structure growth. This is attributed to a lower $\Omega_{\rm m0}$, a consequence of a larger $H_0$, in comparison to $\Lambda$CDM, combined with the added impact of the velocity kick at the transition.

The behavior of spherically symmetric overdensities in a universe dominated by dark energy has been extensively studied~\cite{Eke:1996ds,Lokas:2000cn,Mota:2004pa,Creminelli:2009mu,Pavlidou:2004vq,Horellou:2005qc,Basilakos:2003bi,Tanoglidis:2014lea,Tanoglidis:2016lrj,Pace:2017qxv,Pavlidou:2020afx,Paraskevas:2023itu}. The spherical collapse model is a fundamental tool for understanding the evolution of these overdensities within a Friedmann-Robertson-Walker (FRW) cosmology. These overdensities function as `sub-universes' having mean densities exceeding the background matter density. They are influenced by both DE and the expansion dynamics of the FRW background. Despite the universe's overall expansion, these overdensities evolve relatively independently, akin to the evolution of a closed FRW universe~\cite{Press:1973iz,Bond:1990iw,Cooray:2002dia,Asgari:2023mej,DODELSON2021325}. Initially expanding with the cosmological background, they eventually succumb to local gravitational forces, leading to a `turnaround' phase. After this turnaround, the regions begin to collapse---a process that, according to GR, would theoretically lead to a singularity. However, from an astrophysical standpoint, virialization is commonly understood to occur before the formation of a singularity, resulting in a stable equilibrium state.

The spherical collapse model encompasses three distinct phases:
\begin{enumerate}[nosep,wide]
\item \textit{Expansion Phase:} Initially, the overdense region expands in tandem with the cosmic background.
\item \textit{Turnaround:} Eventually, this spherically symmetric region decouples from the cosmic expansion, reaching its maximum radius. Here, local gravitational forces become dominant, initiating the collapse.
\item \textit{Shell Crossing and Virialization:} In this final stage, shells within the overdensity undergo gravitational oscillations and interactions, leading to an exchange of gravitational potential energy and ultimately resulting in virialization. This stabilizes the system in a state of equilibrium.
\end{enumerate}

In this study, we utilize the spherical collapse model to investigate the formation and evolution of bound cosmic structures within the abrupt $\Lambda_{\rm s}$CDM model. Our primary objective is to assess the impact of replacing the positive cosmological constant in the $\Lambda$CDM model with an abrupt sign-switching cosmological constant. This exploration involves analyzing the effects of (i) a past negative cosmological constant (AdS) phase for $z > z_\dagger \sim 2$, (ii) the abrupt transition itself, characterized by a sudden jump in the expansion rate of the universe—a type II (sudden) singularity—at $z=z_\dagger$, and (iii) the increased expansion rate of the universe in the $\Lambda_{\rm s}$CDM model for $z < z_\dagger$. Notably, aside from the faster expansion rate, the $\Lambda_{\rm s}$CDM and $\Lambda$CDM models are identical for $z < z_\dagger$. Our investigation focuses on the densities and scales of virialized structures, characterized by their virial and turnaround radii, across various turnaround redshifts, $z_{\rm ta}$, while accounting for the sign switch in the cosmological constant in the late universe. We identify three primary transition scenarios based on the timing of $\Lambda_{\rm s}$'s sign switch, i.e., the AdS-dS transition, relative to the evolutionary stage of the overdensities: two occurring before virialization $(a_{\dagger} < a_{\rm vir})$, specifically before turnaround ($a_{\dagger} < a_{\rm ta} < a_{\rm vir}$) and after turnaround ($a_{\rm ta} < a_{\dagger} < a_{\rm vir}$), and a third scenario where the transition occurs post-virialization ($a_{\rm vir} < a_{\dagger}$).

In this paper, we first revisit the spherical collapse model with a cosmological constant of arbitrary sign, employing a semi-Newtonian framework. We derive virialized densities, building upon the methodology presented in earlier work~\cite{Paraskevas:2023itu} (Section~\ref{sec:spherical_collapse_cc}). We then extend our analysis to the $\Lambda_{\rm s}$CDM model~\cite{Akarsu:2019hmw,Akarsu:2021fol,Akarsu:2022typ,Akarsu:2023mfb}, incorporating transitional effects of the cosmological constant into our calculations (Section~\ref{sec3}). Our results include derived virialized densities in the $\Lambda_{\rm s}$CDM model, which we compare with those in the $\Lambda$CDM model. This analysis particularly focuses on the transition marked by a sudden cosmological singularity~\cite{Barrow:2004xh}. Lastly, employing the Newtonian approximation of a bound system within an expanding background~\cite{Baker:2001yc,Nesseris:2004uj,Faraoni:2007es,Perivolaropoulos:2016nhp}, we examine the impact of this sudden singularity on systems that virialized before the transition (Section~\ref{bsystemst}).

\section{\label{sec:spherical_collapse_cc}Spherical Collapse Model in the Presence of Cosmological Constant}

In this section, we focus on calculating virialized densities and radii using the turnaround overdensity ($\delta_{\rm ta}$), in the presence of a cosmological constant~\cite{Tanoglidis:2016lrj,Paraskevas:2023itu}. Building upon the methodology and results presented here, we will analyze the $\Lambda_{\rm s}$CDM model in the next section.

\subsection{Expansion phase}
\subsubsection{Background Universe}

In our analysis, we use $R(t)$ to represent the local scale factor within the spherical overdensity and $R_{\rm p}$ for the physical radius, defined as $R_{\rm p}\equiv R(t)\chi_{0}$, where $\chi_{0}$ is the corresponding comoving radius. The notation $\rho_{\rm m}$ denotes the (pressureless) matter energy density of the spherical overdensity, and $\tilde{\rho}_{\rm m}$ for the matter energy density of the background universe. The overdensity at a given cosmic epoch, characterized by the background scale factor $a$, is described as:
\begin{equation}\label{deltaa}
    \delta(a)=\frac{\rho_{\rm m}(a)-\tilde{\rho}_{\rm m}(a)}{\tilde{\rho}_{\rm m}(a)}\,.
\end{equation}
At the turnaround time, denoted as $t_{\rm ta}$, the background scale factor reaches $a_{\rm ta} \equiv a(t_{\rm ta})$. Simultaneously, the scale factor of the overdense region attains its maximum value, denoted as $R_{\rm ta} \equiv R(t_{\rm ta})$, resulting in its maximum physical size $R_{\rm p,ta} \equiv R_{\rm ta}\chi_0$.

The evolution of the scale factor of the perturbation, $R(t)$, and of the scale factor of the background, $a(t)$, are governed by their respective Friedmann equations. These equations incorporate the effects of spatial curvature, matter density, and the cosmological constant within the spherical overdensity. The Friedmann equation for the background universe is provided in the following form~\cite{DODELSON2021325,Pavlidou:2004vq,Tanoglidis:2014lea,Tanoglidis:2016lrj}:
\begin{equation}\label{afried1} 
    \frac{\dot{a}^2}{a^2} =\frac{8\pi G}{3}\tilde{\rho}_{\rm m0} \qty(a^{-3}+\omega + \xi a^{-2})\,,
\end{equation}
where $\omega$ and $\xi$ are defined as:
\begin{equation}
    \begin{aligned}
    \omega &\equiv \frac{\rho_{\Lambda0}}{\tilde{\rho}_{\rm m0}}=\frac{\Omega_{\Lambda0}}{\Omega_{{\rm m}0}}\,,\\
    \xi &\equiv \frac{\tilde{\rho}_{\text{crit},0}-\tilde{\rho}_{\rm m0}-\rho_{\Lambda0}}{\tilde{\rho}_{{\rm m}0}}=\frac{1}{\Omega_{{\rm m}0}}-1-\omega\,.
    \end{aligned}
\end{equation}

\subsubsection{Overdensity}

Let us define an initial comoving time $t_i$, which is the moment when the scale factors and their time derivatives for both the background universe and the overdense region are the same:
\begin{equation}\label{initcond}
    a_i = R_i\,, \quad \dot{a}_i=\dot{R}_{ i}\,.
\end{equation}
With these initial conditions, we reformulate the Friedmann equations for the local overdensity and the background as follows:
\begin{align}\label{rescale2}
   \frac{\dot{R}^2}{R^2}&= \frac{8\pi G}{3}\rho_{{\rm m},i} \qty(R^{-3}+\frac{\rho_{\Lambda,i}}{\rho_{{\rm m},i}}-\bar{\kappa}R^{-2})\,, \\
    \frac{\dot{a}^2}{a^2}&=\frac{8\pi G}{3}\tilde{\rho}_{{\rm m},i} \qty(a^{-3}+\frac{\rho_{\Lambda,i}}{\tilde{\rho}_{{\rm m},i}} + \bar{\xi} a^{-2})\,.
\end{align}
We define the parameters corresponding to the spatial curvatures of both the spherically overdense region and the cosmological background at time $t_i$ as:
\begin{equation}
\begin{aligned}
    \bar{\kappa} &\equiv \frac{\rho_{{\rm m},i}+\rho_{\Lambda,i}-\rho_{{\rm crit},i}}{\rho_{{\rm m},i}}\,, \\
    \bar{\xi} &\equiv \frac{\tilde{\rho}_{{\rm crit},i} - \tilde{\rho}_{{\rm m},i}-\rho_{\Lambda,i}}{\tilde{\rho}_{{\rm m},i}}\,.
\end{aligned}
\end{equation}
From Eq.~\eqref{initcond}, we infer that the critical densities at time $t_i$ for both the overdense region and the cosmological background are identical:
\begin{equation}
     \rho_{{\rm crit},i}\equiv\frac{3}{8\pi G}\frac{\dot{R}_i^2}{R_i^2}\,, \quad \tilde{\rho}_{{\rm crit},i}\equiv\frac{3}{8\pi G}\frac{\dot{a}_i^2}{a_i^2}\,.
\end{equation}
Assuming that $-\bar{\kappa} = \bar{\xi}$ at that initial moment, and $\rho_{{\rm crit},i} = \tilde{\rho}_{{\rm crit},i}$, this leads to the relationship $\rho_{{\rm m},i}=\tilde{\rho}_{{\rm m},i}$. Subsequently, we can express the overdensity as:
\begin{equation}\label{rhomat}
    \rho_{\rm m}(a) =\tilde{\rho}_{{\rm m},i}R^{-3}(a)\,.
\end{equation}
Eqs.~\eqref{deltaa} and \eqref{rhomat}, evaluated at the moment $t_i$, yield:
\begin{equation}\label{R_a_d}
    R^{-3}(a) = a^{-3} [1 + \delta(a)]\,.
\end{equation}
Given that $\tilde{\rho}_{{\rm m},i}=\tilde{\rho}_{\rm m0}a_{i}^{-3}$ and applying the rescaling transformations $a_i R \rightarrow R$ and ${\bar{\kappa}} / a_i \rightarrow \kappa$, Eq.~\eqref{rescale2} is rewritten as~\cite{Eke:1996ds,Pavlidou:2004vq,Tanoglidis:2016lrj,Pavlidou:2020afx}:
\begin{equation}\label{rfried1}
    \frac{\dot{R}^2}{R^2}=\frac{8\pi G}{3}\tilde{\rho}_{{\rm m}0} \qty(R^{-3}+\omega-\kappa R^{-2})\,,
\end{equation}
which is the Friedmann equation for the local overdensity with scale factor $R(t)$. By dividing Eq.~\eqref{rfried1} by Eq.~\eqref{afried1} and taking the square root, we deduce:
\begin{equation}\label{drda8}
    \dv{R}{a} =\pm \sqrt{\frac{a}{R}\frac{1+\omega R^3-\kappa R}{1+\omega a^3 +\xi a }}\,.
\end{equation}
This equation describes the dynamics of the overdensity, where the positive branch corresponds to expansion (pre-turnaround), and the negative branch indicates contraction (post-turnaround).

\subsection{Turnaround}\label{sec2.2}

We can determine the physical size of the overdensity at the turnaround, denoted as $R_{\rm p, ta}$, using the criterion:
\begin{equation}
    \eval{\dv{R}{a}}_{a=a_{\rm ta}}=0\,.
\end{equation}
This condition consequently establishes the relationship between $\kappa$, $\omega$, and $R_{\rm ta}$:
\begin{equation}\label{kapparmax}
    \kappa= (1 + \omega R^3_{\rm ta})R_{\rm ta}^{-1}\,.
\end{equation}
By rearranging the corresponding terms on each side of the positive branch of Eq.~\eqref{drda8}, which represents the expansion phase of the overdensity, and considering Eq.~\eqref{kapparmax}, we derive:
\begin{equation}
    \begin{aligned}\label{drda3}
    &\int_{0}^{R}\dd R\sqrt{\frac{R}{1+\omega R^3-\qty[(1 + \omega R^3_{\rm ta})R_{\rm ta}^{-1}]R}} \\
    &=\int_{0}^{a} \dd a\sqrt{\frac{a}{1+\omega a^3 + \xi a}}\,.
\end{aligned}
\end{equation}
In Eq.~\eqref{drda3}, employing transformation of variables such that on the left-hand side (LHS) $u=R/ R_{\rm ta}$ and on the right-hand side (RHS) $y=a / a_{\rm ta}$, we obtain:
\begin{equation}\label{drda6}
\begin{aligned}
    &\int_{0}^{1}\dd u\sqrt{\frac{u}{a_{\rm ta}^{-3}(1+\delta_{\rm ta})(1-u) -\omega u(1-u^2)}}\\
    &=\int_{0}^{1} \dd y\sqrt{\frac{y}{a_{\rm ta}^{-3} + \omega y^{3}+\xi a_{\rm ta}^{-2}y}}\,.
\end{aligned}
\end{equation}
Thus, for given values of $\xi,\omega$, and $a_{\rm ta}$, we can calculate the density contrast at the turnaround, denoted as $\delta_{\rm ta}$.

\subsection{Shell crossing and virialization}\label{sec2.3}
\subsubsection{Gravitational potential of the halo}

Drawing from the contributions of~\cite{Lahav:1991wc,Iliev:2001he,Maor:2005hq,Saha:2023zos}, we characterize the gravitational potential energy of a system composed of both dark energy and dark matter, noting that virialization is exclusive to the dark matter component. The gravitational potential energy of the halo system can be represented as:
\begin{equation}\label{potenerg}
    \mathcal{U}_{\rm halo}=\frac{1}{2}\int_{V} \rho_{\rm DM} \Phi_{\rm DM} \, \dd V+\int_{V} \rho_{\rm DM} \Phi_{\rm DE} \, \dd V\,.
\end{equation}
The gravitational potential of dark matter ($\Phi_{\rm DM}$) and dark energy ($\Phi_{\rm DE}$) can be written as (see Appendix~\ref{app:grav_potential_de_dm} for the derivation):
\begin{equation}
\begin{aligned}
    \label{potentials}
    &\Phi_{\rm DM}(r)=\begin{cases}
    -2\pi G \rho_{\rm DM} \qty(R^2-r^2/3) & r<R \\
    -4\pi G \rho_{\rm DM} R^3/3r & r\geq R 
    \end{cases} \\
   &\Phi_{\rm DE}(r)=2 \pi G \rho_{\rm DE} (1+3w_{\rm DE})\frac{r^2}{3}\,.
\end{aligned}
\end{equation}
Considering a homogeneous spherical distribution of dark matter with a physical radius $R_{\rm p}$ and mass $M$, the energy density of DM and DE is given by:
\begin{equation}\label{rho_dm_de}
    \begin{aligned}
    &\rho_{\rm DM} \equiv 3M / 4\pi R_{\rm p}^{3}\,, \\
    &\rho_{\rm DE} \equiv \rho_{\rm DM}\, e_{\rm DE}\qty(\frac{a}{a_{\rm ta}})^{-3(1+w_{\rm DE})}\qty(\frac{R_{\rm p}}{R_{\rm p, ta}})^3\,,
    \end{aligned}
\end{equation}
where $e_{\rm DE}$ represents the ratio of the densities at the turnaround:
\begin{equation}\label{e_de}
    e_{\rm DE} \equiv \frac{\rho_{\rm DE}(a_{\rm ta})}{\rho_{\rm DM}(a_{\rm ta})}\,.
\end{equation}
Thus, the integrals over the volume of $\rho_{\rm DM}\Phi_{\rm DM}$ and of $\rho_{\rm DM}\Phi_{\rm DE}$, considering Eqs.~\eqref{potentials}, \eqref{rho_dm_de}, and \eqref{e_de}, read as follows: 
\begin{equation}\label{intpots}
    \begin{aligned}
    &\frac{1}{2}\int_{V} \rho_{\text{DM}}\Phi_{\text{DM}} \, \dd V=-\frac{3}{5}\frac{GM^2}{R_{\rm p}}\,, \\
    &\int_{V} \rho_{\text{DM}} \Phi_{\text{DE}}\, \dd V=  \\
    &-\frac{3}{5}\frac{GM^2}{R_{\rm p}}\,\Theta_{\rm DE}\, \qty(\frac{a}{a_{\rm ta}})^{-3(1+w_{\rm DE})}\qty(\frac{R_{\rm p}}{R_{\rm p,ta}})^3\,,
    \end{aligned}
\end{equation}
where $\Theta_{\rm DE}\equiv-\frac{1}{2}(1+3w_{\rm DE})e_{\rm DE}$. Consequently, by following Eq.~\eqref{potenerg}, the potential energy of the system can be expressed as:
\begin{equation}\label{u_ds}
    \mathcal{U}_{\text{halo}}=-\frac{3}{5}\frac{GM^2}{R_{\rm p}}\qty[1+\Theta_{\rm DE} \qty(\frac{a}{a_{\rm ta}})^{-3(1+w_{\rm DE})}\qty(\frac{R_{\rm p}}{R_{\rm p, ta}})^3]\,.
\end{equation}

\subsubsection{Virialization condition}

The conservation of energy, from the moment of the turnaround till the virialization stage, implies:
\begin{equation}\label{Energy}
\mathcal{E}\equiv\eval{\mathcal{U}_{\text{halo}}}_{R_{\rm p}=R_{\rm p,ta}}=\eval{\qty(\mathcal{K}_{\text{halo}}+\mathcal{U}_{\text{halo}})}_{R_{\rm p}=R_{{\rm p, vir}}}\,.
\end{equation}
Assuming that the \textit{Virial Theorem} holds, the kinetic energy can be written as:
\begin{equation}\label{virialk}
    \mathcal{K}_{\rm halo} = \frac{R_{\rm p}}{2} \dv{\mathcal{U}_{\rm halo}}{R_{\rm p}}\,,
\end{equation}
and given Eq.~\eqref{virialk}, Eq.~\eqref{Energy} suggests the virialization condition:
\begin{equation}\label{eq017}
    \eval{\mathcal{U}_{\text{halo}}}_{R_{\rm p}=R_{\rm p,ta}} = \eval{\qty(\frac{R_{\rm p}}{2}\dv{\mathcal{U}_{\text{halo}}}{R_{\rm p}}+\mathcal{U}_{\text{halo}})}_{R_{\rm p}=R_{\rm p,vir}}\,.
\end{equation}
For a potential given in Eq.~\eqref{u_ds}, Eq.~\eqref{eq017} can be recast as:
\begin{equation}\label{eq022} 
    4 \Theta_{\rm DE}\alpha_{\rm DE}\eta^3 - 2(1+\Theta_{\rm DE})\eta + 1=0\,,
\end{equation}
where $\eta$ and $\alpha_{\rm DE}$ are defined through:
\begin{equation}\label{eta_cc}
    \eta\equiv \frac{R_{\rm p,vir}}{R_{\rm p, ta}}\,,\quad
    \alpha_{\rm DE}\equiv\left(\frac{a_{\rm vir}}{a_{\rm ta}}\right)^{-3(1+w_{\rm DE})}\,.
\end{equation}
The solution of Eq.~\eqref{eq022} expanded in terms of $\Theta_{\rm DE}$, can be expressed by~\cite{Lahav:1991wc,Iliev:2001he,Battye:2003bm,Maor:2005hq,Saha:2023zos,Wang:1998gt,Weinberg:2002rd,Basilakos:2003bi,Horellou:2005qc,Paraskevas:2023itu}:
\begin{equation}\label{eq024}
    \begin{aligned}
    \eta=&\frac{1}{2} + \frac{1}{4} \Theta_{\rm DE} \qty(-2 + \alpha_{\rm DE}) \\
    &+\frac{1}{8} \Theta_{\rm DE}^2 \qty(-2 + \alpha_{\rm DE}) \qty(-2 + 3 \alpha_{\rm DE}) + \ldots \,.
    \end{aligned}
\end{equation}
In the specific case of a positive cosmological constant ($w_{\Lambda}=-1$) as the dark energy component, implying $\alpha_{\Lambda}=1$ and $\Theta_{\Lambda}=e_{\Lambda}$, Eq.~\eqref{eq022} can be written as:
\begin{equation}
    4e_{\Lambda}\eta^3 - 2(1 + e_{\Lambda})\eta + 1 = 0\,.
\end{equation}
This can be approximated by:
\begin{equation}\label{eta01}
\eta = \frac{1}{2} - \frac{e_{\Lambda}}{4} - \frac{e_{\Lambda}^2}{8} + \ldots \,,
\end{equation}
where the parameter $e_{\Lambda} \equiv \omega a_{\rm ta}^3 / (1+\delta_{\rm ta})$, obtained from Eq.~\eqref{e_de}.

\subsubsection{Density contrast at the virialization}

Given that $t_{\rm vir}=2t_{\rm ta}$ and assuming the collapse is completed at $t=t_{\rm vir}$, Eq.~\eqref{afried1} leads to:
\begin{equation}\label{avircalc}
    \begin{aligned}
    &\int_{0}^{y_{\rm vir}} \dd y \sqrt{\frac{y }{a_{\text{ta}}^{-3}+\omega y^{3}+\xi a_{\text{ta}}^{-2}y}} \\
    =&2 \int_{0}^{1} \dd y \sqrt{\frac{y }{a_{\text{ta}}^{-3}+\omega y^{3}+\xi a_{\text{ta}}^{-2}y}}\,,
    \end{aligned}
\end{equation}
where $y_{\mathrm{vir}} \equiv a_{\mathrm{vir}}/a_{\mathrm{ta}}$. Therefore, given specific values of $\xi$, $\omega$, and $a_{\mathrm{ta}}$, we can use Eq.~\eqref{avircalc} to compute the scale factor at virialization, $a_{\rm vir}$. Subsequently, the value of $\delta_{\mathrm{vir}}$ can be calculated using the following relation:
\begin{equation}\label{deltavir}
    1+\delta_{\rm vir}=(1+\delta_{\rm ta})\qty(\frac{y_{\rm vir}}{\eta})^3\,.
\end{equation}

\section{\texorpdfstring{$\Lambda_{\rm s}$CDM}{LsCDM}: \texorpdfstring{$\Lambda_{\rm s}$}{Ls}-Sign Switch Before Virialization}\label{sec3}

Within the $\Lambda_{\rm s}$CDM~\cite{Akarsu:2021fol,Akarsu:2022typ,Akarsu:2023mfb} framework, the evolution of the background universe is governed by:
\begin{equation}\label{fr1}
    \frac{\dot{a}^2}{a^2}=\frac{8\pi G}{3}\tilde{\rho}_{\rm m0} \left[a^{-3}+\omega_{\rm s}\,\text{sgn}\qty(1/a_{\dagger}-1/a) + \xi_{\rm s} a^{-2}\right]\,,
\end{equation}
whereas the evolution for the overdensity is described by:
\begin{equation}\label{fr2}
    \frac{\dot{R}^2}{R^2}=\frac{8\pi G}{3}\tilde{\rho}_{\rm m0} \left[R^{-3}+\omega_{\rm s}\,\text{sgn}\qty(1/a_{\dagger}-1/a) - \kappa_{\rm s} R^{-2}\right]\,,
\end{equation}
where $\omega_{\rm s}$ and $\xi_{\rm s}$ are parameters defined as:
\begin{align*}
    \omega_{\rm s} &\equiv \frac{\rho_{\Lambda_{\rm s}0}}{\tilde{\rho}_{\rm m0}} = \frac{\Omega_{\Lambda_{\rm s}0}}{\Omega_{\rm m0}}\,,\\
    \xi_{\rm s} &\equiv \frac{\tilde{\rho}_{\text{crit},0}-\tilde{\rho}_{\rm m0}-\rho_{\Lambda_{\rm s}0}}{\tilde{\rho}_{{\rm m}0}}=\frac{1}{\Omega_{{\rm m}0}}-1-\omega_{\rm s}\,. 
\end{align*}
After dividing Eq.~\eqref{fr1} by Eq.~\eqref{fr2}, we obtain:
\begin{equation}\label{drda_lscdm}
    \dv{R}{a} =\pm \sqrt{\frac{a}{R}\frac{1+\omega_{\rm s}\text{sgn}\qty(1/a_{\dagger}-1/a)  R^3-\kappa_{\rm s} R}{1+\omega_{\rm s}\text{sgn}\qty(1/a_{\dagger}-1/a)  a^3 +\xi_{\rm s} a }}\,.
\end{equation}

\subsection{\texorpdfstring{$\Lambda_{\rm s}$}{Ls}-sign switch before the turnaround \texorpdfstring{($a_{\dagger} < a_{\rm ta} < a_{\rm vir}$)}{Lg}}\label{tbeforet}
\subsubsection{Density contrast at the turnaround}

Assuming that the $\Lambda_{\rm s}$-sign switch transition occurs during the expansion phase of the overdensity, we integrate Eq.~\eqref{drda_lscdm} as follows:
\begin{equation}\label{dRda01}
    \begin{aligned}
    &\int_{0}^{R}\dd R \sqrt{\frac{R}{1+\omega_{\rm s} \,\text{sgn}\qty(1/a_{\dagger}-1/a) R^3-\kappa_{\rm s} R}} \\
    =&\int_{0}^{a}\dd a \sqrt{\frac{a}{1+\omega_{\rm s} \text{sgn}\qty(1/a_{\dagger}-1/a)a^3+\xi_{\rm s} a}}\:.
    \end{aligned}
\end{equation}
By implementing the change of variables $u = R / R_{\mathrm{ta}}$ to the LHS of Eq.~\eqref{dRda01}, and $y = a / a_{\mathrm{ta}}$ to the RHS of Eq.~\eqref{dRda01}, and integrating until the turnaround moment, we obtain the following system of equations:
\begin{align}\label{equat1}
    &\int_{0}^{y_{\dagger}} \dd y \sqrt{\frac{y }{a_{\rm ta}^{-3}-\omega_{\rm s} y^{3}+\xi_{\rm s}a_{\rm ta}^{-2}y}} \nonumber \\
    =&\int_{0}^{u_{\dagger}} \dd u\sqrt{\frac{u}{a^{-3}_{\rm ta}(1+\delta_{\rm ta})(1-u)-\omega_{\rm s} u(1+u^2)}}\,, \\ \label{equat2}
    &\int_{y_{\dagger}}^{1} \dd y \sqrt{\frac{y }{a_{\rm ta}^{-3}+\omega_{\rm s} y^{3}+\xi_{\rm s}a_{\rm ta}^{-2}y}} \nonumber \\
    =&\int_{u_{\dagger}}^{1} \dd u\sqrt{\frac{u}{a^{-3}_{\rm ta}(1+\delta_{\rm ta})(1-u)-\omega_{\rm s} u(1-u^2)}}\,.
\end{align}
Here, we have denoted $y_{\dagger} \equiv a_{\dagger}/a_{\mathrm{ta}}$, $u_{\dagger} \equiv R_{\dagger}/R_{\mathrm{ta}}$, and $\kappa_{\mathrm{s}} = (1 + \omega_{\mathrm{s}} R^3_{\mathrm{ta}})R_{\mathrm{ta}}^{-1}$, given that the cosmological constant is positive at the moment of turnaround. Thus, for a given $y_{\dagger}$, we can derive the corresponding values of $u_{\dagger}$ and $\delta_{\mathrm{ta}}$ that satisfy Eqs.~\eqref{equat1} and \eqref{equat2} simultaneously.
\subsubsection{Density contrast at the virialization}

In such a case, the $\Lambda_{\rm s}$-sign switch transition occurs prior to the turnaround moment. As a result, at the turnaround, the cosmological constant has already been positive, ensuring that the collapse proceeds with a positive cosmological constant throughout, similar to the $\Lambda$CDM case. At the moment of turnaround:
\begin{equation}
      \frac{\rho_{\Lambda_{\rm s}}(a_{\rm ta})}{\rho_{\rm m}(a_{\rm ta})}={\rm sgn}\qty(1/a_{\dagger} - 1/a_{\rm ta})e_{\Lambda_{\rm s}} \equiv e_{\Lambda_{\rm s}}\,,
\end{equation}
where we have defined:
\begin{equation}
e_{\Lambda_{\rm s}}\equiv \omega_{\rm s}a_{\rm ta}^3 / (1+\delta_{\rm ta})\,.
\end{equation}
The virialization condition results in:
\begin{equation}
    4e_{\Lambda_{\rm s}}\eta^3 - 2(1 + e_{\Lambda_{\rm s}})\eta + 1 = 0\,,
\end{equation}
which can be approximated by:
\begin{equation}\label{eta02ls}
\eta = \frac{1}{2} - \frac{e_{\Lambda_{\rm s}}}{4} - \frac{e_{\Lambda_{\rm s}}^2}{8} + \ldots \,.
\end{equation}
Assuming that the collapse is completed at $t=t_{\rm vir}$ (with $t_{\rm vir} \simeq 2\, t_{\rm ta}$), the Friedmann equation for the background universe (for $y_{\dagger}<1$) is by given:
\begin{equation}\label{avircalc001}
    \begin{aligned}
    &\int_{0}^{y_{\dagger}}\dd y \sqrt{\frac{y }{a_{\text{ta}}^{-3}-\omega_{\rm s} y^{3}+\xi_{\rm s}a_{\text{ta}}^{-2}y}} \\
    +&\int_{y_{\dagger}}^{y_{\rm vir}}\dd y \sqrt{\frac{y}{a_{\text{ta}}^{-3}+\omega_{\rm s} y^{3}+\xi_{\rm s} a_{\text{ta}}^{-2}y}}  \\
    =&2 \Bigg(\int_{0}^{y_{\dagger}}\dd y \sqrt{\frac{y }{a_{\text{ta}}^{-3}-\omega_{\rm s} y^{3}+\xi_{\rm s} a_{\text{ta}}^{-2}y}} \\
    +&\int_{y_{\dagger}}^{1}\dd y \sqrt{\frac{y }{a_{\text{ta}}^{-3}+\omega_{\rm s} y^{3}+\xi_{\rm s} a_{\text{ta}}^{-2}y}}\Bigg)\,.
    \end{aligned}
\end{equation}
From Eq.~\eqref{avircalc001}, we deduce:
\begin{equation}\label{avircalc10}
    \begin{aligned}
    &\int_{1}^{y_{\rm vir}} \dd y \sqrt{\frac{y }{a_{\text{ta}}^{-3}+\omega_{\rm s} y^{3}+\xi_{\rm s} a_{\text{ta}}^{-2}y}} \\
    =&\int_{0}^{y_{\dagger}} \dd y \sqrt{\frac{y }{a_{\text{ta}}^{-3}-\omega_{\rm s} y^{3}+\xi_{\rm s}a_{\text{ta}}^{-2}y}} \\
    +&\int_{y_{\dagger}}^{1} \dd y \sqrt{\frac{y }{a_{\text{ta}}^{-3}+\omega_{\rm s} y^{3}+\xi_{\rm s}a_{\text{ta}}^{-2}y}}\,.
    \end{aligned}
\end{equation}
Thus, it becomes feasible to deduce the value of $a_{\rm vir}$ in terms of $a_{\rm ta}$, and subsequently $\delta_{\rm vir}$ from Eq.~\eqref{deltavir}. 

\subsection{\texorpdfstring{$\Lambda_{\rm s}$}{Ls}-sign switch after the turnaround \texorpdfstring{($a_{\rm ta} <a_{\dagger} < a_{\rm vir}$)}{Lg}}\label{taftert}

\subsubsection{Density contrast at the turnaround}

Given that the transition occurs after the turnaround (i.e., $a_{\dagger} > a_{\rm ta}$), the expansion phase persists with a negative cosmological constant throughout. The value of $\delta_{\rm ta}$ is determined using the following equation:
\begin{equation}\label{postdeltata}
    \begin{aligned}
    &\int_{0}^{1}\dd u\sqrt{\frac{u}{a_{\rm ta}^{-3}(1+\delta_{\rm ta})(1-u) +\omega_{\rm s} u(1-u^2)}} \\
    =&\int_{0}^{1} \dd y\sqrt{\frac{y}{a_{\rm ta}^{-3} - \omega_{\rm s} y^{3}+\xi_{\rm s} a_{\rm ta}^{-2}y}}\,,
    \end{aligned}
\end{equation}
which corresponds to Eq.~\eqref{drda6}, with the substitution of $\omega\to -\omega_{\rm s}$\footnote{We have also assumed that $a_{\rm ta}\leq a_{{\rm m}\Lambda_{\rm s}}\equiv \qty(\Omega_{\rm m0} / \Omega_{\Lambda_{\rm s}0})^{\frac{1}{3}}$, i.e., the moment where $H^2(a_{{\rm m}\Lambda_{\rm s}})=0$.}. Note that have used the relation $\kappa_{\mathrm{s}} = (1 - \omega_{\mathrm{s}} R^3_{\mathrm{ta}})R_{\mathrm{ta}}^{-1}$, where the cosmological constant is negative at the moment of turnaround.
 
\subsubsection{Effect of the Type II Singularity: Free particle in the Hubble flow}\label{freepartv}

We assume that a certain amount of kinetic energy is induced in each free particle that experienced the $\Lambda_{\rm s}$-sign switch transition event. Considering the RW metric, 
\begin{equation}\label{metric0}
    \dd s^2=-\dd t^2+a^2\left[\dd \chi^2+\chi^2(\dd \theta^2+\sin^2\theta\, \dd \phi^2)\right] \:,
\end{equation}
the physical distance is given by $r(t) = a(t)\chi$ with $\chi$ being the comoving coordinate. The geodesic equation, representing the physical radial coordinate of a free particle with a constant comoving coordinate in a RW metric, is expressed as~\cite{Baker:2001yc}:
\begin{equation}\label{rgeod}
     \Ddot{r}-\frac{\Ddot{a}}{a}r=0\:.
\end{equation}
To elucidate the consequences of the type II (sudden) singularity that occurs at the $\Lambda_{\rm s}$-sign switch transition (see Appendix~\ref{app:sudden_cosmo_sing}), it is imperative to evaluate the integral of Eq.~\eqref{rgeod} over an interval surrounding the transition moment. Consider, in particular, the time interval $t$ given by $t \in [t_{\dagger} - \varepsilon, t_{\dagger} + \varepsilon]$, where $\varepsilon$ is a positive infinitesimal. Proceeding with this approach and given that $r=a\chi$:
\begin{equation}\label{intgeod}
    \int^{t_{\dagger}+\varepsilon}_{t_{\dagger}-\varepsilon}\,\dd t\,\Ddot{r}-\int^{t_{\dagger}+\varepsilon}_{t_{\dagger}-\varepsilon}\,\dd t\,\frac{\Ddot{a}}{a}r=0\implies \eval{\dot{r}}^{t_{\dagger}+\varepsilon}_{t_{\dagger}-\varepsilon}=\eval{(Hr)}^{t_{\dagger}+\varepsilon}_{t_{\dagger}-\varepsilon}\:.
\end{equation}
Consider the velocity difference\footnote{Throughout the text, we define the left and right limits of a function as; $f^{(-)} \equiv \lim_{\varepsilon \to 0} f(t_{\dagger} - \varepsilon)$ and $f^{(+)} \equiv \lim_{\varepsilon \to 0} f(t_{\dagger} + \varepsilon)$, respectively.}, around the moment of singularity $t_{\dagger}$:
\begin{equation}\label{deltav}
    \delta V \equiv \dot{r}^{(+)} - \dot{r}^{(-)} = H^{(+)} r(t_{\dagger})-H^{(-)} r(t_{\dagger})\:.
\end{equation}
Given the continuity of the physical distance $r(t)$, we derive the velocity impulse as
\begin{equation}\label{impulse}
   \delta V  = \delta H\, r_{\dagger}\:.
\end{equation}
Here, we denote the discontinuous increase in the Hubble parameter, resulting from the sign switch of the cosmological constant, as derived from the Friedmann equation:
\begin{equation}
    \begin{aligned}
    \delta H\equiv& H^{(+)}-H^{(-)} \\
    =&\sqrt{\frac{8\pi G\tilde{\rho}_{{\rm m}0}}{3}} \Bigg[\Big(a_{\dagger}^{-3}+\omega_{\rm s}+\xi_{\rm s} a_{\dagger}^{-2}\Big)^{\frac{1}{2}} \\
    -&\Big(a_{\dagger}^{-3}-\omega_{\rm s}+\xi_{\rm s} a_{\dagger}^{-2}\Big)^{\frac{1}{2}}\Bigg] \:.
    \end{aligned}
\end{equation}

\begin{figure}[t!]
\includegraphics[width = \columnwidth]{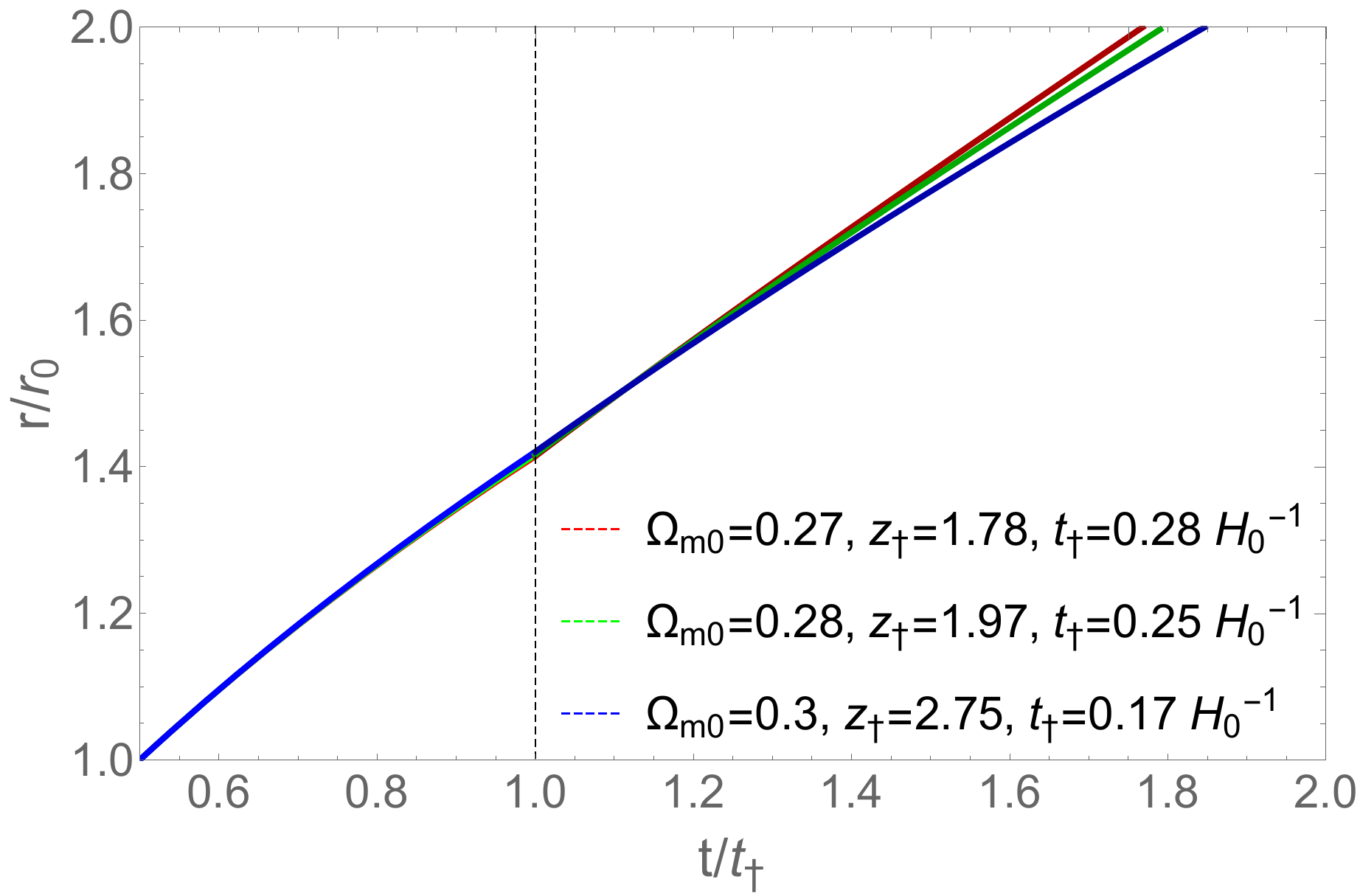}
\caption{Utilizing the joint boundary conditions at the transition moment $t/t_{\dagger}=1$, an analytical solution to the geodesic equation for a free particle within the $\Lambda_{\rm s}$CDM cosmological framework is derived. The solution of geodesic Eq.~\eqref{rfreeparticle}, which maintains the continuity of the physical distance $r(t)$ and incorporates the velocity kick, Eq.~\eqref{impulse}. The initial conditions $ r(t/t_{\dagger}=0.5)=r_0$ and $\dot{r}(t/t_{\dagger}=0.5)=r_0/t_{\dagger}$ are imposed. We observe that the effect of the kick (at $t/t_{\dagger}=1$) is insignificant.}
\label{free_particle}
\end{figure}

The solution to the geodesic equation for a free particle, as denoted in Eq.~\eqref{rgeod}, within the $\Lambda_{\rm s}$CDM framework can be derived using joint boundary conditions at the transition moment i.e. $r(t)$ continuous at $t_{\dagger}$ and also Eq.~\eqref{impulse}. These conditions include the continuity of $r(t)$ at $t_\dagger$ and conditions described in Eq.~\eqref{impulse}, which accounts for the velocity kick. With a specified set of initial conditions for $r$ and $\dot{r}$, we are able to derive the analytical solution Eq.~\eqref{rfreeparticle}(Fig.~\ref{free_particle}). Considering that the physical distance $r = a\chi$ (where $a(t)$ is the scale factor, as shown in Eq.~\eqref{scalefactor}), follows the exact same form as $a(t)$, assuming a constant $\chi$ (see Appendix \ref{app:scale_factor_lscdm} for detailed derivation of the scale factor in the $\Lambda_{\rm s}$CDM model).

\subsubsection{Effect of the Type II Singularity: Spherical collapse model}

A free particle in constant comoving coordinates will infall due to the contraction of the spherical overdensity entrained by the spacetime geometry. Following the same reasoning as in Section~\ref{freepartv}, we introduce the corresponding velocity kick to a particle in the spherical overdensity as:
\begin{equation}\label{impulse1}
\Delta V \equiv\dot{R}^{(+)}_{\rm p,\dagger} - \dot{R}^{(-)}_{\rm p,\dagger}\,,
\end{equation}
where $\dot{R}_{\rm p}\equiv \chi_0 a H\frac{\dd R}{\dd a}$. 
During the collapsing phase, we implement the negative branch of Eq.\eqref{drda_lscdm} in Eq.\eqref{impulse1}, which results in:
\begin{equation}\label{deltaV}
\begin{aligned}
    \Delta V=-\chi_{0}a_{\dagger}\Bigg(&H^{(+)}\sqrt{\frac{a_{\dagger}}{R_{\dagger}}\frac{1+\omega_{\rm s}  R_{\dagger}^3-\kappa_{\rm s} R_{\dagger}}{1+\omega_{\rm s} a_{\dagger}^3 +\xi_{\rm s}a_{\dagger} }}\\
    &-H^{(-)}\sqrt{\frac{a_{\dagger}}{R_{\dagger}}\frac{1-\omega_{\rm s}  R_{\dagger}^3-\kappa_{\rm s} R_{\dagger}}{1-\omega_{\rm s} a_{\dagger}^3 +\xi_{\rm s}a_{\dagger}}}\Bigg)\,.
\end{aligned}
\end{equation}

\subsubsection{Effect of the Type II Singularity: Modified virialization condition}

\begin{figure}[t]
\centering
\includegraphics[width = \columnwidth]{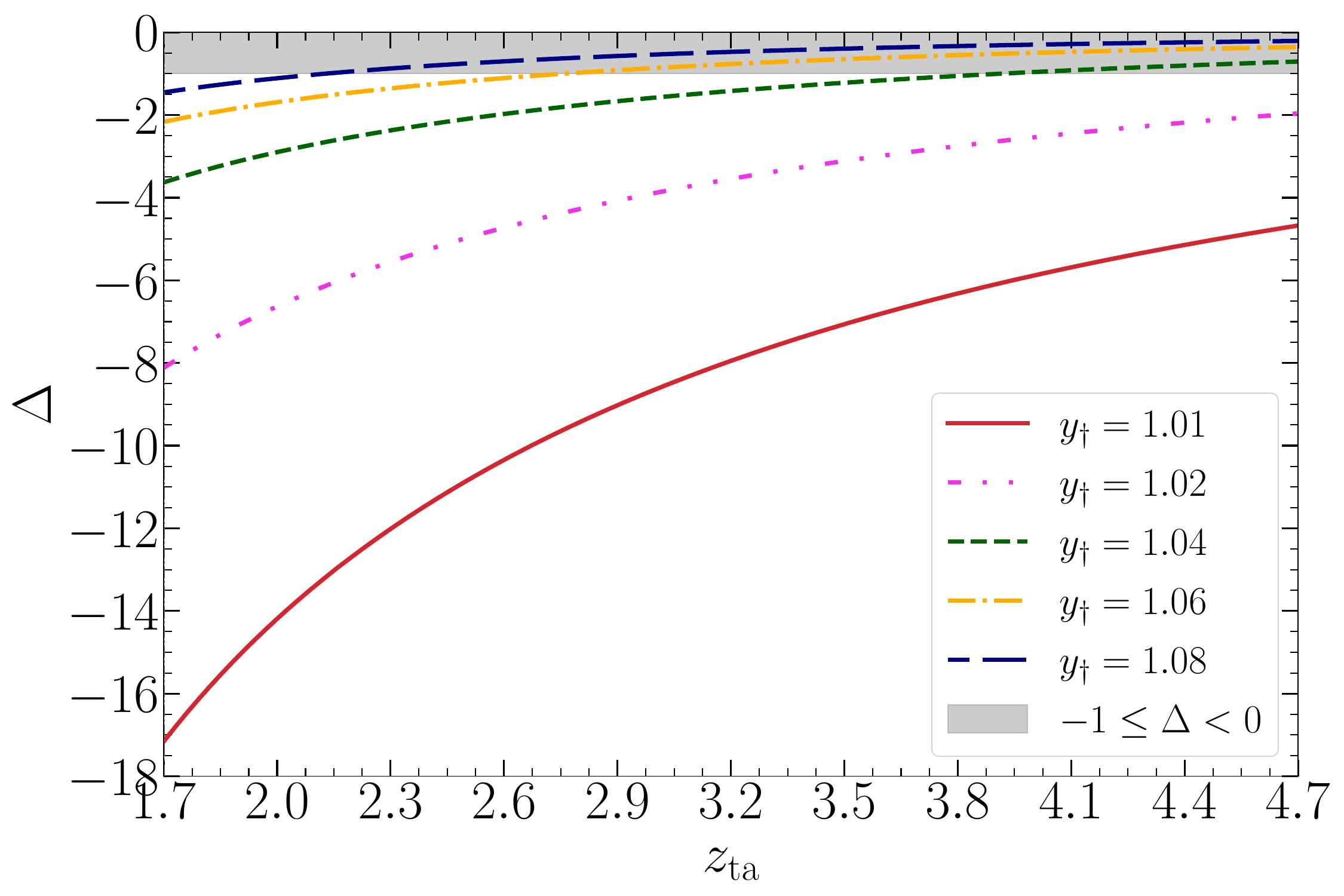}
\caption{Variation of $\Delta$ with respect to $z_{\mathrm{ta}}$, obtained by keeping $y_{\dagger} \in [1.01, 1.02, 1.04, 1.06, 1.08]$ constant. The collapse of the overdense region and the expansion of the universe exert opposing effects. Consequently, the emergence of the sudden singularity effectively dissipates kinetic energy, ($\Delta < 0$) as demonstrated above. The shaded gray area represents the region of $-1 \leq \Delta < 0$, where only the kinetic energy is extracted from the overdensity. Meanwhile, for $\Delta < -1$, the overdensity expands once more to a slightly larger value of a new physical radius, before it collapses again and eventually virializes.}
\label{fig:delta0}
\end{figure}

Assume the transition occurs during the collapse of the overdensity (where $a_{\dagger}>a_{\rm ta}$), at the turnaround moment:
\begin{equation}
      \frac{\rho_{\Lambda_{\rm s}}(a_{\rm ta})}{\rho_{\rm m}(a_{\rm ta})}={\rm sgn}\qty(1/a_{\dagger} - 1/a_{\rm ta})e_{\Lambda_{\rm s}} = -e_{\Lambda_{\rm s}}\,,
\end{equation} 
and the potential energy of the system, at the moment when the scale factor attains the value $a_{\rm ta}$ and the cosmological constant is negative, is substituted according to the Eq.~\eqref{intpots}, as:
\begin{equation}\label{potvir}
    \mathcal{U}_{\text{halo}}=-\frac{3}{5}\frac{GM^2}{R_{\rm p}}\qty[1-e_{\Lambda_{\rm s}}\left(\frac{R_{\rm p}}{R_{\rm p,ta}}\right)^3]\,.
\end{equation}
Energy conservation is upheld right until the brink of the singularity's emergence at $a =a_{\dagger}$. If $\varepsilon > 0$ is an infinitesimally small positive quantity, and $t_{\dagger}-\varepsilon$ represents the moment just before the singularity, then, given the continuity of $R_{\rm p}$, when $\varepsilon \to 0$, we obtain:
\begin{equation}\label{eq3.6}
   \mathcal{K}^{(-)}= \eval{\mathcal{U}_{\text{halo}}}_{R_{\rm p,ta}}-\eval{\mathcal{U}_{\text{halo}}}_{R_{\rm p,\dagger}}\,.
\end{equation}
Immediately after, energy conservation continues to be valid but incorporates an energy impulse attributed to the singularity. Specifically, the velocity $V$ of a spherically symmetric shell is given by $V = \frac{\dot{R}}{R} R_{\text{p}}$. This implies that for $t_{\dagger} + \varepsilon$, due to a velocity kick, $V$ transitions to $V + \Delta V$ (Eq.~\eqref{deltaV}). Consequently, this impulse affects the kinetic energy in the following manner: 
\begin{equation}\label{eq3.5}
    \mathcal{K}^{(+)}=\mathcal{K}^{(-)}\left[1+\Delta\right]^2\,,
\end{equation}
where we have defined $\Delta$ as:
\begin{equation}
\begin{aligned}
    \Delta \equiv &\frac{\Delta V}{\qty|V^{(-)}|}= \frac{\Delta V}{\qty|\chi_{0} a_{\dagger} H^{(-)} \dv{R}{a}^{(-)}|} \\
    =&1-\frac{H^{(+)}}{H^{(-)}}\sqrt{\frac{\qty(1+\omega_{\rm s} R_{\dagger}^3-\kappa_{\rm s} R_{\dagger})\qty(1-\omega_{\rm s} a_{\dagger}^3 +\xi_{\rm s}a_{\dagger})}{\qty(1+\omega_{\rm s} a_{\dagger}^3 +\xi_{\rm s}a_{\dagger})\qty(1-\omega_{\rm s}  R_{\dagger}^3-\kappa_{\rm s} R_{\dagger})}}\,. 
\end{aligned}
\end{equation}
From immediately post-singularity up to the point of virialization, the conservation of energy, along with the Virial theorem, gives:
\begin{equation}
     \eval{\frac{R_{\rm p, vir}}{2}\dv{\mathcal{U}_{\rm halo}}{R_{\rm p}}}_{R_{\rm p,vir}}+\eval{\mathcal{U}_{\text{halo}}}_{R_{\rm p,vir}}=\mathcal{K}^{(+)}+\eval{\mathcal{U}_{\rm halo}}_{R_{\rm p,\dagger}}\,.
\end{equation}
Eqs.~\eqref{eq3.5} and~\eqref{eq3.6} combined imply the following modified \textit{virialization condition}:
\begin{align}\label{virialization}
    &\eval{\frac{R_{\rm p,vir}}{2}\dv{\mathcal{U}_{\rm halo}}{R_{\rm p}}}_{R_{\rm p,vir}}+\eval{\mathcal{U}_{\text{halo}}}_{R_{\rm p,vir}} \nonumber \\
    =&\qty(\eval{\mathcal{U}_{\text{halo}}}_{R_{\rm p,ta}}-\eval{\mathcal{U}_{\text{halo}}}_{R_{\rm p,\dagger}})\qty(1+\Delta)^2+\eval{\mathcal{U}_{\text{halo}}}_{R_{\rm p,\dagger}}\,.
\end{align}
Implementing the potential energy, given by Eq.~\eqref{potvir}, into Eq.~\eqref{virialization} and taking into account that the singularity takes place subsequent to the turnaround moment, i.e., $a_{\dagger } > a_{\rm ta}$, we derive:
\begin{equation}\label{Upoteq}
\begin{aligned}
    -\frac{1 - y_{\rm vir}^{-3 (1 + w)}\Theta \,\eta^3}{10 R_{\rm p,vir}} 
+\frac{ 1 - y_{\dagger}^{-3 (1 + w)} \Theta u_{\dagger}^3}{5 R_{\rm p,\dagger}}\\
+\frac{3 \Theta y_{\rm vir}^{-3 (1 + w)}  R^2_{\rm p,vir} }{10 R_{\rm ta}^3} \\
-(1 + \Delta)^2 \left(-\frac{ 1 - \Theta}{5 R_{\rm ta}} + \frac{1 - y_{\dagger}^{-3 (1 + w)} \Theta\, u_{\dagger}^3}{5 R_{\rm p,\dagger}}\right)=0\,.
\end{aligned}
\end{equation}
Subsequently, for $\Lambda_{\rm s}$ ($w_{\Lambda_{\rm s}}=-1$), Eq.~\eqref{Upoteq} implies that:
\begin{equation}\label{eqtr}
\begin{aligned}
    &\Delta_0 (-1 + u_{\dagger}) + u_{\dagger}\qty(1-\frac{1}{2 \eta}) \\
    +&u_{\dagger}e_{\Lambda_{\rm s}} \qty[-1 + 2 \eta^2+\Delta_0 (-1 + u_{\dagger}^2)]=0\,,
\end{aligned}
\end{equation}
where we have defined the following dimensionless parameters:
\begin{equation*}
    \Delta_{0} \equiv \Delta(2+\Delta)\,, \quad u_{\dagger} \equiv \frac{R_{\dagger}}{R_{\rm ta}}\,.
\end{equation*}
Note that when $\Delta = 0$, Eq.~\eqref{eq022} is recovered, corresponding to a collapse with a negative cosmological constant. While Eq.~\eqref{eqtr} admits an analytical solution, its complexity can hinder a straightforward physical interpretation. As such, we resort to an approximate solution. At first order in $\Delta_{0}$ and values of $u_{\dagger}$ close to $1$, we obtain:
\begin{equation}\label{posteta}
    \begin{aligned}
    \eta =& \frac{1}{2}\qty[1 + \frac{\Delta_{0}(1-u_{\dagger})}{u_{\dagger}}]+\frac{e_{\Lambda_{\rm s}}}{4} \qty[1 + 2 \Delta_{0} (1 - u_{\dagger}^2)] \\
     &- \frac{e_{\Lambda_{\rm s}}^2}{8}\qty[1 + \frac{7 \Delta_{0} (1 - u_{\dagger})}{u_{\dagger}}]+\ldots\:.
    \end{aligned}
\end{equation}
This approximation remains valid for $|\Delta| \ll 1$. In cases where this condition is not met, it becomes necessary to resort to the analytical solutions of Eq.~\eqref{eqtr} (Figure~\ref{fig:delta0}). The collapse of an overdense region and the expansion of the universe exert opposing effects. The impact of the singularity on the overdensity is significant when $y_{\dagger} \simeq 1$ and becomes smaller for values that are further away, as well as for larger values of $z_{\rm ta}$, as shown in Fig.~\ref{fig:main_results}. Although the sudden singularity, extracts kinetic energy (i.e., when $-1 \leq \Delta < 0$) from the collapsing overdensity, it is notable that near the turnaround moment, the velocity of the contracting overdensity approaches zero. Consequently, the `velocity brake' (occurring when $\Delta < -1$) is sufficient to reverse the direction, effectively inducing a velocity kick. This kick expands the shell once more to a slightly larger value of a new physical radius, before it collapses again and eventually virializes.

Additionally, we can assume a brief period of time after the turnaround but before the shell-crossing, where we can still apply the negative branch of the Eq.~\eqref{drda_lscdm}. Thus, by integrating Eq.~\eqref{drda_lscdm} throughout the period where the scale factor ranges $a_{\rm ta}$ to $a_{\dagger}$, we obtain the following equation:
\begin{equation}\label{postdeltatas1}
    \begin{aligned}
    &\int_{u_{\dagger}}^{1}\dd u\sqrt{\frac{u}{a_{\rm ta}^{-3}(1+\delta_{\rm ta})(1-u) +\omega_{\rm s} u(1-u^2)}} \\
    =&\int_{1}^{y_{\dagger}} \dd y\sqrt{\frac{y}{a_{\rm ta}^{-3} - \omega_{\rm s} y^{3}+\xi_{\rm s} a_{\rm ta}^{-2}y}}\,.
    \end{aligned}
\end{equation}
We evaluate $u_{\dagger}$ using Eq.~\eqref{postdeltatas1} and after that we incorporate $\eta$, as calculated in Eq.~\eqref{posteta}.  

Subsequently, the value of $R_{\mathrm{vir}}$ is is ascertained in accordance with Eq.\eqref{rvir}, leading to the determination of the anticipated ratios, as discussed in Section~\ref{contrast}.

Assuming that $t_{\rm vir} \simeq 2\, t_{\rm ta}$ and that the collapse is completed when $t = t_{\rm vir}$, Eq.~\eqref{fr1} yields $(y_{\dagger} > 1$):
\begin{equation}\label{avircalc2}
    \begin{aligned}
    &\int_{0}^{y_{\dagger}}\dd y \sqrt{\frac{y }{a_{\text{ta}}^{-3}-\omega_{\rm s} y^{3}+\xi_{\rm s} a_{\text{ta}}^{-2}y}} \\
    +&\int_{y_{\dagger}}^{y_{\rm vir}} \dd y \sqrt{\frac{y }{a_{\text{ta}}^{-3}+\omega_{\rm s} y^{3}+\xi_{\rm s}a_{\text{ta}}^{-2}y}} \\
    =&2\int_{0}^{1}\dd y \sqrt{\frac{y }{a_{\text{ta}}^{-3}-\omega_{\rm s} y^{3}+\xi_{\rm s}a_{\text{ta}}^{-2}y}}\,,
    \end{aligned}
\end{equation}
from which we obtain:
\begin{equation}\label{avircalc20}
    \begin{aligned}
    &\int_{1}^{y_{\dagger}}\dd y \sqrt{\frac{y }{a_{\text{ta}}^{-3}-\omega_{\rm s} y^{3}+\xi_{\rm s}a_{\text{ta}}^{-2}y}} \\
    +&\int_{y_{\dagger}}^{y_{\rm vir}}\dd y \sqrt{\frac{y }{a_{\text{ta}}^{-3}+\omega_{\rm s} y^{3}+\xi_{\rm s} a_{\text{ta}}^{-2}y}} \\
    =&\int_{0}^{1}\dd y \sqrt{\frac{y }{a_{\text{ta}}^{-3}-\omega_{\rm s} y^{3}+\xi_{\rm s}a_{\text{ta}}^{-2}y}}\,.
    \end{aligned}
\end{equation}
It then becomes feasible to deduce the value of $a_{\rm vir}$ in terms of $a_{\rm ta}$, through Eq.~\eqref{avircalc} and subsequently $\delta_{\rm vir}$ from Eq.~\eqref{deltavir}.

\subsection{Contrasting \texorpdfstring{$\Lambda_{\rm s}$CDM}{LsCDM} with Standard \texorpdfstring{$\Lambda$CDM}{LCDM}: Insights into the Physical Outcomes}\label{contrast}

Based on Eqs.~\eqref{R_a_d} and~\eqref{rho_dm_de}, the physical radius of the overdensity at turnaround is expressed as:
\begin{equation}\label{rpta}
    R_{\rm p,ta}=\qty[\frac{3 M}{4 \pi (1+\delta_{\rm ta})\tilde{\rho}_{{\rm m}0}}]^{\frac{1}{3}}a_{\rm ta}\:.
\end{equation}
Furthermore, using the definition of $\eta$ from Eq.~\eqref{eta_cc}, we can write the virialized physical radius of the overdensity as:
\begin{equation}\label{rvir}
   R_{\rm p,vir} = (1+z_{\rm ta})^{-1} \qty[\frac{3 M}{4 \pi (1+\delta_{\rm ta})\tilde{\rho}_{\rm m0}}]^{\frac{1}{3}}\eta\,.
\end{equation}
Consequently, the ratio of the virialized matter density in the $\Lambda_{\rm s}$CDM model to that in the $\Lambda$CDM model~\cite{DODELSON20211,Planck:2018vyg} reads:
\begin{equation}
\label{rvirrat0}
    \frac{(\rho_{\rm vir})_{ \Lambda_{\rm s}}}{(\rho_{\rm vir})_{\Lambda}} \equiv \qty[\frac{(R_{\rm p,vir})_{\Lambda_{\rm s}}}{(R_{\rm p, vir})_{\Lambda}}]^{-3} = \frac{(1+\delta^{\Lambda_{\rm s}}_{\rm ta})}{(1+\delta^{\Lambda}_{\rm ta}) }\qty[\frac{\eta(e_{\Lambda_{\rm s}})}{\eta(e_{\Lambda})}]^{-3}\:.
\end{equation}
\begin{figure*}[t]
\centering
\includegraphics[width=0.46\textwidth]{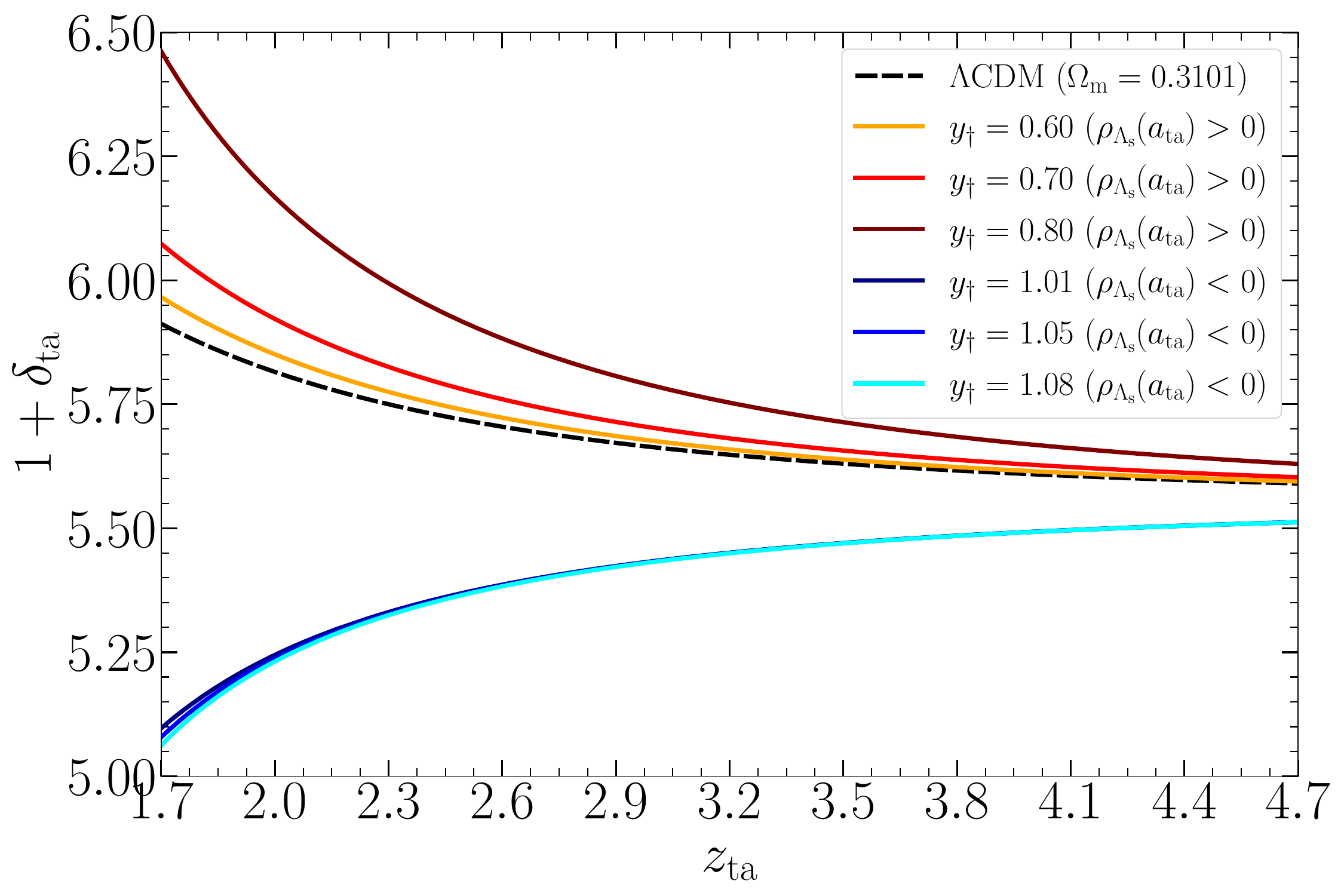}
\includegraphics[width=0.46\textwidth]{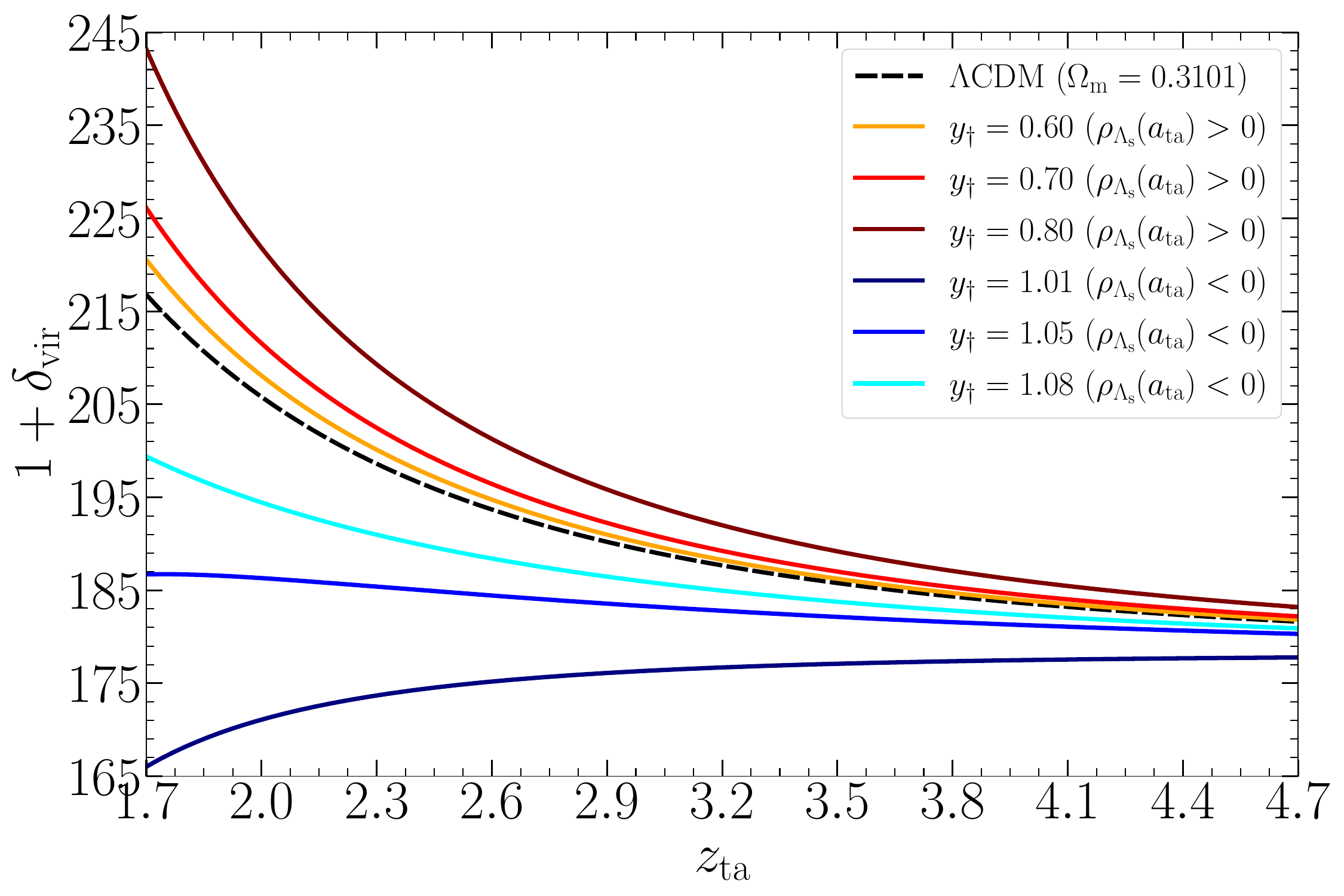}
\includegraphics[width=0.46\textwidth]{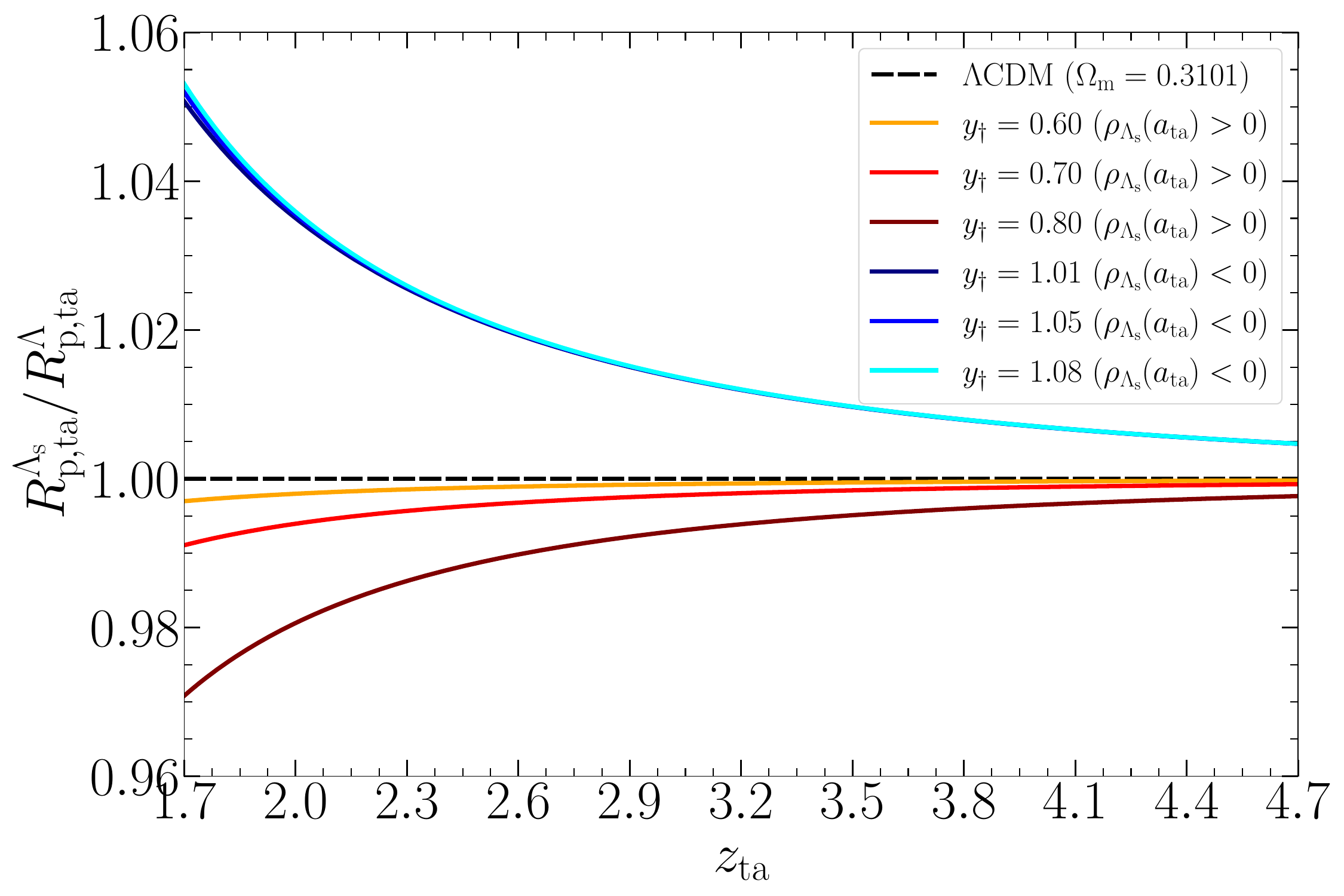}
\includegraphics[width=0.46\textwidth]{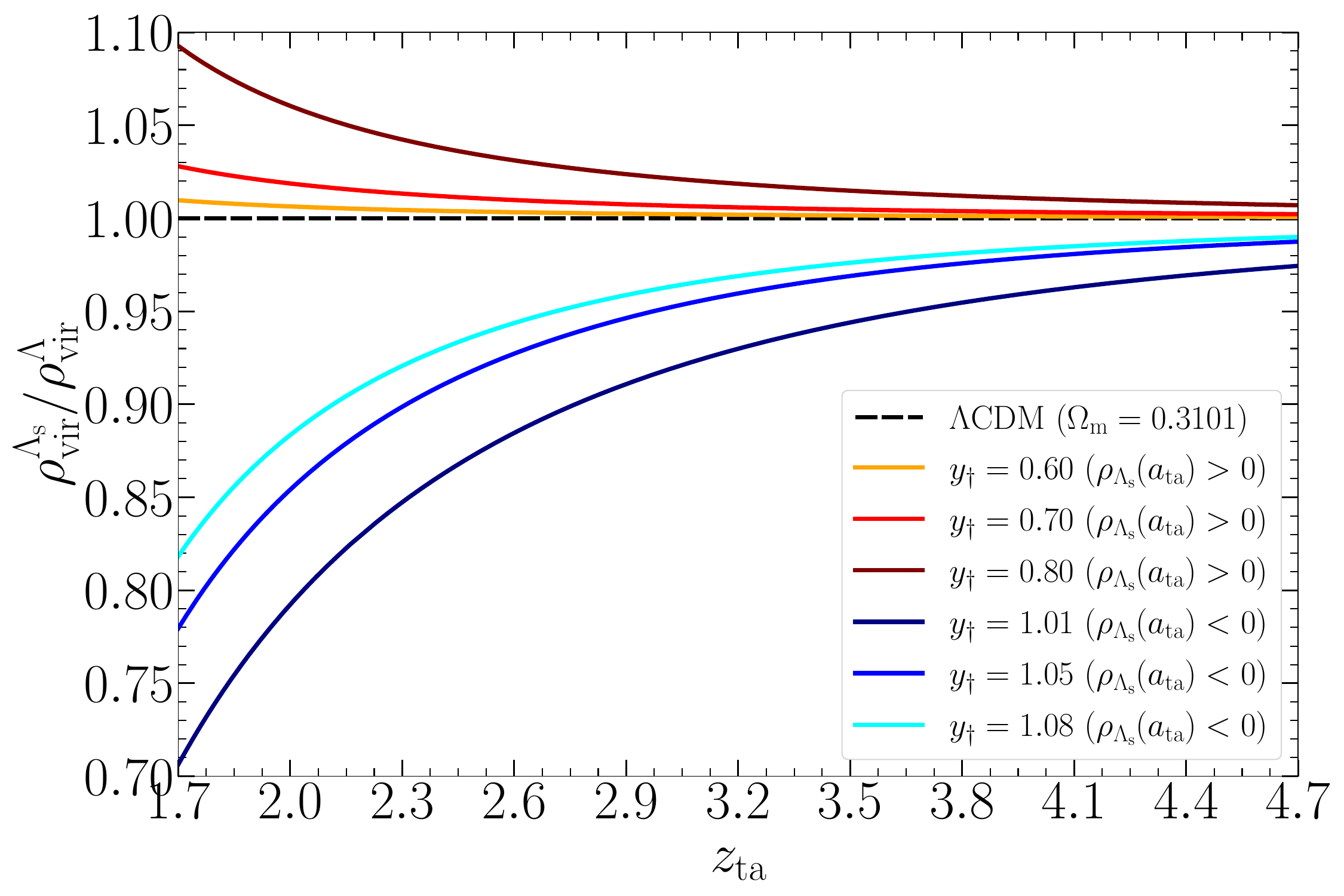}
\includegraphics[width=0.46\textwidth]{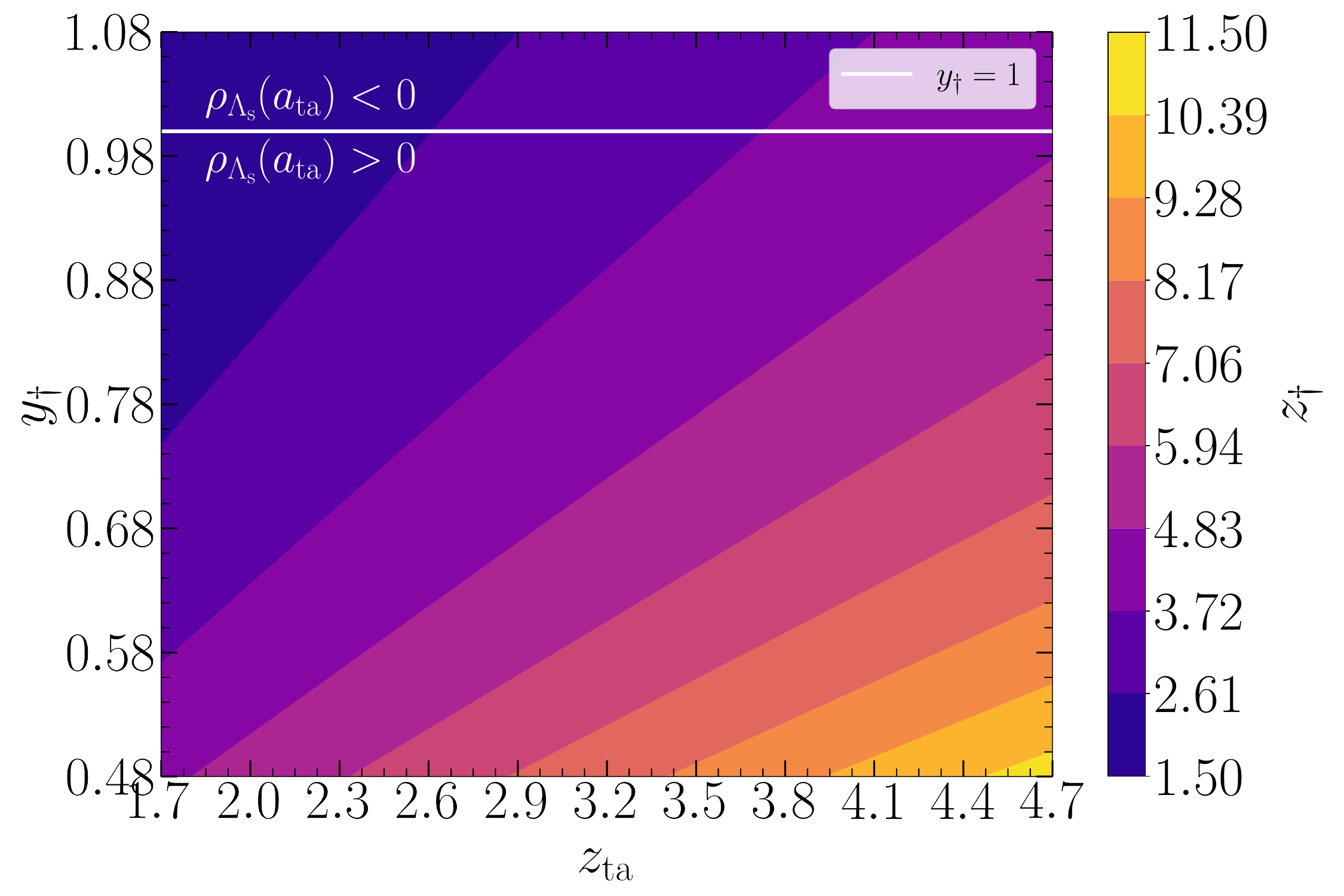}
\includegraphics[width=0.46\textwidth]{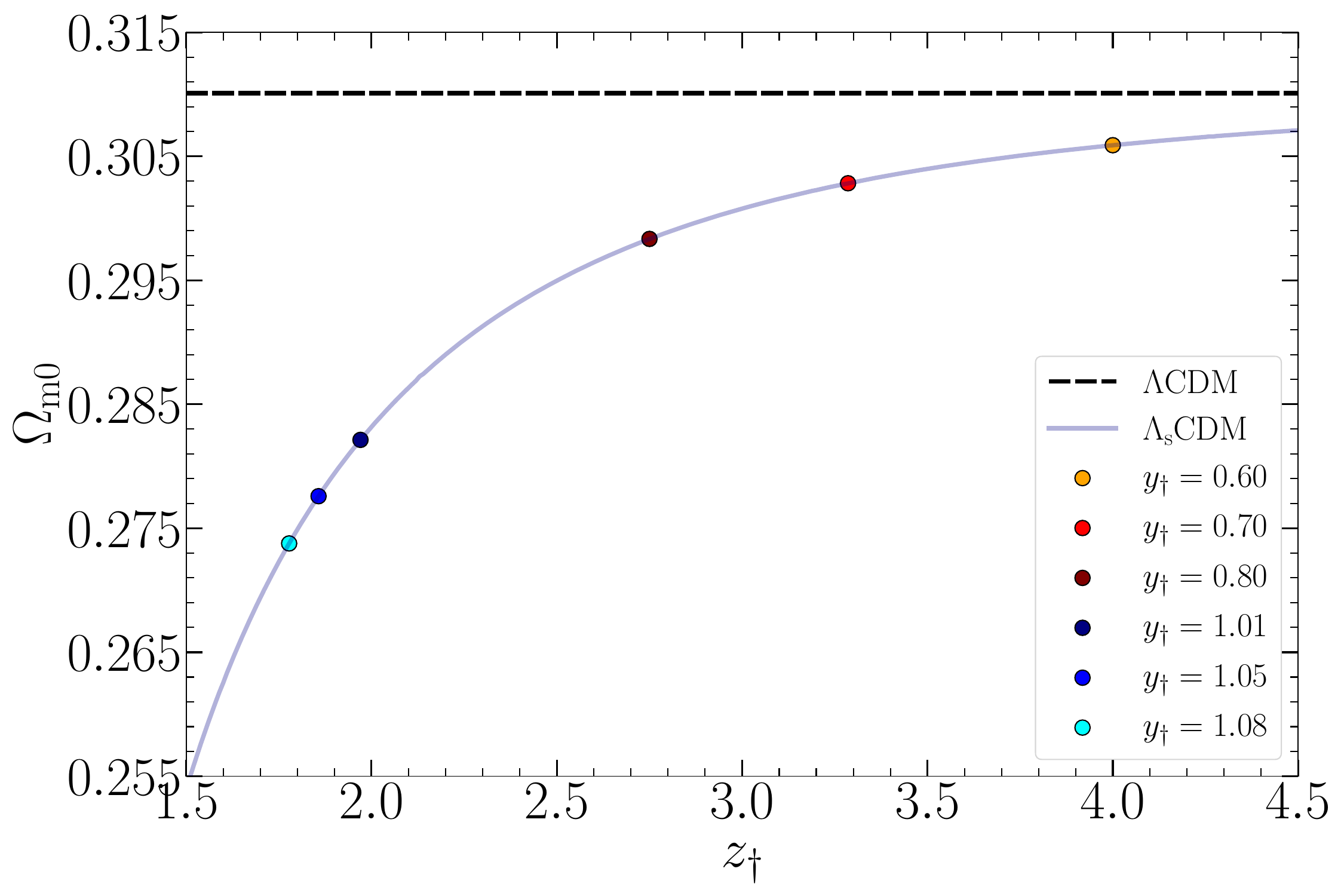}
\caption{Numerical analysis performed to calculate $\delta_{\rm ta}$, $\delta_{\rm vir}$, $\rho_{\rm vir}$, and $R_{\rm p, ta}$ for post-turnaround (shades of blue) and pre-turnaround (shades of red) cases, where the cosmological parameters are derived from the analysis given in the Appendix~\ref{app:deter_cosmo_param}. Each of the figure obtained by varying $1.7 \leq z_{\rm ta} \leq 4.7$, for constant $y_{\dagger} \equiv a_{\dagger} / a_{\rm ta}$. The corresponding $z_{\dagger}$ value for a given $z_{\rm ta}$ can be calculated via $1+z_{\dagger}=\qty(1+z_{\rm ta})/y_{\dagger}$. Even though we have used the same physical matter density parameter throughout the analysis, the $\Omega_{\rm m0}$ parameters will be different for both models. For a given $z_{\dagger}$, one can easily find the corresponding $\Omega_{\rm m0}$ for the $\Lambda_{\rm s}$CDM model (see Eq.~\eqref{eq:fitting}).}
\label{fig:main_results}
\end{figure*}
\begin{figure}[t]
\centering
\includegraphics[width=0.5\textwidth]{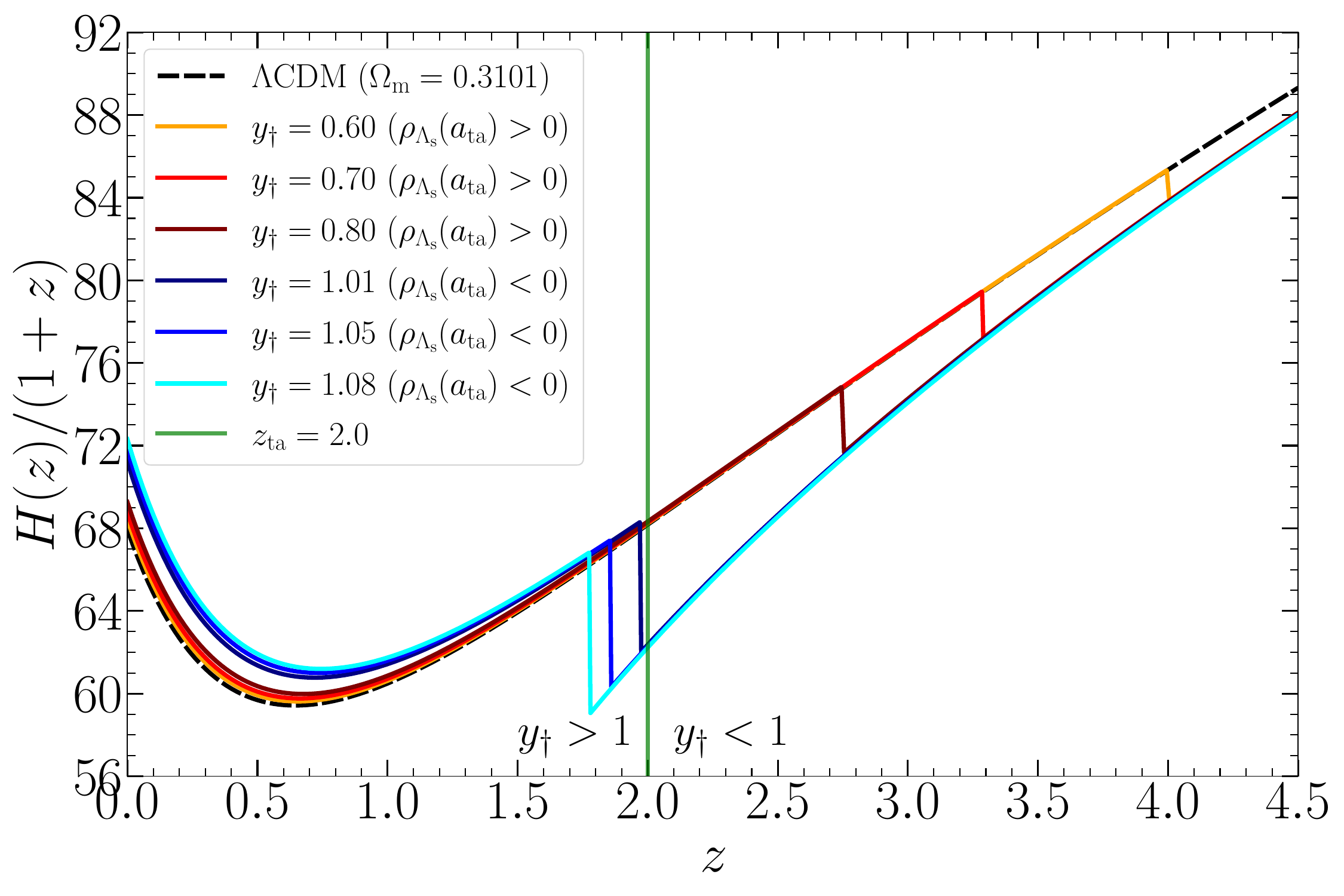}
\caption{$\dot{a} \equiv H(z) / (1+z)$ vs $z$ plotted for constant $y_{\dagger} \equiv a_{\dagger} / a_{\rm ta}$, by fixing the turnaround redshift at $2$ (i.e., $z_{\rm ta}=2$). We have applied that method described in the Appendix~\ref{app:deter_cosmo_param} to determine the cosmological parameters. For a fixed $a_{\rm ta}$, as $y_{\dagger} \rightarrow 0$, the scale factor of the transition approaches to $a_{\dagger} \rightarrow 0$ and the $\Lambda_{\rm s}$CDM model dynamics becomes similar to the $\Lambda$CDM. Meanwhile, as $y_{\dagger} \rightarrow 1/a_{\rm ta}$, the transition approaches today. Also note that, before the transition the expansion rate of the universe is smaller ($\dot{a}^{\Lambda_{\rm s}{\rm CDM}} < \dot{a}^{\Lambda{\rm CDM}}$) but after the transition it's faster ($\dot{a}^{\Lambda_{\rm s}{\rm CDM}} > \dot{a}^{\Lambda{\rm CDM}}$) than the $\Lambda$CDM universe.}
\label{fig:hz}
\end{figure}

The physical effects of the $\Lambda_{\rm s}$-sign switch (AdS-dS) transition can primarily be understood by considering the timing of the transition relative to the turnaround moment, distinguishing between the pre-turnaround ($z_\dagger>z_{\rm ta}$) and post-turnaround ($z_{\rm ta}<z_\dagger$) $\Lambda_{\rm s}$-sign switch transitions. These effects are demonstrated in Fig.~\ref{fig:main_results}, where, for a realistic assessment, the cosmological parameters for both models, namely, $\Lambda_{\rm s}$CDM (for various $z_\dagger$ cases) and $\Lambda$CDM, are chosen to ensure consistency with the Planck-CMB power spectra, as detailed in Appendix~\ref{app:deter_cosmo_param}.

\subsubsection{Pre-turnaround \texorpdfstring{$\Lambda_{\rm s}$}{Ls}-sign switch transition (\texorpdfstring{$z_\dagger>z_{\rm ta}$}{Ls})}

We first consider the case $y_{\dagger} < 1$, where the $\Lambda_{\rm s}$-sign switch transition occurs before the turnaround, i.e., $z_\dagger>z_{\rm ta}$. Our findings indicate that if this transition happens before turnaround, the density contrast at turnaround ($\delta_{\rm ta}$) will be higher than in the $\Lambda$CDM model, as shown in the top-left panel of Fig.~\ref{fig:main_results}. The rationale behind this is as follows: Because the transition has already occurred, the positive cosmological constant has already begun to influence the curvature of the halo as described by Eq.~\ref{kapparmax} ($\omega_{\rm s}>0$). On the other hand, as illustrated in Fig.~\ref{fig:hz} plotted, which is plotted by choosing $z_{\rm ta}=2$ (noting that different values of $z_{\rm ta}$ do not change the trends in the plots), a larger $y_\dagger<1$ results in the overdensity evolving under the negative cosmological constant's influence for a longer period. This implies that for larger $y_\dagger<1$ values, the negative cosmological constant's slowing-down effect on the overdensity's expansion, due to its induced gravitational attraction, lasts longer, leading to denser structures. Consequently, we observe generally higher $\delta_{\rm ta}$ values for the $\Lambda_{\rm s}$CDM model compared to the $\Lambda$CDM model in the top-left panel of Fig.~\ref{fig:main_results}, with this difference increasing for larger $y_\dagger$ values, as long as $y_\dagger<1$. We note that the expansion rate of the background universe, $\dot{a}=\frac{H(z)}{1+z}$, around $z_{\rm ta}$ is almost identical across various $y_\dagger$ values for a given $z_{\rm ta}$, as seen in Fig.~\ref{fig:hz} plotted by choosing $z_{\rm ta}=2$ as an example. Thus, the expansion rate of the universe at or around the turnaround moment does not intervene in and influence our discussions on the value of $\delta_{\rm ta}$ within the $\Lambda_{\rm s}$CDM framework. And of course, since $\delta_{\rm ta}$ and $R_{\rm p, ta}$ are interrelated through Eq.~\ref{rpta}, a higher $\delta_{\rm ta}$ corresponds to a smaller $R_{\rm p, ta}$, and vice versa, for a given value of $z_{\rm ta}$ or $a_{\rm ta}$. Finally, the collapsing phase of the overdensity proceeds under the influence of a positive cosmological constant until virialization, similar to the standard $\Lambda$CDM model. Therefore, evidently, the larger values of $\delta_{\rm ta}$ achieved in the $\Lambda_{\rm s}$CDM model, compared to $\Lambda$CDM model, give rise to corresponding larger values of $\delta_{\rm vir}$ and $\rho_{\rm vir}^{\Lambda_{\rm s}}$, as seen in the top- and middle-right panels of Fig.~\ref{fig:main_results}, respectively.

\subsubsection{Post-turnaround \texorpdfstring{$\Lambda_{\rm s}$}{Ls}-sign switch transition (\texorpdfstring{$z_{\rm ta}>z_\dagger$}{Ls})}

We now consider the case $y_{\dagger} > 1$, where the $\Lambda_{\rm s}$-sign switch transition occurs after the turnaround, i.e., $z_{\rm ta}>z_\dagger$, focusing on the phase during which halo is collapsing, but well before shell-crossing, when the halo is still homogeneous and isotropic. In this phase, the density contrast at the turnaround, $\delta_{\rm ta}$, will be lower compared to that in the $\Lambda$CDM model, with this effect more pronounced at lower values of $z_{\rm ta}$, as seen in the top-left panel of Fig.~\ref{fig:main_results}. The rationale behind this is as follows: In the $\Lambda$CDM model, the overdensity experiences a positive cosmological constant throughout its evolution. In contrast, in the post-turnaround $\Lambda_{\rm s}$-sign switch transition case of the $\Lambda_s$CDM model, the overdensity experiences a negative cosmological constant until and for some time afer reaching the turnaround radius. This implies that less matter energy density is required for the overdensity to achieve turnaround due to the enhancing gravitational attraction effects of the negative cosmological constant (in contrast to the positive cosmological constant), leading to less curvature for the overdensity. We note that the lower the $z_{\rm ta}$, the larger difference in $\delta_{\rm ta}$ between the $\Lambda_{\rm s}$CDM and $\Lambda$CDM models. This understandable, as at lower redshifts the cosmological constant is more dominant in both models, but it is negative in the $\Lambda_{\rm s}$CDM and positive in the $\Lambda$CDM model. Specifically, the lower the $z_{\rm ta}$, the higher the $\delta_{\rm ta}$ in the $\Lambda$CDM model, while the lower $z_{\rm ta}$ is, the lower the $\delta_{\rm ta}$ in the $\Lambda_{\rm s}$CDM model.

Note that $\delta_{\rm ta}$ is almost identical for a given $z_{\rm ta}$ value, with only a barely visible differences at lower $z_{\rm ta}$ values for various $y_\dagger$ values. This explained by the fact that for $z>z_{\rm ta}$, the expansion rate of the universe for different $y_\dagger$ values is almost the same, allowing the overdensities to evolve through nearly identical background universe dynamics until turnaround is achieved, as seen in Fig.~\ref{fig:hz} plotted by choosing $z_{\rm ta}=2$ as an example. Note that for different $y_\dagger>1$ values, we have used the same physical matter density value and fixed the angular size of the sound horizon at the last-scattering surface to ensure consistency with CMB-Planck spectra (see Appendix~\ref{app:deter_cosmo_param}), resulting in slightly different values for $\Omega_{\rm m0}$, corresponding to slightly different $H_0$ values. This implies, for a given $z_{\rm ta}$, slightly different matter density parameters/expansion rates of the universe at the times when the over density is evolving towards the turnaround, and the time taken to reach the turnaround radius would be slightly different. The $\Omega_{\rm m0}$ can be read from the bottom panels of Fig.~\ref{fig:main_results}; for $y_\dagger=\{1.01, 1.05, 1.08\}$ values we used, we have $z_\dagger=\{1.97, 1.86, 1.78\}$ and $\Omega_{\rm m0}=\{0.2821, 0.2777, 0.2739\}$. In other words, $\dot{a}$ is almost identical across all $\Lambda_{\rm s}$CDM cases for $y_\dagger>1$ for $z>z_{\rm ta}$ and thus is not expected cause significant variations in $\delta_{\rm ta}$ for different $y_\dagger>1$ values. In these cases, the overdensity experiences the negative cosmological constant for a nearly identical duration, see Fig.~\ref{fig:hz}. The minor variations in $\delta_{\rm ta}$ are attributed to slightly different values of $\Omega_{\rm m0}$, implying slightly different expansion rates of the universe and hence a slightly different passage of time taken until turnaround is achieved. In the middle-left panel of Fig.~\ref{fig:hz}, we plot the ratio of the physical radius of the halo at the turnaround in the $\Lambda_{\rm s}$CDM model to that in the $\Lambda$CDM model, viz., $R_{\rm p,ta}^{\Lambda_{\rm s}} / R_{\rm p,ta}^{\Lambda}$.\footnote{The results are independent of the halo's mass, as demonstrated in Eq.~\eqref{rpta}.} We observe that $R_{\rm p,ta}^{\Lambda_{\rm s}} / R_{\rm p,ta}^{\Lambda}>1$, being larger for lower values of $z_{\rm ta}$, and is almost the identical for different $y_\dagger>1$ values with a barely visible difference for small $z_{\rm ta}$ values. This aligns with expectations when considering Eq.(\ref{kapparmax}) for $\omega\to -\omega_{\rm s}$, implying the higher values of $R_{\rm p, ta}$ correspond to lower values of $\delta_{\rm ta}$.


Finally, in the top-right panel of Fig.~\ref{fig:main_results}, we plot the density contrast at the moment of virialization, $\delta_{\rm vir}$, for the cases where the $\Lambda_{\rm s}$-sign switch transition occurs during the collapsing phase, before the halo virializes. It is conceivable that the $\delta_{\rm vir}$ values in the $\Lambda_{\rm s}$CDM are lower than in the $\Lambda$CDM model, similar to the situation with $\delta_{\rm ta}$. However, we immediately see that, although this expectation holds, unlike with the situation for $\delta_{\rm ta}$, the $\delta_{\rm vir}$ values differ significantly for different $y_\dagger$ values and become more pronounced for lower $z_{\rm ta}$ values. This phenomenon is not surprising and can roughly be explained as follows: As seen in Fig.~\ref{fig:hz} (plotted by choosing $z_{\rm ta}=2$ as an example), the larger the value of $y_\dagger$ the greater the difference between $z_\dagger$ and $z_{\rm ta}$, implying that $\Lambda_{\rm s}$-sign switch transition occurs at later in the collapsing phase of the overdensity for the larger $y_\dagger>1$ values. That is, given that $\delta_{\rm ta}$ is almost the same for a given $z_{\rm ta}$ for different values of $y_\dagger$, for larger $y_\dagger$ values, the $\Lambda_{\rm s}$-sign switch transition occurs when the overdensity is more compact, thereby it is conceivable that the rapid increase in the universe's expansion rate at the transition will have less influence on the collapsing overdensity for larger $y_\dagger>1$ values. A more precise, but also more concise, explanation is as follows: We observe that the impact of the singularity at $z_\dagger$ on the overdensity is considerable when $y_{\dagger} \simeq 1$ and diminishes for values further from 1 or for larger $z_{\rm ta}$ values. The singularity extracts kinetic energy (i.e., when $-1 \leq \Delta < 0$), leading to a 'velocity brake' in the collapse. Conversely, when $\Delta < -1$, it induces a velocity kick, reversing the direction of the collapse and re-expanding the shell to a slightly larger new physical radius before its eventual collapse and virialization. This effect, observable in Figure~\ref{fig:delta0}, suggests that higher $\Delta$ values result in lower virialized densities in the halo.

\section{The Impact of Background Expansion on Bound Systems}\label{bsystemst}
\subsection{Newtonian approximation of a bound system in an expanding background}

Numerous studies have been devoted to the exploring the impacts of cosmic expansion on bound systems~\cite{Einstein:1945id,Nesseris:2004uj,Nandra:2011ug,Nandra:2011ui,Bouhmadi-Lopez:2014jfa,Antoniou:2016obw,Perivolaropoulos:2016nhp}. In the Newtonian limit, the vicinity of a point mass $M$ situated within a dynamically expanding background is characterized as follows:
\begin{equation}\label{metric}
\begin{aligned}
      \dd s^2&=-\left(1-\frac{2GM}{a\,\chi}\right)\dd t^2\\
      &+a^2\qty[\dd \chi^2+\chi^2(\dd \theta^2+\sin^2\theta\, \dd \phi^2)]\,,  
\end{aligned}
\end{equation}
where $t$ is the comoving time and $a\equiv a(t)$. Interpreting the gravitational field of the point mass as a \textit{weak field}, and by admitting \textit{low velocities}, the geodesic equations can be differentiated with respect to the coordinate time $t$, as:
\begin{equation}
\ddot{x}^{\mu}+\Gamma^{\mu}_{\nu\sigma}\dot{x}^{\nu}\dot{x}^{\sigma}=0\,,
\end{equation}
where $\,\dot{}\equiv\dv*{t}$. For the geodesic equation in the $\chi$-coordinate, considering $\theta = \pi / 2$ in the equatorial plane, we derive:
\begin{equation}\label{chic}
\frac{G M}{\chi(t)^2 a(t)^3} - \chi(t)  \dot{\phi}(t)^2 + \frac{2 \dot{a}(t) \dot{\chi}(t)}{a(t)} + \ddot{\chi}(t) = 0\,.
\end{equation}
Similarly, the geodesic equation for $x^3$, taking $\theta=\pi / 2$ again, yields:
\begin{equation}
    \frac{2 \dot{a}(t)  \dot{\phi}(t)}{a(t)}  + \frac{2 \dot{\phi}(t) \dot{\chi}(t)}{\chi} + \ddot{\phi}(t) = 0\implies \dv{t}\left[(a\chi)^2 \dot{\phi}\right]=0\,.
\end{equation}
Defining $L = (a\chi)^2 \dot{\phi}$ as the angular momentum per unit mass, we express the geodesic equation from Eq.~\eqref{chic} in terms of the physical radial coordinate $r=a\chi$ as follows:
\begin{equation}\label{geodesiceq}
    \Ddot{r}-\frac{\Ddot{a}}{a}r-\frac{L^2}{r^3}+\frac{GM}{r^2}=0\,.
\end{equation}
Considering a moment in time, $t_{\rm 0}$, where expansion can be disregarded (i.e., $\dot{r}(t_{\rm 0}) = 0$), we define the physical radius of an orbit as $r_0 \equiv r(t_{\rm 0})$. The angular speed at this moment, neglecting expansion, is described as:
\begin{equation}\label{eqphi}
   \eval{\dot{\phi}^2}_{t=t_{\rm 0}} = \frac{L^2}{r_0^4} \equiv \omega_0^2 \equiv \frac{GM}{r_0^3}\,.
\end{equation}
Given the condition outlined in Eq.~\eqref{eqphi}, this leads to the conclusion that $L^2 = GMr_0$. While a circular orbit with a constant physical radius does not exist at all times, $r_0$ represents the radius at a specific instant. Under the condition that the initial angular speed is significant enough to overlook cosmic expansion, it approximates a stable Newtonian circular orbit, assuming $L^2 \neq 0$ and $\dot{r} = 0$ at all times.

\begin{figure}[t]
\centering
    \includegraphics[width=0.5\textwidth]{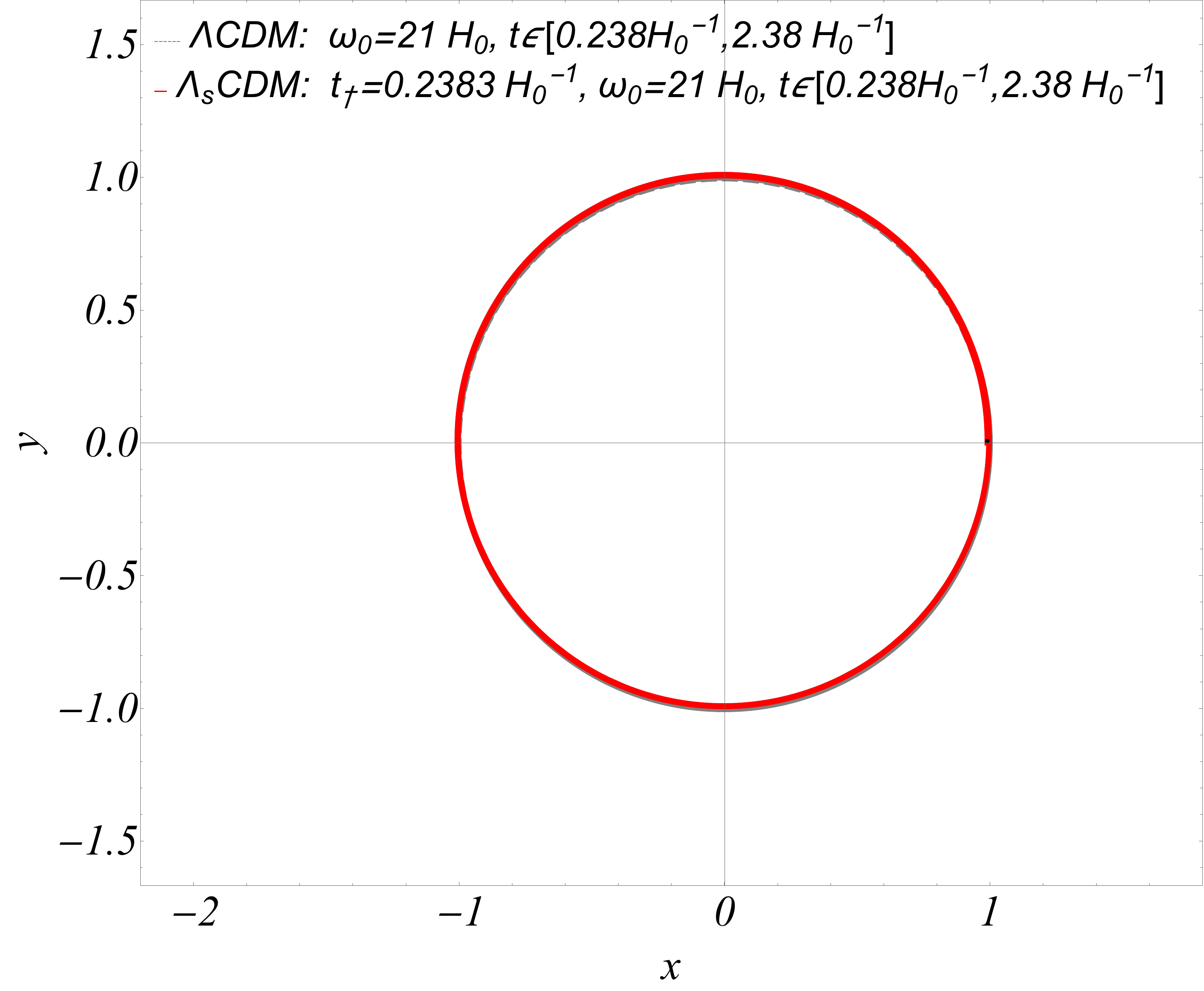}
    \includegraphics[width=0.5\textwidth]{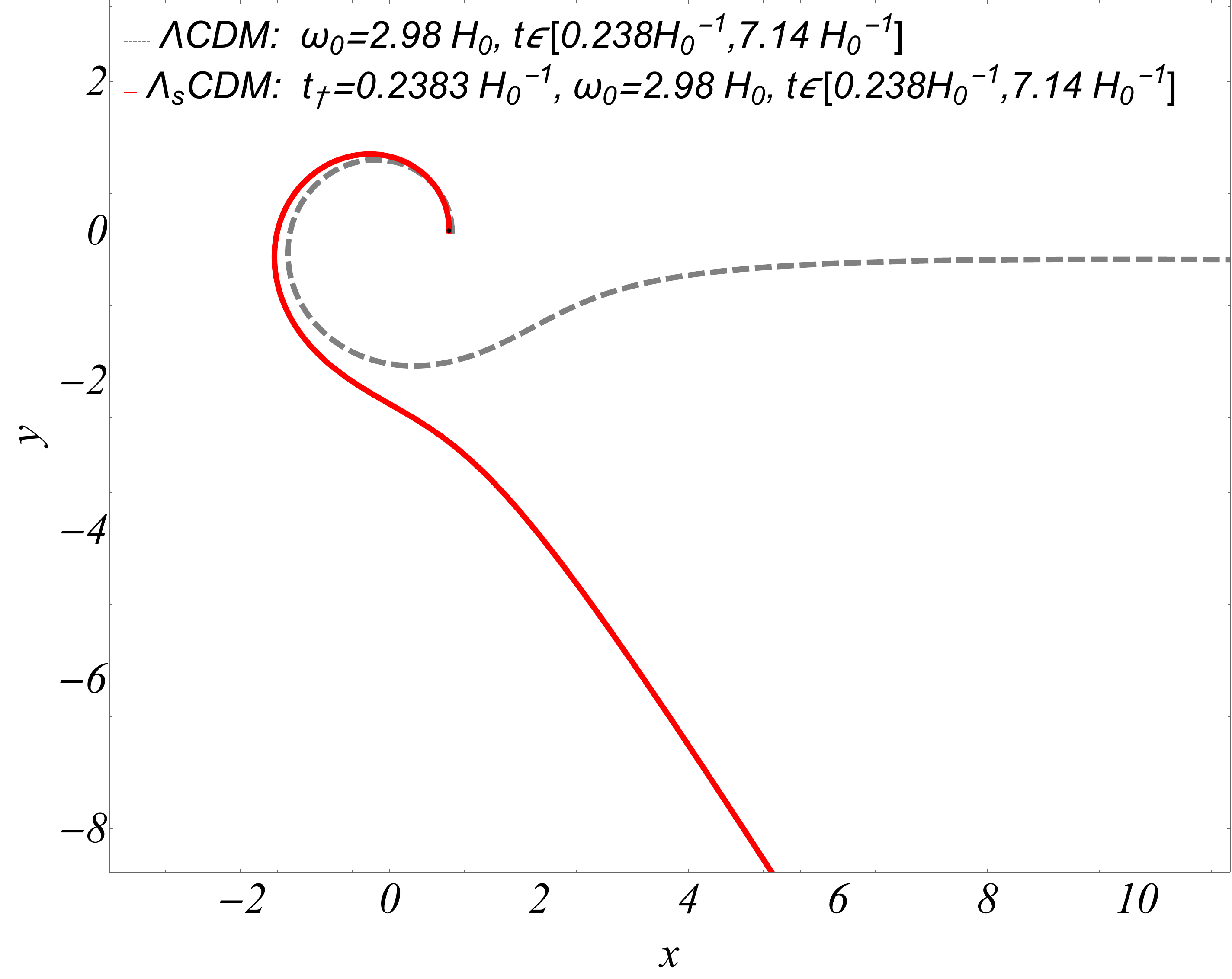}
\caption{A particle in orbit, embedded in $\Lambda$CDM (gray) and $\Lambda_{\rm s}$CDM (red), with $t_{\dagger} = 0.2383 H_{0}^{-1}$ for $\Omega_{\rm m0} = 0.29$, and $t_{\rm init} = 0.238 H_{0}^{-1}$. The initial angular velocity set as $\omega_{0} = 21 H_{0}$ (top-panel) and $\omega_{0} = 2.98 H_{0}$ (bottom-panel) (See also Table~\ref{tab:scale}). We initiate a circular orbit against an expanding background i.e. with a rescaled radius $\bar{r}(t_{\rm init}) = s$ and $\dot{r}(t_{\rm init})=0$. The $s$ is determined by solving Eq.~\eqref{quasistable}. Subsequently, we numerically solve Eq.~\eqref{rescaledg}, in each case. The \textit{black dot} represents the position of the particle in orbit in the rescaled $x-y$ plane at the moment the singularity.}
\label{fig_b01}
\end{figure}

\begin{figure}
\centering
    \includegraphics[width=0.5\textwidth]{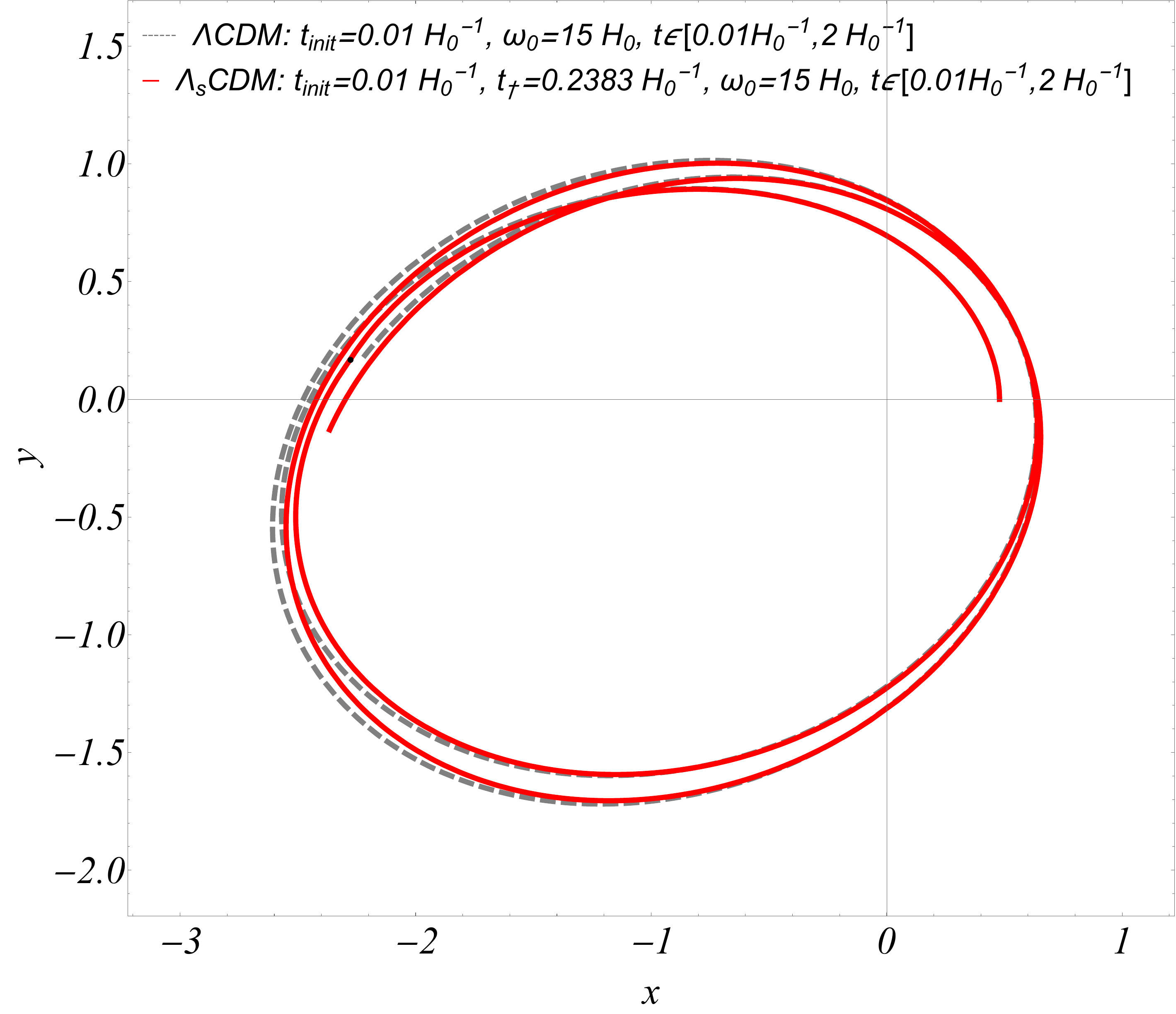}
    \includegraphics[width=0.5\textwidth]{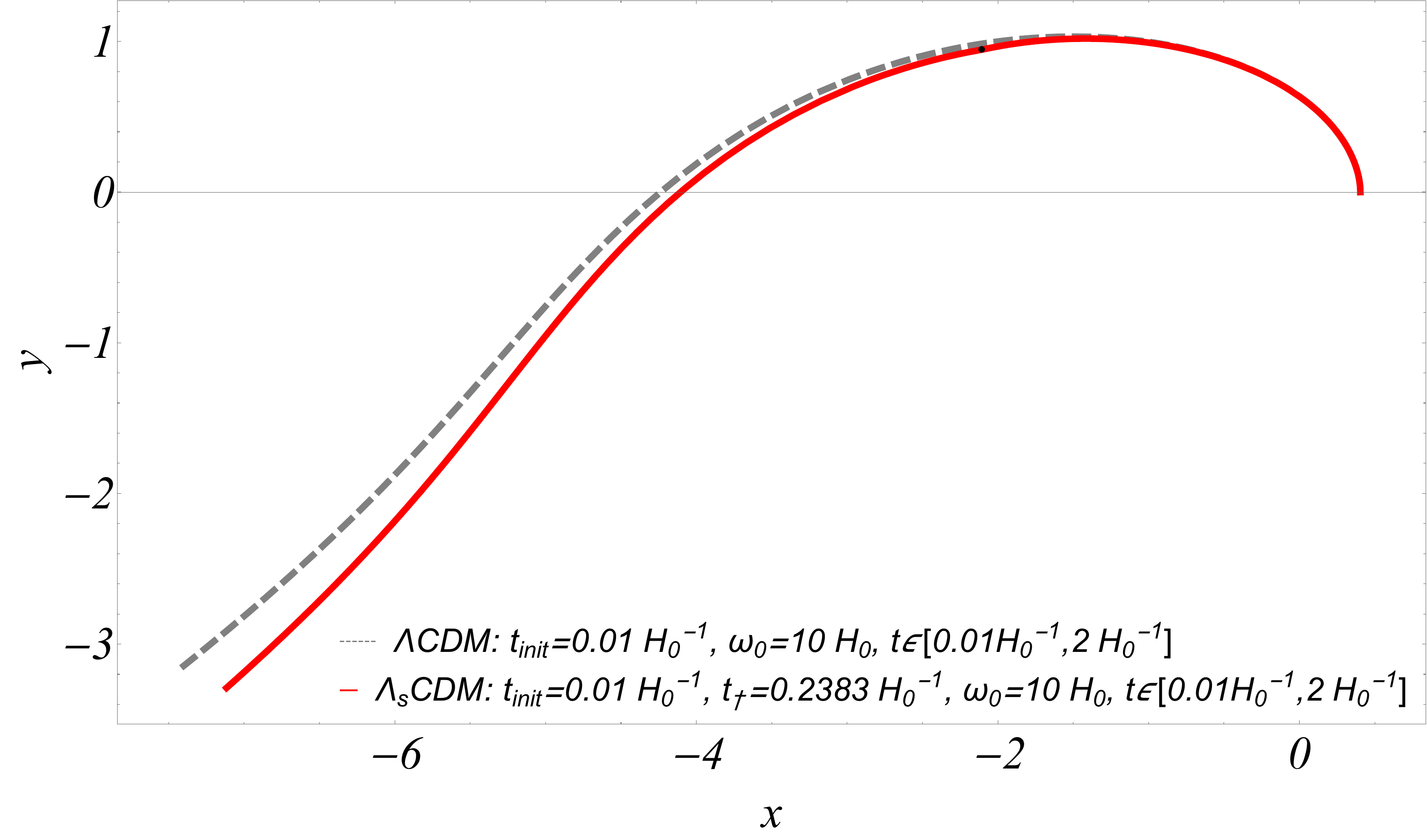}
\caption{A particle in orbit, embedded in $\Lambda$CDM (gray) and $\Lambda_{\rm s}$CDM (red), with $t_{\dagger} = 0.2383 H_{0}^{-1}$ for $\Omega_{\rm m0} = 0.29$ and $t_{\rm init} = 0.01 H_{0}^{-1}$. The initial angular velocity is set as $\omega_{0} = 15 H_{0}$ (top-panel) and $\omega_{0} = 10 H_{0}$ (bottom-panel) (See also Table \ref{tab:scale}). We initiate a circular orbit against an expanding background i.e. with a rescaled radius $\bar{r}(t_{\rm init}) = s$ and $\dot{r}(t_{\rm init})=0$. The $s$ is determined by solving Eq.~\eqref{quasistable}. Subsequently, we numerically solve Eq.~\eqref{rescaledg}, in each case. The \textit{black dot} represents the position of the particle in orbit in the rescaled $x-y$ plane at the moment the singularity.}
\label{fig_b}
\end{figure}

Additionally, considering a rescaling $\dv*{t} = t_{\rm init}^{-1}\dv*{\bar{t}}$ and $'\equiv\dv*{\bar{t}}$ (where $t_{\rm init}$ is the initial cosmic time selected for the system), we derive from Eq.~\eqref{geodesiceq}:
\begin{equation}
\label{rescaledg}
    \bar{r}'' = \frac{\bar{\omega}_0^2}{\bar{r}^3} - \frac{\bar{\omega}_0^2}{\bar{r}^2} + \frac{a''}{a} \bar{r}\,,
\end{equation}
where
\begin{equation}
    \begin{aligned}
    \frac{a''}{a}&=-\frac{1}{2} (H_0 t_{\text{init}})^2 \Omega_{{\rm m}0} \Big[a^{-3} - 2 \omega_{\rm s} a^{-1}\delta\qty(1/a_{\dagger}-1/a) \\
    &-2\omega_{\rm s}{\rm sgn}\qty(1/a_{\dagger}-1/a)\Big]\,,
    \end{aligned}
\end{equation}
and
\begin{equation}
  \bar{r} \equiv \frac{r}{r_0}, \quad \bar{\omega}_0 \equiv \omega_0 t_{\text{init}}, \quad \bar{t} \equiv \frac{t}{t_{\text{init}}}, \quad \dot{\phi}\equiv\frac{\bar{\omega}_0}{\bar{r}^2}\:.
\end{equation}
The final step before proceeding to solve Eq.~\eqref{rescaledg}, involves initiating from an orbit with a rescaled radius, $\bar{r} (t_{\rm init}) = s$, taking into account the expanding background. This is achieved by setting the right-hand side of Eq.\eqref{rescaledg} to zero and solving for $s$:
\begin{equation}\label{quasistable}
    \bar{\omega}_{0}^2(1-s) + \frac{a''}{a}\bigg{|}_{t_{\rm init}}s^4 = 0\,.
\end{equation}
Subsequently, we numerically solve the geodesic equation, as presented in Eq.~\eqref{rescaledg}, for a bound system. Numerical examples of bound orbits within the $\Lambda_{\rm s}$CDM mode are demonstrated in Fig.~\ref{fig_b01} and Fig~\ref{fig_b}. 

\begin{table}[t!]
\caption{Clusters of galaxies form massive, virialized structures. Nevertheless, this is not the case for super-clusters. Assuming an initial angular speed defined by $\omega_0^2 = GM / r_0^3$, we derive some very rough typical scales for both galaxy clusters and super-clusters
\cite{Kravtsov:2012zs, Gao:2019tfj,Harvey:2020gsy,Fernandez:2021jlm,Sankhyayan:2023tii}.}
\begin{ruledtabular}
\begin{tabular}{lcccr} 
\textbf{System} & $M [M_{\odot}]$ & $R[\rm Mpc]$ &  $T_{0}\,[H_0^{-1}]$ &  $\omega_{0}\,[H_0]$ \\
\colrule
Cluster & $\sim 10^{15}$ & $\sim 2$ &  $\sim 1$ & $\sim 10$ \\
Supercluster & $\sim 10^{16}$ & $\sim 100$ & $\sim 100$ & $\sim 0.01$ \\
\end{tabular}
\end{ruledtabular}
\label{tab:scale}
\end{table}

\subsection{Physical outcomes}

We consider a point particle with a path parameterized by the proper time $\tau$. In this setting, the $x^0$-geodesic equation is:
\begin{equation}\label{x0coord}
\begin{aligned}
    \dv{\tau}\qty[\dv{t}{\tau}\qty(1-\frac{2 GM}{ a\chi})] &= \\
    -H(a\chi) \dv{t}{\tau}&\qty[\frac{L^2}{(a\chi)^3} + \frac{1}{a\chi}\qty(a\dv{\chi}{\tau})^2]\,.  
\end{aligned}
\end{equation}
For the first integral of a timelike geodesic, described by $\dd s^2 = -\dd \tau^2$ and assuming $\theta = \pi / 2$, we redefine the metric from Eq.~\eqref{metric} as:
\begin{equation}\label{metricconst}
    -1 = -\qty(1-\frac{2 GM}{ a\chi})\qty(\dv{t}{\tau})^2 + a^2\qty(\dv{\chi}{\tau})^2 + \frac{L^2}{(a\chi)^2} \,.
\end{equation}
Assuming a sufficiently small time interval where every cosmic bound system maintains a physical radius $r = a\chi$ and adheres to the approximation $H(a\chi) \ll 1$, particle in orbit with this radius will conform to this approximation. If we also assume a small peculiar velocity, the second term in Eq.~\eqref{x0coord} is deemed higher order, simplifying the equation to:
\begin{equation}\label{ener}
    \dv{t}{\tau}\qty(1-\frac{2 GM}{ a\chi}) = k\equiv\rm const.
\end{equation}
Upon integrating Eq.~\eqref{ener} into Eq.~\eqref{metricconst}, we derive:
\begin{equation}
    -1 = -\qty(1+\frac{2GM}{ a\chi})k^2 + a^2\qty(\dv{\chi}{\tau})^2 + \frac{L^2}{(a\chi)^2}\,.
\end{equation}
In the regime of small peculiar velocities ($\dd t\simeq\dd \tau$), this results in an expression for the quasi-energy:
\begin{equation}\label{energynb}
E \equiv \frac{k^2-1}{2} = \frac{1}{2}\dot{r}^2-\frac{GM}{r}  - Hr\dot{r} + \frac{1}{2}H^2r^2 + \frac{L^2}{2r^2}\,.
\end{equation}
Consider a particle initially moving in an orbit at $r(t_{\rm init}) = s\, r_0$ (with $\dot{r}(t_{\rm init}) = 0$) in a bound system. During a short time interval around $t_{\dagger}$, if the approximations $\bar{r}(t_{\rm init}) \simeq 1$ and $\dot{\bar{r}} \simeq 0$ are valid, then Eq.~\eqref{energynb} simplifies to:
\begin{equation}\label{energyeqcr}
  E=\frac{1}{2}(H+\delta H)^2r_{0}^2+\frac{(\omega_{0}\,r_{0})^2}{2}-(\omega_{0}\,r_{0})^2 \,.
\end{equation}
Given these approximations, consider any particle in a  system with mass $M$, that initiates its orbit at a time $t_{\rm init}$, where $t_{\rm init} \in (t_{\dagger} - T, t_{\dagger} )$ and $T$ is sufficiently small. The particle has an initial angular momentum $L$, and $\dot{r}(t_{\rm init}) = 0$. Also, the particle's initial physical radius is approximately $r(t_{\rm init}) \simeq r_0\simeq r(t_{\dagger})$. We define the \textit{critical angular speed}, $\omega_{\rm crit}$, at the moment $t_{\dagger}$, \textit{iff} the total energy $E = 0$, according to the equation
\begin{equation}\label{energy02}
  \omega_{\rm crit}=H+\delta H \,.
\end{equation}
Given that the singularity is relatively weak, it alone cannot dissociate any bound system before the continual expansion of the universe does. Thus, the dissociation of bound orbits is primarily driven by cosmic expansion, with the singularity inducing a minor perturbation that slightly increases the Hubble value at a specific instant. This perturbation contributes minimally to the dissociation of a bound orbit. Notably, the timing of initiating a bound orbit significantly influences its evolution, as demonstrated in Figure~\ref{fig_b01}, where we initiate an orbit near the moment the singularity. We calculate the critical angular speed, $\omega_{\rm crit}$, to be approximately $3$ for $\Lambda_{\rm s}$CDM and $2.9$ for $\Lambda$CDM.

Given the brief period when the cosmological constant is negative in the $\Lambda_{\rm s}$CDM compared to its positive phase, variations in orbits relative to those equivalent orbits in the $\Lambda$CDM model suggest a scenario in the $\Lambda_{\rm s}$CDM model where orbits with sufficiently low angular speed may dissociate, unlike their counterparts in the $\Lambda$CDM model. However, this dynamic changes when orbits are initiated far in the past, away from the singularity. In such cases, the negative cosmological constant in the $\Lambda_{\rm s}$CDM model aids in maintaining the orbit's binding over a significant period, preventing the singularity from weakening the gravitational attraction enough to cause more dissociation than in the $\Lambda$CDM model.

The critical initial angular speed, $\omega_{\mathrm{crit}}$, approximated from Eq.~\eqref{energy02}, acts as a threshold. For bound systems with an angular speed $\omega_{0}$---the speed in a Newtonian bound system disregarding cosmic expansion---dissociation due to the cosmic expansion occurs if and only if $\omega_{\mathrm{crit}} > \omega_{0}$. This approximation holds mainly for orbits beginning at moments near $t_{\dagger}$. The results from Eq.~\eqref{energy02} align well with those obtained numerically from the geodesic equation Eq.~\eqref{geodesiceq}, as shown in Fig.~\ref{fig_b01}. However, this method does not accurately approximate numerical results for orbits starting well before the moment under study; such an example is shown in Fig.~\ref{fig_b}.

\section{Conclusion}

The $\Lambda_{\rm s}$CDM model~\cite{Akarsu:2019hmw,Akarsu:2021fol,Akarsu:2022typ,Akarsu:2023mfb} has emerged as a highly promising approach to addressing major cosmological tensions within the standard $\Lambda$CDM model and its canonical extensions, such as the $H_0$ and $S_8$ tensions. This achievement is realized through the introduction of a single parameter, $z_{\dagger}$, to the six-parameter base $\Lambda$CDM model. This parameter determines the timing of an abrupt AdS-dS transition, changing from $\Lambda_{\rm s}=-\Lambda_{\rm s0}$ for $z>z_\dagger$ to its late-time positive value $\Lambda_{\rm s}=\Lambda_{\rm s0}$ for $z<z_\dagger$. Examining the implications of adopting the $\Lambda_{\rm s}$CDM model framework for the universe's evolution, especially on the formation and evolution of bound cosmic structures, is crucial. The switch to the $\Lambda_{\rm s}$CDM model is anticipated to impact bound cosmic structures for three primary reasons: (i) the negative cosmological constant (AdS) phase for $z > z_\dagger \sim 2$, (ii) the abrupt transition itself, marked by a sudden jump in the universe's expansion rate—a type II (sudden) singularity—at $z=z_\dagger$, and (iii) the increased expansion rate compared to the $\Lambda$CDM model for $z < z_\dagger$. Despite the faster expansion rate, the $\Lambda_{\rm s}$CDM and $\Lambda$CDM models are otherwise identical for $z < z_\dagger$. All these aspects warrant thorough investigation as their potential effects can be used to test the unique predictions of the model. In this paper, we analyze the non-linear evolution of a spherical overdensity within the $\Lambda_{\rm s}$CDM cosmology. We begin by revisiting the dynamics of spherical collapse within the $\Lambda$CDM framework and then integrate the physical effects of the AdS-dS transition into the spherical collapse model. This integration is accomplished by adjusting the Friedmann equations for the $\Lambda_{\rm s}$CDM model. Furthermore, using energy considerations, we make predictions about the eventual state of the halo, dependent on the timing of the transition relative to the turnaround. Specifically:
\begin{itemize}[nosep,wide]
    \item If the turnaround occurs after the transition, in what we refer to as the \textit{pre-turnaround transition}, halos that form and undergo this transition exhibit virialized overdensities exceeding those predicted by the Planck/$\Lambda$CDM model, particularly at lower turnaround redshifts. This observation can be attributed to the increased $\delta_{\mathrm{ta}}$ values resulting from a period with a negative cosmological constant, which facilitates the formation of denser structures. In both the $\Lambda$CDM and $\Lambda_{\mathrm{s}}$CDM models under pre-turnaround transition scenarios, the expansion of the universe reaches the turnaround radius with a positive cosmological constant. This results in greater curvature and matter overdensity at the turnaround radius compared to the post-turnaround case.
    \item If the turnaround occurs before the transition, in what we refer to as \textit{post-turnaround transition}, halos experiencing this transition typically achieve virialization at lower overdensities compared to those predicted by Planck/$\Lambda$CDM. This observation can be attributed to the negative cosmological constant at the turnaround, necessitating a lower matter overdensity at this moment, which results in reduced curvature at turnaround. Consequently, in this scenario, overdensities attain a larger maximum physical radius owing to their diminished matter overdensity and curvature.
\end{itemize}

In the abrupt $\Lambda_{\rm s}$CDM model, we have observed that the Hubble parameter displays a discontinuity at a specific past redshift, $z = z_{\dagger}$. This discontinuity leads to a type II (sudden) singularity at $z = z_{\dagger}$, as discussed in Appendix \ref{app:sudden_cosmo_sing} and supported by~\cite{Barrow:2004xh}. Despite its mild nature, this singularity imparts a velocity kick to particles. We have shown that smoothing out the abrupt behavior effectively eliminates this singularity, even when the transition occurs very rapidly. Hence, our findings regarding the velocity kick pertain to the extreme version of the $\Lambda_{\rm s}$CDM model. However, we demonstrate that, even in this case, we ascertain that the singularity's impact on Newtonian bound virialized systems is minimal, thereby not threatening the model's viability in this context. Notably, the singularity, being relatively weak, does not lead to the dissociation of large bound systems before this is done by the universe's continuous background expansion. For instance, large clusters or super-clusters, corresponding to $\omega_0 < 10 H_{0}$, will be dissociated by the background expansion but remain practically unaffected by the singularity (as illustrated in Fig.~\ref{fig_b01}-\ref{fig_b}). The expansion in both the $\Lambda_{\rm s}$CDM and $\Lambda$CDM models tends to induce dissociation of bound systems at scales of large clusters and above at similar levels. Therefore, the presence of unbound orbits is primarily driven by universal expansion, with the singularity merely causing an increase in the Hubble expansion rate at a specific moment. Interestingly, the negative cosmological constant in the $\Lambda_{\rm s}$CDM model tends to enhance the stability of bound systems due to its attractive gravity effects.

The outcomes of our analysis open intriguing avenues for future research. Broadening the scope, a natural extension could involve generalizing the spherical collapse model to accommodate a more diverse range of sudden cosmological singularities. Another promising direction is the investigation of the impact of the $\Lambda$-sign switch transition on gravitational waves traversing the sudden cosmological singularity, as discussed in~\cite{Paraskevas:2023aae}. Moreover, delving into the physical mechanisms that induce the sign switch of the cosmological constant, as explored in Refs.~\cite{Anchordoqui:2023woo,Alexandre:2023nmh,Akarsu:2024qsi}, remains a significant area of interest. The $\Lambda_{\mathrm{s}}$CDM model, with its potential to address the Hubble and $S_8$ tensions, may also influence early universe structure growth due to its period of negative cosmological constant. Our current study assumes a uniform transition in a homogeneous universe. However, slight inhomogeneities could lead to timing variations in this transition, as suggested in \cite{Akarsu:2022typ}. Such variations might result in different regions of the universe experiencing the cosmological constant's sign switch at distinct redshifts, with potential implications for galaxy formation. This scenario could encompass sudden singularities of varying intensities and even halos formed entirely under a negative cosmological constant. Recent observations from the JWST hint at more intense early galaxy formation, potentially aligning with our model's implications. Although our results show only minor deviations from the $\Lambda$CDM model for structures formed at higher redshifts, they underscore the necessity for further exploration. Particularly, the prospect of halos forming entirely under a negative cosmological constant could present a different narrative and warrants detailed investigation.

\section*{Data Availability Statement}

The Mathematica (v12) and python files used for the production of the figures and for derivation of the main results of the analysis can be found at \href{https://github.com/camarman/transition-dynamics-lscdm}{camarman/transition-dynamics-lscdm} GitHub repository under the MIT license.

\begin{acknowledgments}
\"{O}.A. acknowledges the support by the Turkish Academy of Sciences in scheme of the Outstanding Young Scientist Award (T\"{U}BA-GEB\.{I}P). \"{O}.A. and A.\c{C}. are supported in part by TUBITAK Grant No.~122F124. This research was supported by COST Action CA21136 - Addressing observational tensions in cosmology with systematics and fundamental physics (CosmoVerse), supported by COST (European Cooperation in Science and Technology). This project was also supported by the Hellenic Foundation for Research and Innovation (H.F.R.I.), under the ``First call for H.F.R.I. Research Projects to support Faculty members and Researchers and the procurement of high-cost research equipment Grant" (Project Number: 789).
\end{acknowledgments}

\appendix

\section{Demonstration of Type II Singularity}
\label{app:sudden_cosmo_sing}

We have studied the effects an abrupt transition described by the signum function. This description leads to a Type II (sudden) singularity, characterized by:

\begin{equation}
    t = t_{\dagger}, \quad a = a_{\dagger} < \infty, \quad \rho_{\mathrm{tot}}(a_{\dagger}) < \infty, \quad |P_{\mathrm{tot}}(a_{\dagger})| \rightarrow \infty,
\end{equation}
with the following characteristics: the scale factor, $a(t)$, is continuous and non-zero at the moment $t_{\dagger}$; the first derivative of the scale factor, $\dot{a}$, is discontinuous at $t_{\dagger}$; and its second derivative $\ddot{a}$ diverges at $t_{\dagger}$~\cite{Barrow:2004xh}. We prove this argument, by considering that $\ddot{a} / {a} =\dot{H}+H^2$, and by implementing Friedmann equation 
\begin{equation} H^2=\frac{8\pi G}{3}\tilde{\rho}_{\rm m0} \left[a^{-3}+\omega_{\rm s}\,\text{sgn}\qty(1/a_{\dagger}-1/a) \right],\end{equation}
we obtain:
\begin{align}\label{ddota}
    \frac{\ddot{a}}{a}&=-\frac{4\pi G}{3}\rho_{\rm m0}\Big[a^{-3}-2\omega_{\rm s}{\rm sgn}(1/a_{\dagger} - 1/a) \\ 
    &-2\omega_{\rm s}a^{-1}\delta(1/a_{\dagger} - 1/a)\Big].\nonumber
\end{align}
Where $\delta$ represents the Dirac delta function. It is evident that at the precise moment of the singularity, $\frac{\ddot{a}}{a}\to \infty$. Given that, $P_{\rm tot}= -\frac{1}{8\pi G} \left[ 2\frac{\ddot{a}}{a} + \left( \frac{\dot{a}}{a} \right)^2 \right]$, at the moment of transition, $P_{\rm tot}\to -\infty$. This is a feature that can arise in universes with any spatial curvature.
Additionally, by resembling a smooth transition, the energy density of $\Lambda_{\rm s}$ will be written as follows:
\begin{equation}
   \rho_{\Lambda_{\rm s}}(a) = \rho_{\Lambda_{\rm s}0} \frac{\tanh\qty[\sigma(a-a_{\dagger})]}{\tanh\qty[\sigma(1-a_{\dagger})]}\,,
\end{equation}
for $\rho_{\Lambda_{\rm s}0}>0$, where $a_\dagger<1$ and $\sigma>0$\footnote{One must set $\sigma \gg 1$ to approximate the signum function.} being a parameter determining the rapidity of the transition.
\begin{figure}[tbp]
\centering
\includegraphics[width = 0.45\textwidth]{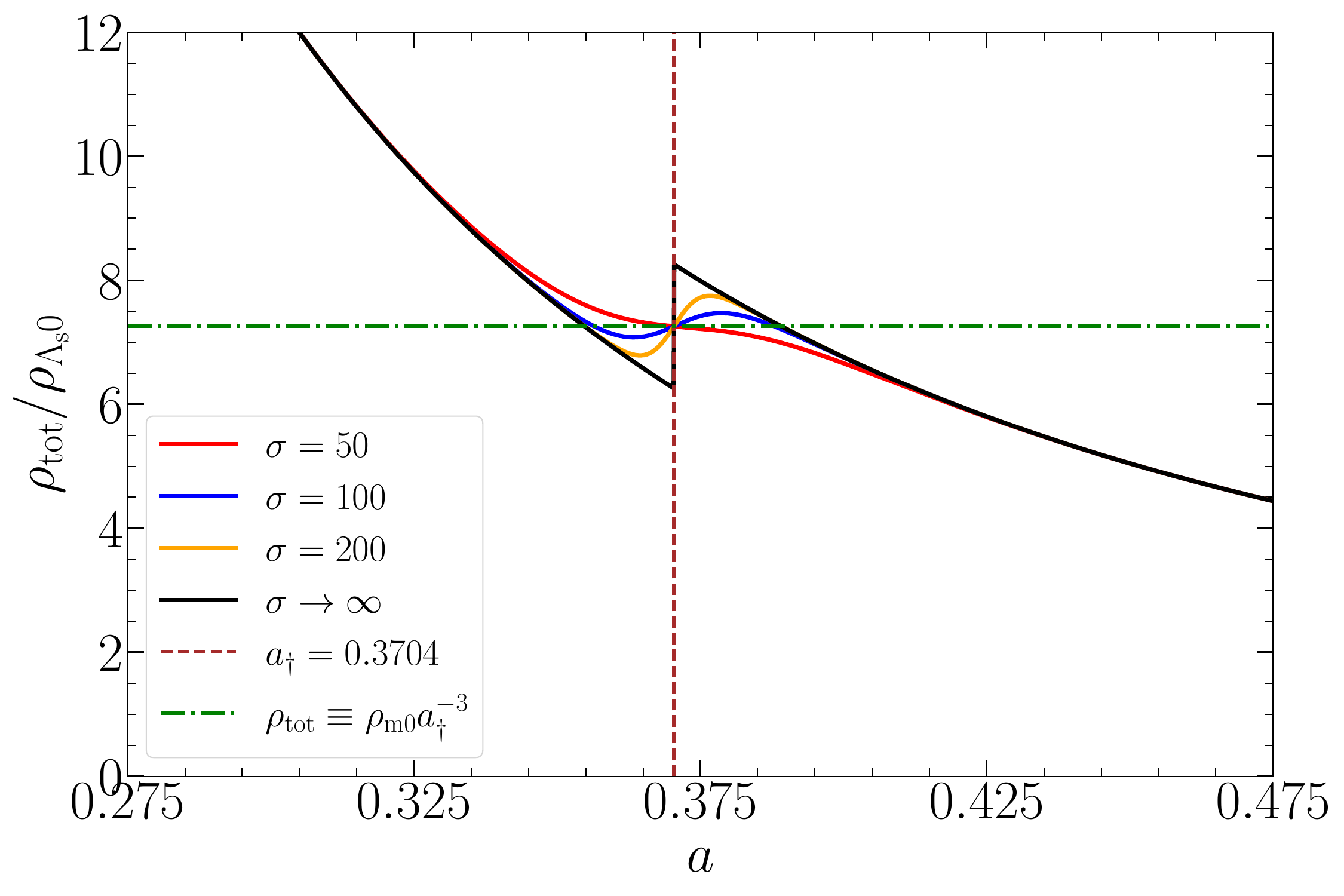}
\includegraphics[width = 0.45\textwidth]{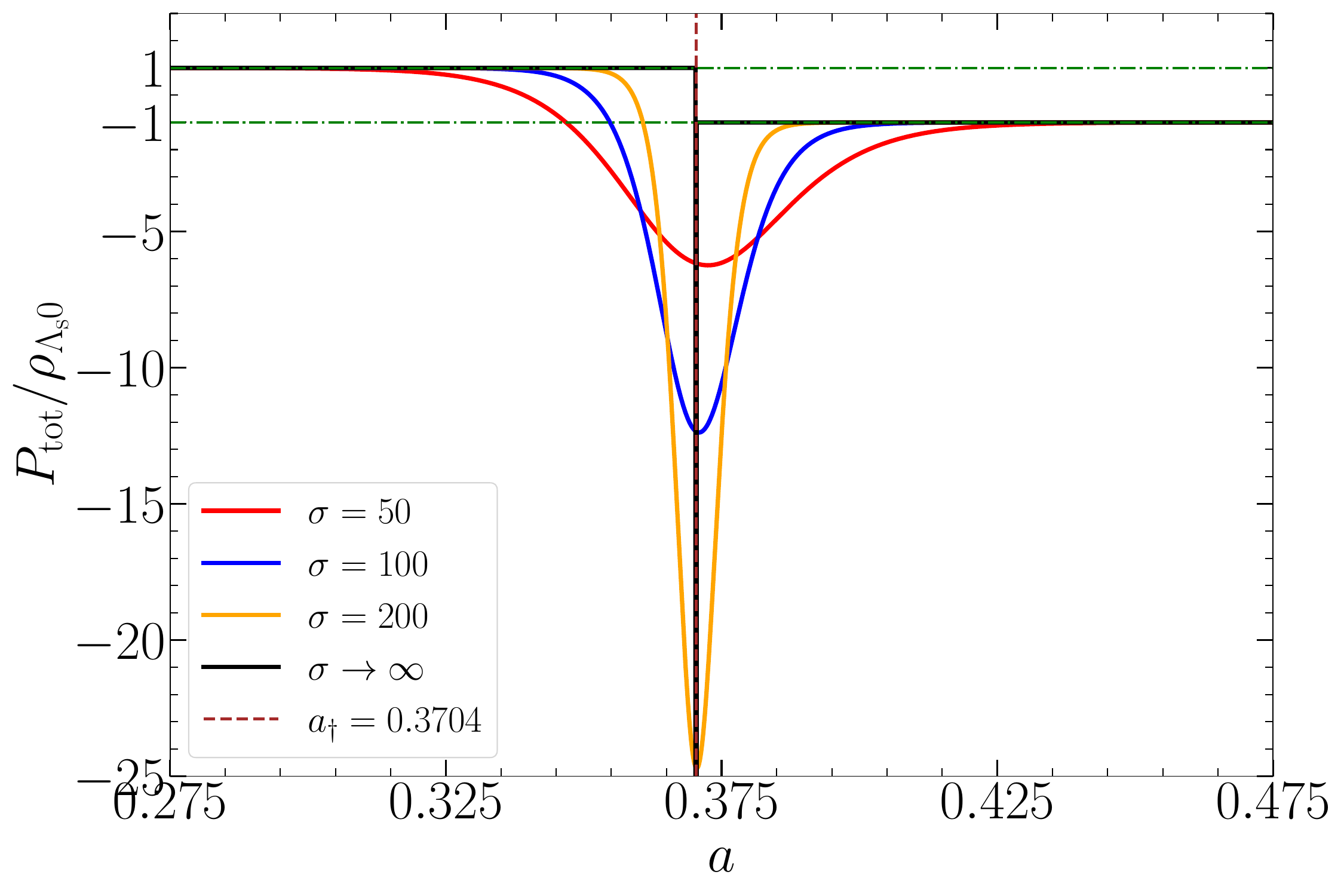}
\caption{Total density and total pressure of the $\Lambda_{\rm s}$CDM universe (see Eq.~\eqref{eq:smooth_lscdm}) with respect to scale factor. For $\sigma \rightarrow \infty$, we observe that $\rho_{\rm tot}(a_{\dagger}) \rightarrow {\rm const.} >0$ while $P_{\rm tot}(a_{\dagger}) \rightarrow -\infty$.}
\label{fig:pressure}
\end{figure}
Under this parametrization, the total energy density and total pressure of the universe, containing only dust and $\Lambda_{\rm s}$, can be written as:
\begin{equation}
    \begin{aligned}
    \label{eq:smooth_lscdm}
    \rho_{\rm tot}(a) &= \rho_{\rm m0}a^{-3} + \rho_{\Lambda_{\rm s}0} \frac{\tanh\qty[\sigma(a-a_{\dagger})]}{\tanh\qty[\sigma(1-a_{\dagger})]}\,, \\
    P_{\rm tot}(a) &= -\rho_{\Lambda_{\rm s}0}\qty[\frac{\tanh\qty[\sigma(a-a_{\dagger})]}{\tanh\qty[\sigma(1-a_{\dagger})]} + \sigma\frac{a}{3}\frac{\sech^2\qty[\sigma(a-a_{\dagger})]}{\tanh\qty[\sigma(1-a_{\dagger})]}]\,.
\end{aligned}
\end{equation}
Upon examining the characteristics of $\rho_{\rm tot}(a)$ and $P_{\rm tot}(a)$ at $a=a_{\dagger}$, we find:
\begin{equation}\label{rhodeensmooth}
    \begin{aligned}
    \rho_{\rm tot}(a_{\dagger}) &= \rho_{\rm m0}a_{\dagger}^{-3}\,, \\
    P_{\rm tot}(a_{\dagger}) &= -\rho_{\Lambda_{\rm s}0}\frac{a}{3}\frac{\sigma}{\tanh\qty[\sigma(1-a_{\dagger})]}\,.
\end{aligned}
\end{equation}
Notice that $\rho_{\rm tot}(a_{\dagger})$ does not depend on $\sigma$, and $P_{\rm tot}(a_{\dagger})$ is negative but finite for finite values of $\sigma$.

The ``smoothed-out energy density of $\Lambda_{\rm s}$'' ( Eq.(\ref{rhodeensmooth})) reduces, by taking $\sigma \rightarrow \infty$, to an abrupt (sudden) AdS $\rightarrow$ dS transition, which we have studied in the current paper:
\begin{equation}
    \lim_{\sigma \rightarrow \infty}\rho_{\Lambda_{\rm s}}(a) = \rho_{\Lambda_{\rm s}0} {\rm sgn}[a-a_{\dagger}]\,.
\end{equation}
In this case, we observe that the total pressure diverges to negative infinity:
\begin{equation}
    \lim_{\sigma \rightarrow \infty}P_{\rm tot}(a_{\dagger}) = -\infty\,,
\end{equation}
while the total energy density remains positive and finite. Thus, this behavior, which occurs at the limit of $\sigma \rightarrow \infty$, is characterized by a type II (sudden) cosmological singularity~\cite{Barrow:2004xh,Nojiri:2005sx}. Note that at the moment when the singularity occurs, the equation of state parameter $w_{\Lambda_{\mathrm{s}}}$ is undefined; this is a feature of sudden singularities. But even for finite values of $\sigma$, i.e., by smoothing out $\rho_{\Lambda_{\mathrm{s}}}$ (ensuring continuity at $\rho_{\Lambda_{\mathrm{s}}}$, which implies it should obtain the zero value at the moment of the transition), where a sudden cosmological singularity won't occur, even then, given that we have assumed the continuity equation $\dot{\rho}+3H(\rho+p)=0$ together with an equation of state, we conclude that the right and left limits of the equation of state parameter at the moment of transition will be $w^{(\pm)}\to \pm\infty$. Therefore, the equation of state parameter cannot be defined even for finite $\sigma$ (see \cite{Ozulker:2022slu}).


\section{Gravitational potentials for dark matter and dark energy}
\label{app:grav_potential_de_dm}

The Poisson equation for barotropic fluid, described by $P = w \rho$, is expressed as:
\begin{equation}\label{Poiss}
    \nabla^2 \Phi = 4\pi G \rho(1+3 w)\:.
\end{equation}
In a theoretical framework where a homogeneous energy density is spherically distributed within a radius $R$, and the gravitational potential is denoted as $\Phi\equiv\Phi(r)$, the general solution is applied:
\begin{equation}\label{grav_poten}
    \Phi(r)=2 \pi G \rho (1+3w)\frac{r^2}{3}-\frac{\mathcal{C}}{r}+\mathcal{D}\:.
\end{equation}
The gravitational potential of dark matter, $\Phi_{\text{DM}}$, is determined by assuming $\rho(r>R)=0$. As a result, the general solution is reformulated as:
\begin{equation}\label{gensol}
    \begin{aligned}
    \Phi_{\rm I}(r>R) &= 2 \pi G \rho (1+3w)\frac{r^2}{3}-\frac{\mathcal{C}}{r}+\mathcal{D}\:, \\
    \Phi_{\rm II}(r<R) &= \frac{\mathcal{C}}{r}+\mathcal{D}\:.
    \end{aligned}
\end{equation}
The boundary conditions are defined as follows: (a) $\Phi_{\rm I}(r=0)<\infty$, (b) $\Phi_{\rm II}(r\to \infty)=0$, (c) $\Phi_{\rm I}(r=R)=\Phi_{\rm II}(r=R)$, and (d) $\eval{\dv*{\Phi_{\rm I}}{r}}_{r=R}=\eval{\dv*{\Phi_{\rm II}}{r}}_{r=R}$. Applying the general solution from Eq.~\eqref{gensol} for dark matter density $\rho_{\text{DM}}$ (assuming $w_{\rm DM}=0$) within a sphere of radius $R$, we obtain:
\begin{equation}\label{potentialdm}
\Phi_{\rm DM}(r)=\begin{cases}
    -2\pi G \rho_{\rm DM} \qty(R^2-r^2/3) & r\leq R \\ \\
    -4\pi G \rho_{\rm DM} R^3/3r & r\geq R
    \end{cases}
\end{equation}
Meanwhile, the gravitational potential of dark energy, $\Phi_{\text{DE}}$, is determined by imposing uniform energy density across the universe as boundary conditions: (a) $\Phi(r=0)<\infty$ and (b) $\Phi(r=0)=0$. Thus, applying these conditions to Eq.~\eqref{grav_poten}, we deduce:
\begin{equation}\label{potentialde}
   \Phi_{\text{DE}}(r)= 2 \pi G \rho_{\text{DE}} (1+3w_{\rm DE})\frac{r^2}{3}\,.
\end{equation}

\section{Scale factor of the \texorpdfstring{$\Lambda_{\rm s}$CDM}{LsCDM} model}
\label{app:scale_factor_lscdm}

To elucidate the scale factor in terms of comoving time, we employ the Friedmann equation, articulated as:
\begin{equation}\label{friedmann}
    \frac{H^2}{H_0^2}=\Omega_{\rm m0}a^{-3}+\Omega_{\Lambda_{\rm s}0} \text{sgn}\qty(1/a_{\dagger}-1/a)\:.
\end{equation}

The scale factor maintains continuity and can be characterized by integrating Eq.~\eqref{friedmann}, as demonstrated below:
\begin{equation}
    \dot{a}=H_{0}\qty[\Omega_{\rm m0}a^{-1}+\Omega_{\Lambda_{\rm s}0}a^{2} \text{sgn}\qty(1/a_{\dagger}-1/a)]^{\frac{1}{2}}\:.
\end{equation}
In particular for $a<a_{\dagger}$ the integration is given by:
\begin{equation}\label{aminus}
    \int_{0}^{a} \frac{\dd a}{\qty(\Omega_{\rm m0}a^{-1}-\Omega_{\Lambda_{\rm s}0}a^{2})^{\frac{1}{2}}}=H_0 t\:.
\end{equation}
By changing variable $a=y^{\frac{2}{3}}\qty(\frac{1-\Omega_{\Lambda_{\rm s}0}}{\Omega_{\Lambda_{\rm s}0}})^{\frac{1}{3}}$ and given that $\dd a=\frac{2}{3}\qty(\frac{1-\Omega_{\Lambda_{\rm s}0}}{\Omega_{\Lambda_{\rm s}0}})^{\frac{1}{3}}y^{-\frac{1}{3}}\dd y$, Eq.~\eqref{aminus} implies:
\begin{equation}
    a(t)=\qty(\frac{1-\Omega_{\Lambda_{\rm s}0}}{\Omega_{\Lambda_{\rm s}0}})^{\frac{1}{3}}\sin^{\frac{2}{3}}\left(\frac{3}{2}\sqrt{\Omega_{\Lambda_{\rm s}0}}H_0 t\right)\,.
\end{equation}
Subsequently, if $a>a_{\dagger}$, then the integration proceeds as:
\begin{equation}\label{aplus}
    \begin{aligned}
    &\int_{0}^{a_{\dagger}} \frac{\dd a}{\qty(\Omega_{{\rm m}0}a^{-1}-\Omega_{\Lambda_{\rm s}0}a^2)^{\frac{1}{2}}} \\
    +&\int_{a_{\dagger}}^{a} \frac{\dd a}{\qty(\Omega_{{\rm m}0}a^{-1}+\Omega_{\Lambda_{\rm s}0}a^2)^{\frac{1}{2}}}=H_0 t\,.
    \end{aligned}
\end{equation}
In that case, Eq.~\eqref{aplus} implies:
\begin{equation}\label{scalefactor}
    \begin{aligned}
    a(t)=&\qty(\frac{1-\Omega_{\Lambda_{\rm s}0}}{\Omega_{\Lambda_{\rm s}0}})^{\frac{1}{3}} \\
    \times&\sinh^{\frac{2}{3}}\Bigg[\frac{3}{2}\sqrt{\Omega_{\Lambda_{\rm s}0}}H_0 (t-t_{\dagger}) \\
    +&\sinh^{-1}\Big[\sin\big(\frac{3}{2}\sqrt{\Omega_{\Lambda_{\rm s}0}}H_0 \, t_{\dagger}\big)\Big]\Bigg]\,.
    \end{aligned}
\end{equation}
 Subsequently, we get:
\begin{equation}
    \frac{\ddot{a}}{a} = 
    \begin{cases} 
        -\frac{4}{9}\mathcal{A}_{1}^{2}\left[1+\frac{1}{1-\cos\left(2 \mathcal{A}_{1} t\right)}\right] & t < t_{\dagger} \\ \\
       -\frac{4}{9}\mathcal{A}_{1}^{2}\left[-2+\frac{1}{\sinh\qty[\mathcal{A}_{1} (t-t_{\dagger})+\mathcal{A}_{2}]}\right] & t > t_{\dagger}
    \end{cases}
\end{equation}
where we have denoted: 
\begin{align*}
    \mathcal{A}_{1} &\equiv \frac{3}{2}H_0 \sqrt{(1-\Omega_{{\rm m}0})}\,, \\
    \mathcal{A}_{2} &\equiv \sinh^{-1}\qty[\sin\qty(\mathcal{A}_{1} \,t_{\dagger})]\:.
\end{align*}
Next, the general solutions to the free particle geodesic Eq.~\eqref{rgeod}, are outlined separately for the two distinct periods of comoving time, $t \geq t_{\dagger}$ and $t \leq t_{\dagger}$, as follows:
\begin{align}
\label{rfreeparticle}
    r(t) =
    \begin{cases} \mathcal{C}_1 \left[-1 + \cos\left(\mathcal{A}_{1} t\right)^2 \right]^{\frac{1}{4}} P_{\frac{1}{6}}^{\frac{1}{6}}\left[\cos\left(\mathcal{A}_{1} t\right)\right] \\
    + \mathcal{C}_2 \left[ -1 + \cos\left(\mathcal{A}_{1} t\right)^2 \right]^{\frac{1}{4}} Q_{\frac{1}{6}}^{\frac{1}{6}}\left[\cos\left(\mathcal{A}_{1} t\right)\right]& t \leq t_{\dagger} \\ \\
    \mathcal{B}\,\mathcal{D}_1 \,{}_2F_1\left[-\frac{1}{6}, \frac{1}{3}, \frac{5}{6}, \tanh^2\left[\mathcal{A}_{1} (t - t_{\dagger}) + \mathcal{A}_{2}\right]\right] \\
    +\mathcal{B} \,\mathcal{D}_2\, \tanh^{\frac{2}{3}}\left[\mathcal{A}_{1} (t - t_{\dagger}) + \mathcal{A}_{2}\right]  & t \geq t_{\dagger}
    \end{cases}
\end{align}
where we have denoted
\begin{equation*}
    \mathcal{B}\equiv\qty[-1 + \tanh^2\qty(\mathcal{A}_{1} (t - t_{\dagger}) + \mathcal{A}_{2})]^{-1/2} \,,
\end{equation*}
and $P_{l}^{m}$, $Q_{l}^{m}$ represents the associated Legendre functions of first and second kind, respectively. Meanwhile, ${}_2F_1$ is the hyper-geometric function \cite{arfken2012mathematical}. The constants are determined through the appropriate boundary conditions.

\section{Determining cosmological parameters}
\label{app:deter_cosmo_param}

In order to determine the cosmological parameters for our analysis, we will follow a method used in~\cite{Knox:2019rjx, Shah:2021onj, Vagnozzi:2019ezj}. The locations of peaks (i.e., peak spacing) in the CMB power spectrum\footnote{Also known as the ``acoustic scales''}, $l_{\rm A}$, is a well-measured quantity and it is defined as: 
\begin{equation}\label{eq:12}
l_{\rm A} \equiv \pi (1+z_*) \frac{D_{\rm A}(z_*)}{r_s^*} = \frac{\pi}{\theta_s^*}\:,
\end{equation}
where $\theta_s^* \equiv r_s^* / d_A(z_*)$ represents the angular size of the sound horizon at the last-scattering surface~\cite{WMAP:2008lyn, Komatsu_2011, Wang:2007mza}. Here, $r_s^*$ represents the comoving sound horizon at the last-scattering surface:
\begin{equation}\label{eq:13}
r_s^* = \displaystyle\int_{z_{*}}^{\infty}c_s(z)\frac{\dd{z}}{H(z)}\:,
\end{equation}
where $c_s(z)$ is the sound speed of the photon-baryon fluid:
\begin{equation}\label{eq:14}
c_s(z)=\frac{c}{\sqrt{3\qty(1+\frac{3\omega_{\rm b}}{4\omega_{\rm \gamma}(1+z)})}}\:,
\end{equation}
and $z_*$ is the redshift at the last-scattering surface, which can be approximated analytically via~\cite{Hu:1995en}:
\begin{equation}\label{eq:15}
\begin{aligned}
&z_* = 1048\qty[1+0.00124(\omega_{\rm b})^{-0.738}][1+g_1(\omega_{\rm m})^{g_2}]\,,\\
&g_1 = 0.0783(\omega_{\rm b})^{-0.238}\qty[1+39.5(\omega_{\rm b})^{0.763}]^{-1}\,,\\
&g_2 = 0.560[1+21.1(\omega_{\rm b})^{1.81}]^{-1}\,.
\end{aligned}
\end{equation}
Meanwhile, the proper angular diameter distance to the last-scattering surface defined as:
\begin{equation}\label{eq:16}
D_A(z_{*}) \equiv \frac{d_A(z_*)}{1+z_*} = \frac{c}{1+z_*}\displaystyle\int_ {0}^{z_{*}} \frac{\dd{z}}{H(z)}\:.
\end{equation}
In what follows, we will start the calculations by assuming that the acoustic scale and the physical density parameter for matter in the $\Lambda$CDM universe will be equal to the $\Lambda_{\rm s}$CDM\footnote{CMB distance priors, $l_A$ and $R$, (viz. $\omega_{\rm m}$) are actually derived parameters by fitting a cosmological model to the CMB power spectra~\cite{Wang:2013mha, Huang:2015vpa, Chen:2018dbv}. Thus, the underlying cosmology would change the distance priors \cite{Elgaroy:2007bv}. Since, $\Lambda_{\rm s}$CDM does not change the physics of the early universe, we can assume that it will not cause a significant variation in $l_A$ or in $\omega_{\rm m}$.} (i.e., $l_{\rm A}^{\Lambda} \simeq l_{\rm A}^{\Lambda_{\rm s}}$ and $\omega^{\Lambda}_{\rm m} \simeq \omega^{\Lambda_{\rm s}}_{\rm m}$). Since $\Lambda_{\rm s}$CDM does not change $N_{\rm eff}$ or the physics of the early universe, we can further assume that $\omega_{\rm r}^{\Lambda} \simeq \omega_{\rm r}^{\Lambda_{\rm s}}$ and $\omega_{\rm b}^{\Lambda} \simeq \omega_{\rm b}^{\Lambda_{\rm s}}$.

Under these conditions, $z_*$ and $c_s(z)$ will be the same for $\Lambda$CDM and $\Lambda_{\rm s}$CDM models, which can be seen via Eqs.~\eqref{eq:14} and \eqref{eq:15}. By combining Eqs.~\eqref{eq:12}, \eqref{eq:13} and \eqref{eq:16}, we can write:
\begin{equation}\label{eq:17}
    \theta_s^* = \frac{\displaystyle\int_{z_{*}}^{\infty}\frac{c_s(z)\dd{z}}{\sqrt{\omega_{\rm m}(1+z)^3 + \omega_{\rm r}(1+z)^4 + \qty(h_0^2-\omega_{\rm m}-\omega_{\rm r})f_{\rm DE}(z)}}}{\displaystyle\int_{0}^{z_*}\frac{c\dd{z}}{\sqrt{\omega_{\rm m}(1+z)^3 + \omega_{\rm r}(1+z)^4 + \qty(h_0^2-\omega_{\rm m}-\omega_{\rm r})f_{\rm DE}(z)}}}\,.
\end{equation}
Since the value of the parameter $\theta_s^*$ is fixed by Planck CMB observations almost model independently, we can constrain $h_0$ for any given $f_{\rm DE}(z) \equiv \rho_{\rm DE}(z) / \rho_{\rm DE,0}$ (viz. $f_{\Lambda}(z) \equiv 1$ and $f_{\Lambda_{\rm s}}(z) \equiv {\rm sgn}(z_{\dagger}-z)$), provided that the pre-recombination universe remains as in the standard cosmological model.

\begin{table}[t!]
\caption{\label{table:model_param} Plik Best Fit values taken from the Planck-2018 dataset~\cite{Planck:2018vyg}. We have defined the physical radiation density parameter as the sum of the physical photon and neutrino density parameters; $\omega_{\rm r} \equiv \omega_{\rm \gamma} + \omega_{\rm n} = 2.469 \times 10^{-5} \times \qty[1 + \frac{7}{8} \qty(\frac{4}{11})^{4/3}N_{\rm eff}]$ with $N_{\rm eff}=3.046$ for standard model of particle physics.}
\begin{ruledtabular}
\begin{tabular}{lr}
\textbf{Parameter}  &  \textbf{Value} \\ \colrule
$\omega_{\rm b}$ & $0.022383$ \\
$\omega_{\rm m}$ & $0.143140$ \\
$\omega_{\rm r}$ & $4.177\times 10^{-5}$ \\
$100\theta_s^*$ & $1.041085$ \\
\end{tabular}
\end{ruledtabular}
\end{table}

To simplify the above procedure, one can use the fitting formulae given below to calculate the $\Omega_{\rm m0}$ for the $\Lambda_{\rm s}$CDM model, expressed in terms of the transition redshift, $z_{\dagger}$:
\begin{equation}\label{eq:fitting}
    \Omega_{\rm m0}(z_{\dagger}) = c_0 + c_1z_{\dagger}^{-1}+ c_2z_{\dagger}^{-2} + c_3z_{\dagger}^{-3}\:.
\end{equation}
By assuming the Table~\ref{table:model_param} parameters, we obtain the constants of the equation as $(c_0, c_1, c_2, c_3) = (0.3093,0.0155,-0.0994,-0.0722)$, which correctly finds $\Omega_{\rm m0}$ up to order of $10^{-4}$ for $1.5 \leq z_{\dagger} \leq 11.5$.

\newpage
\bibliography{main}

\providecommand{\noopsort}[1]{}\providecommand{\singleletter}[1]{#1}%
\begin{thebibliography}{177}%
\makeatletter
\providecommand \@ifxundefined [1]{%
 \@ifx{#1\undefined}
}%
\providecommand \@ifnum [1]{%
 \ifnum #1\expandafter \@firstoftwo
 \else \expandafter \@secondoftwo
 \fi
}%
\providecommand \@ifx [1]{%
 \ifx #1\expandafter \@firstoftwo
 \else \expandafter \@secondoftwo
 \fi
}%
\providecommand \natexlab [1]{#1}%
\providecommand \enquote  [1]{``#1''}%
\providecommand \bibnamefont  [1]{#1}%
\providecommand \bibfnamefont [1]{#1}%
\providecommand \citenamefont [1]{#1}%
\providecommand \href@noop [0]{\@secondoftwo}%
\providecommand \href [0]{\begingroup \@sanitize@url \@href}%
\providecommand \@href[1]{\@@startlink{#1}\@@href}%
\providecommand \@@href[1]{\endgroup#1\@@endlink}%
\providecommand \@sanitize@url [0]{\catcode `\\12\catcode `\$12\catcode `\&12\catcode `\#12\catcode `\^12\catcode `\_12\catcode `\%12\relax}%
\providecommand \@@startlink[1]{}%
\providecommand \@@endlink[0]{}%
\providecommand \url  [0]{\begingroup\@sanitize@url \@url }%
\providecommand \@url [1]{\endgroup\@href {#1}{\urlprefix }}%
\providecommand \urlprefix  [0]{URL }%
\providecommand \Eprint [0]{\href }%
\providecommand \doibase [0]{http://dx.doi.org/}%
\providecommand \selectlanguage [0]{\@gobble}%
\providecommand \bibinfo  [0]{\@secondoftwo}%
\providecommand \bibfield  [0]{\@secondoftwo}%
\providecommand \translation [1]{[#1]}%
\providecommand \BibitemOpen [0]{}%
\providecommand \bibitemStop [0]{}%
\providecommand \bibitemNoStop [0]{.\EOS\space}%
\providecommand \EOS [0]{\spacefactor3000\relax}%
\providecommand \BibitemShut  [1]{\csname bibitem#1\endcsname}%
\let\auto@bib@innerbib\@empty
\bibitem [{\citenamefont {Dodelson}\ and\ \citenamefont {Schmidt}(2021{\natexlab{a}})}]{DODELSON20211}%
  \BibitemOpen
  \bibfield  {author} {\bibinfo {author} {\bibfnamefont {S.}~\bibnamefont {Dodelson}}\ and\ \bibinfo {author} {\bibfnamefont {F.}~\bibnamefont {Schmidt}},\ }\href {\doibase https://doi.org/10.1016/B978-0-12-815948-4.00007-3} {\emph {\bibinfo {title} {Modern Cosmology}}},\ \bibinfo {edition} {second edition}\ ed.\ (\bibinfo  {publisher} {Academic Press},\ \bibinfo {year} {2021})\ pp.\ \bibinfo {pages} {1--19}\BibitemShut {NoStop}%
\bibitem [{\citenamefont {Riess}\ \emph {et~al.}(1998)\citenamefont {Riess} \emph {et~al.}}]{SupernovaSearchTeam:1998fmf}%
  \BibitemOpen
  \bibfield  {author} {\bibinfo {author} {\bibfnamefont {A.~G.}\ \bibnamefont {Riess}} \emph {et~al.} (\bibinfo {collaboration} {Supernova Search Team}),\ }\href {\doibase 10.1086/300499} {\bibfield  {journal} {\bibinfo  {journal} {Astron. J.}\ }\textbf {\bibinfo {volume} {116}},\ \bibinfo {pages} {1009} (\bibinfo {year} {1998})},\ \Eprint {http://arxiv.org/abs/astro-ph/9805201} {arXiv:astro-ph/9805201} \BibitemShut {NoStop}%
\bibitem [{\citenamefont {Perlmutter}\ \emph {et~al.}(1999)\citenamefont {Perlmutter} \emph {et~al.}}]{SupernovaCosmologyProject:1998vns}%
  \BibitemOpen
  \bibfield  {author} {\bibinfo {author} {\bibfnamefont {S.}~\bibnamefont {Perlmutter}} \emph {et~al.} (\bibinfo {collaboration} {Supernova Cosmology Project}),\ }\href {\doibase 10.1086/307221} {\bibfield  {journal} {\bibinfo  {journal} {Astrophys. J.}\ }\textbf {\bibinfo {volume} {517}},\ \bibinfo {pages} {565} (\bibinfo {year} {1999})},\ \Eprint {http://arxiv.org/abs/astro-ph/9812133} {arXiv:astro-ph/9812133} \BibitemShut {NoStop}%
\bibitem [{\citenamefont {Aghanim}\ \emph {et~al.}(2020)\citenamefont {Aghanim} \emph {et~al.}}]{Planck:2018vyg}%
  \BibitemOpen
  \bibfield  {author} {\bibinfo {author} {\bibfnamefont {N.}~\bibnamefont {Aghanim}} \emph {et~al.} (\bibinfo {collaboration} {Planck}),\ }\href {\doibase 10.1051/0004-6361/201833910} {\bibfield  {journal} {\bibinfo  {journal} {Astron. Astrophys.}\ }\textbf {\bibinfo {volume} {641}},\ \bibinfo {pages} {A6} (\bibinfo {year} {2020})},\ \bibinfo {note} {[Erratum: Astron.Astrophys. 652, C4 (2021)]},\ \Eprint {http://arxiv.org/abs/1807.06209} {arXiv:1807.06209 [astro-ph.CO]} \BibitemShut {NoStop}%
\bibitem [{\citenamefont {Aiola}\ \emph {et~al.}(2020)\citenamefont {Aiola} \emph {et~al.}}]{ACT:2020gnv}%
  \BibitemOpen
  \bibfield  {author} {\bibinfo {author} {\bibfnamefont {S.}~\bibnamefont {Aiola}} \emph {et~al.} (\bibinfo {collaboration} {ACT}),\ }\href {\doibase 10.1088/1475-7516/2020/12/047} {\bibfield  {journal} {\bibinfo  {journal} {JCAP}\ }\textbf {\bibinfo {volume} {12}},\ \bibinfo {pages} {047} (\bibinfo {year} {2020})},\ \Eprint {http://arxiv.org/abs/2007.07288} {arXiv:2007.07288 [astro-ph.CO]} \BibitemShut {NoStop}%
\bibitem [{\citenamefont {Alam}\ \emph {et~al.}(2021)\citenamefont {Alam} \emph {et~al.}}]{eBOSS:2020yzd}%
  \BibitemOpen
  \bibfield  {author} {\bibinfo {author} {\bibfnamefont {S.}~\bibnamefont {Alam}} \emph {et~al.} (\bibinfo {collaboration} {eBOSS}),\ }\href {\doibase 10.1103/PhysRevD.103.083533} {\bibfield  {journal} {\bibinfo  {journal} {Phys. Rev. D}\ }\textbf {\bibinfo {volume} {103}},\ \bibinfo {pages} {083533} (\bibinfo {year} {2021})},\ \Eprint {http://arxiv.org/abs/2007.08991} {arXiv:2007.08991 [astro-ph.CO]} \BibitemShut {NoStop}%
\bibitem [{\citenamefont {Asgari}\ \emph {et~al.}(2021)\citenamefont {Asgari} \emph {et~al.}}]{KiDS:2020suj}%
  \BibitemOpen
  \bibfield  {author} {\bibinfo {author} {\bibfnamefont {M.}~\bibnamefont {Asgari}} \emph {et~al.} (\bibinfo {collaboration} {KiDS}),\ }\href {\doibase 10.1051/0004-6361/202039070} {\bibfield  {journal} {\bibinfo  {journal} {Astron. Astrophys.}\ }\textbf {\bibinfo {volume} {645}},\ \bibinfo {pages} {A104} (\bibinfo {year} {2021})},\ \Eprint {http://arxiv.org/abs/2007.15633} {arXiv:2007.15633 [astro-ph.CO]} \BibitemShut {NoStop}%
\bibitem [{\citenamefont {Abbott}\ \emph {et~al.}(2022)\citenamefont {Abbott} \emph {et~al.}}]{DES:2021wwk}%
  \BibitemOpen
  \bibfield  {author} {\bibinfo {author} {\bibfnamefont {T.~M.~C.}\ \bibnamefont {Abbott}} \emph {et~al.} (\bibinfo {collaboration} {DES}),\ }\href {\doibase 10.1103/PhysRevD.105.023520} {\bibfield  {journal} {\bibinfo  {journal} {Phys. Rev. D}\ }\textbf {\bibinfo {volume} {105}},\ \bibinfo {pages} {023520} (\bibinfo {year} {2022})},\ \Eprint {http://arxiv.org/abs/2105.13549} {arXiv:2105.13549 [astro-ph.CO]} \BibitemShut {NoStop}%
\bibitem [{\citenamefont {Weinberg}(1989)}]{Weinberg:1988cp}%
  \BibitemOpen
  \bibfield  {author} {\bibinfo {author} {\bibfnamefont {S.}~\bibnamefont {Weinberg}},\ }\href {\doibase 10.1103/RevModPhys.61.1} {\bibfield  {journal} {\bibinfo  {journal} {Rev. Mod. Phys.}\ }\textbf {\bibinfo {volume} {61}},\ \bibinfo {pages} {1} (\bibinfo {year} {1989})}\BibitemShut {NoStop}%
\bibitem [{\citenamefont {Carroll}\ \emph {et~al.}(1992)\citenamefont {Carroll}, \citenamefont {Press},\ and\ \citenamefont {Turner}}]{Carroll:1991mt}%
  \BibitemOpen
  \bibfield  {author} {\bibinfo {author} {\bibfnamefont {S.~M.}\ \bibnamefont {Carroll}}, \bibinfo {author} {\bibfnamefont {W.~H.}\ \bibnamefont {Press}}, \ and\ \bibinfo {author} {\bibfnamefont {E.~L.}\ \bibnamefont {Turner}},\ }\href {\doibase 10.1146/annurev.aa.30.090192.002435} {\bibfield  {journal} {\bibinfo  {journal} {Ann. Rev. Astron. Astrophys.}\ }\textbf {\bibinfo {volume} {30}},\ \bibinfo {pages} {499} (\bibinfo {year} {1992})}\BibitemShut {NoStop}%
\bibitem [{\citenamefont {Sahni}\ and\ \citenamefont {Starobinsky}(2000)}]{Sahni:1999gb}%
  \BibitemOpen
  \bibfield  {author} {\bibinfo {author} {\bibfnamefont {V.}~\bibnamefont {Sahni}}\ and\ \bibinfo {author} {\bibfnamefont {A.~A.}\ \bibnamefont {Starobinsky}},\ }\href {\doibase 10.1142/S0218271800000542} {\bibfield  {journal} {\bibinfo  {journal} {Int. J. Mod. Phys. D}\ }\textbf {\bibinfo {volume} {9}},\ \bibinfo {pages} {373} (\bibinfo {year} {2000})},\ \Eprint {http://arxiv.org/abs/astro-ph/9904398} {arXiv:astro-ph/9904398} \BibitemShut {NoStop}%
\bibitem [{\citenamefont {Peebles}\ and\ \citenamefont {Ratra}(2003)}]{Peebles:2002gy}%
  \BibitemOpen
  \bibfield  {author} {\bibinfo {author} {\bibfnamefont {P.~J.~E.}\ \bibnamefont {Peebles}}\ and\ \bibinfo {author} {\bibfnamefont {B.}~\bibnamefont {Ratra}},\ }\href {\doibase 10.1103/RevModPhys.75.559} {\bibfield  {journal} {\bibinfo  {journal} {Rev. Mod. Phys.}\ }\textbf {\bibinfo {volume} {75}},\ \bibinfo {pages} {559} (\bibinfo {year} {2003})},\ \Eprint {http://arxiv.org/abs/astro-ph/0207347} {arXiv:astro-ph/0207347} \BibitemShut {NoStop}%
\bibitem [{\citenamefont {Padmanabhan}(2003)}]{Padmanabhan:2002ji}%
  \BibitemOpen
  \bibfield  {author} {\bibinfo {author} {\bibfnamefont {T.}~\bibnamefont {Padmanabhan}},\ }\href {\doibase 10.1016/S0370-1573(03)00120-0} {\bibfield  {journal} {\bibinfo  {journal} {Phys. Rept.}\ }\textbf {\bibinfo {volume} {380}},\ \bibinfo {pages} {235} (\bibinfo {year} {2003})},\ \Eprint {http://arxiv.org/abs/hep-th/0212290} {arXiv:hep-th/0212290} \BibitemShut {NoStop}%
\bibitem [{\citenamefont {Bull}\ \emph {et~al.}(2016)\citenamefont {Bull} \emph {et~al.}}]{Bull:2015stt}%
  \BibitemOpen
  \bibfield  {author} {\bibinfo {author} {\bibfnamefont {P.}~\bibnamefont {Bull}} \emph {et~al.},\ }\href {\doibase 10.1016/j.dark.2016.02.001} {\bibfield  {journal} {\bibinfo  {journal} {Phys. Dark Univ.}\ }\textbf {\bibinfo {volume} {12}},\ \bibinfo {pages} {56} (\bibinfo {year} {2016})},\ \Eprint {http://arxiv.org/abs/1512.05356} {arXiv:1512.05356 [astro-ph.CO]} \BibitemShut {NoStop}%
\bibitem [{\citenamefont {Di~Valentino}\ \emph {et~al.}(2021{\natexlab{a}})\citenamefont {Di~Valentino} \emph {et~al.}}]{DiValentino:2020zio}%
  \BibitemOpen
  \bibfield  {author} {\bibinfo {author} {\bibfnamefont {E.}~\bibnamefont {Di~Valentino}} \emph {et~al.},\ }\href {\doibase 10.1016/j.astropartphys.2021.102605} {\bibfield  {journal} {\bibinfo  {journal} {Astropart. Phys.}\ }\textbf {\bibinfo {volume} {131}},\ \bibinfo {pages} {102605} (\bibinfo {year} {2021}{\natexlab{a}})},\ \Eprint {http://arxiv.org/abs/2008.11284} {arXiv:2008.11284 [astro-ph.CO]} \BibitemShut {NoStop}%
\bibitem [{\citenamefont {Di~Valentino}\ \emph {et~al.}(2021{\natexlab{b}})\citenamefont {Di~Valentino}, \citenamefont {Mena}, \citenamefont {Pan}, \citenamefont {Visinelli}, \citenamefont {Yang}, \citenamefont {Melchiorri}, \citenamefont {Mota}, \citenamefont {Riess},\ and\ \citenamefont {Silk}}]{DiValentino:2021izs}%
  \BibitemOpen
  \bibfield  {author} {\bibinfo {author} {\bibfnamefont {E.}~\bibnamefont {Di~Valentino}}, \bibinfo {author} {\bibfnamefont {O.}~\bibnamefont {Mena}}, \bibinfo {author} {\bibfnamefont {S.}~\bibnamefont {Pan}}, \bibinfo {author} {\bibfnamefont {L.}~\bibnamefont {Visinelli}}, \bibinfo {author} {\bibfnamefont {W.}~\bibnamefont {Yang}}, \bibinfo {author} {\bibfnamefont {A.}~\bibnamefont {Melchiorri}}, \bibinfo {author} {\bibfnamefont {D.~F.}\ \bibnamefont {Mota}}, \bibinfo {author} {\bibfnamefont {A.~G.}\ \bibnamefont {Riess}}, \ and\ \bibinfo {author} {\bibfnamefont {J.}~\bibnamefont {Silk}},\ }\href {\doibase 10.1088/1361-6382/ac086d} {\bibfield  {journal} {\bibinfo  {journal} {Class. Quant. Grav.}\ }\textbf {\bibinfo {volume} {38}},\ \bibinfo {pages} {153001} (\bibinfo {year} {2021}{\natexlab{b}})},\ \Eprint {http://arxiv.org/abs/2103.01183} {arXiv:2103.01183 [astro-ph.CO]} \BibitemShut {NoStop}%
\bibitem [{\citenamefont {Peebles}(2022)}]{Peebles:2022akh}%
  \BibitemOpen
  \bibfield  {author} {\bibinfo {author} {\bibfnamefont {P.~J.~E.}\ \bibnamefont {Peebles}},\ }\href {\doibase 10.1016/j.aop.2022.169159} {\bibfield  {journal} {\bibinfo  {journal} {Annals Phys.}\ }\textbf {\bibinfo {volume} {447}},\ \bibinfo {pages} {169159} (\bibinfo {year} {2022})},\ \Eprint {http://arxiv.org/abs/2208.05018} {arXiv:2208.05018 [astro-ph.CO]} \BibitemShut {NoStop}%
\bibitem [{\citenamefont {Perivolaropoulos}\ and\ \citenamefont {Skara}(2022)}]{Perivolaropoulos:2021jda}%
  \BibitemOpen
  \bibfield  {author} {\bibinfo {author} {\bibfnamefont {L.}~\bibnamefont {Perivolaropoulos}}\ and\ \bibinfo {author} {\bibfnamefont {F.}~\bibnamefont {Skara}},\ }\href {\doibase 10.1016/j.newar.2022.101659} {\bibfield  {journal} {\bibinfo  {journal} {New Astron. Rev.}\ }\textbf {\bibinfo {volume} {95}},\ \bibinfo {pages} {101659} (\bibinfo {year} {2022})},\ \Eprint {http://arxiv.org/abs/2105.05208} {arXiv:2105.05208 [astro-ph.CO]} \BibitemShut {NoStop}%
\bibitem [{\citenamefont {Abdalla}\ \emph {et~al.}(2022)\citenamefont {Abdalla} \emph {et~al.}}]{Abdalla:2022yfr}%
  \BibitemOpen
  \bibfield  {author} {\bibinfo {author} {\bibfnamefont {E.}~\bibnamefont {Abdalla}} \emph {et~al.},\ }\href {\doibase 10.1016/j.jheap.2022.04.002} {\bibfield  {journal} {\bibinfo  {journal} {JHEAp}\ }\textbf {\bibinfo {volume} {34}},\ \bibinfo {pages} {49} (\bibinfo {year} {2022})},\ \Eprint {http://arxiv.org/abs/2203.06142} {arXiv:2203.06142 [astro-ph.CO]} \BibitemShut {NoStop}%
\bibitem [{\citenamefont {Vagnozzi}(2023)}]{Vagnozzi:2023nrq}%
  \BibitemOpen
  \bibfield  {author} {\bibinfo {author} {\bibfnamefont {S.}~\bibnamefont {Vagnozzi}},\ }\href {\doibase 10.3390/universe9090393} {\bibfield  {journal} {\bibinfo  {journal} {Universe}\ }\textbf {\bibinfo {volume} {9}},\ \bibinfo {pages} {393} (\bibinfo {year} {2023})},\ \Eprint {http://arxiv.org/abs/2308.16628} {arXiv:2308.16628 [astro-ph.CO]} \BibitemShut {NoStop}%
\bibitem [{\citenamefont {Hu}\ and\ \citenamefont {Wang}(2023)}]{Hu:2023jqc}%
  \BibitemOpen
  \bibfield  {author} {\bibinfo {author} {\bibfnamefont {J.-P.}\ \bibnamefont {Hu}}\ and\ \bibinfo {author} {\bibfnamefont {F.-Y.}\ \bibnamefont {Wang}},\ }\href {\doibase 10.3390/universe9020094} {\bibfield  {journal} {\bibinfo  {journal} {Universe}\ }\textbf {\bibinfo {volume} {9}},\ \bibinfo {pages} {94} (\bibinfo {year} {2023})},\ \Eprint {http://arxiv.org/abs/2302.05709} {arXiv:2302.05709 [astro-ph.CO]} \BibitemShut {NoStop}%
\bibitem [{\citenamefont {Akarsu}\ \emph {et~al.}(2024{\natexlab{a}})\citenamefont {Akarsu}, \citenamefont {Colg\'ain}, \citenamefont {Sen},\ and\ \citenamefont {Sheikh-Jabbari}}]{Akarsu:2024qiq}%
  \BibitemOpen
  \bibfield  {author} {\bibinfo {author} {\bibfnamefont {O.}~\bibnamefont {Akarsu}}, \bibinfo {author} {\bibfnamefont {E.~O.}\ \bibnamefont {Colg\'ain}}, \bibinfo {author} {\bibfnamefont {A.~A.}\ \bibnamefont {Sen}}, \ and\ \bibinfo {author} {\bibfnamefont {M.~M.}\ \bibnamefont {Sheikh-Jabbari}},\ }\href@noop {} {\  (\bibinfo {year} {2024}{\natexlab{a}})},\ \Eprint {http://arxiv.org/abs/2402.04767} {arXiv:2402.04767 [astro-ph.CO]} \BibitemShut {NoStop}%
\bibitem [{\citenamefont {Riess}\ \emph {et~al.}(2022)\citenamefont {Riess} \emph {et~al.}}]{Riess:2021jrx}%
  \BibitemOpen
  \bibfield  {author} {\bibinfo {author} {\bibfnamefont {A.~G.}\ \bibnamefont {Riess}} \emph {et~al.},\ }\href {\doibase 10.3847/2041-8213/ac5c5b} {\bibfield  {journal} {\bibinfo  {journal} {Astrophys. J. Lett.}\ }\textbf {\bibinfo {volume} {934}},\ \bibinfo {pages} {L7} (\bibinfo {year} {2022})},\ \Eprint {http://arxiv.org/abs/2112.04510} {arXiv:2112.04510 [astro-ph.CO]} \BibitemShut {NoStop}%
\bibitem [{\citenamefont {Burger}\ \emph {et~al.}(2023)\citenamefont {Burger} \emph {et~al.}}]{Burger:2023qef}%
  \BibitemOpen
  \bibfield  {author} {\bibinfo {author} {\bibfnamefont {P.~A.}\ \bibnamefont {Burger}} \emph {et~al.}\ }(\bibinfo {year} {2023})\ \Eprint {http://arxiv.org/abs/2309.08602} {arXiv:2309.08602 [astro-ph.CO]} \BibitemShut {NoStop}%
\bibitem [{\citenamefont {Chen}\ \emph {et~al.}(2022)\citenamefont {Chen}, \citenamefont {Vlah},\ and\ \citenamefont {White}}]{Chen:2021wdi}%
  \BibitemOpen
  \bibfield  {author} {\bibinfo {author} {\bibfnamefont {S.-F.}\ \bibnamefont {Chen}}, \bibinfo {author} {\bibfnamefont {Z.}~\bibnamefont {Vlah}}, \ and\ \bibinfo {author} {\bibfnamefont {M.}~\bibnamefont {White}},\ }\href {\doibase 10.1088/1475-7516/2022/02/008} {\bibfield  {journal} {\bibinfo  {journal} {JCAP}\ }\textbf {\bibinfo {volume} {02}},\ \bibinfo {pages} {008} (\bibinfo {year} {2022})},\ \Eprint {http://arxiv.org/abs/2110.05530} {arXiv:2110.05530 [astro-ph.CO]} \BibitemShut {NoStop}%
\bibitem [{\citenamefont {Adil}\ \emph {et~al.}(2023{\natexlab{a}})\citenamefont {Adil}, \citenamefont {Akarsu}, \citenamefont {Malekjani}, \citenamefont {Colg\'ain}, \citenamefont {Pourojaghi}, \citenamefont {Sen},\ and\ \citenamefont {Sheikh-Jabbari}}]{Adil:2023jtu}%
  \BibitemOpen
  \bibfield  {author} {\bibinfo {author} {\bibfnamefont {S.~A.}\ \bibnamefont {Adil}}, \bibinfo {author} {\bibfnamefont {O.}~\bibnamefont {Akarsu}}, \bibinfo {author} {\bibfnamefont {M.}~\bibnamefont {Malekjani}}, \bibinfo {author} {\bibfnamefont {E.~O.}\ \bibnamefont {Colg\'ain}}, \bibinfo {author} {\bibfnamefont {S.}~\bibnamefont {Pourojaghi}}, \bibinfo {author} {\bibfnamefont {A.~A.}\ \bibnamefont {Sen}}, \ and\ \bibinfo {author} {\bibfnamefont {M.~M.}\ \bibnamefont {Sheikh-Jabbari}},\ }\href {\doibase 10.1093/mnrasl/slad165} {\  (\bibinfo {year} {2023}{\natexlab{a}}),\ 10.1093/mnrasl/slad165},\ \Eprint {http://arxiv.org/abs/2303.06928} {arXiv:2303.06928 [astro-ph.CO]} \BibitemShut {NoStop}%
\bibitem [{\citenamefont {Karwal}\ and\ \citenamefont {Kamionkowski}(2016)}]{Karwal:2016vyq}%
  \BibitemOpen
  \bibfield  {author} {\bibinfo {author} {\bibfnamefont {T.}~\bibnamefont {Karwal}}\ and\ \bibinfo {author} {\bibfnamefont {M.}~\bibnamefont {Kamionkowski}},\ }\href {\doibase 10.1103/PhysRevD.94.103523} {\bibfield  {journal} {\bibinfo  {journal} {Phys. Rev. D}\ }\textbf {\bibinfo {volume} {94}},\ \bibinfo {pages} {103523} (\bibinfo {year} {2016})},\ \Eprint {http://arxiv.org/abs/1608.01309} {arXiv:1608.01309 [astro-ph.CO]} \BibitemShut {NoStop}%
\bibitem [{\citenamefont {Poulin}\ \emph {et~al.}(2019)\citenamefont {Poulin}, \citenamefont {Smith}, \citenamefont {Karwal},\ and\ \citenamefont {Kamionkowski}}]{Poulin:2018cxd}%
  \BibitemOpen
  \bibfield  {author} {\bibinfo {author} {\bibfnamefont {V.}~\bibnamefont {Poulin}}, \bibinfo {author} {\bibfnamefont {T.~L.}\ \bibnamefont {Smith}}, \bibinfo {author} {\bibfnamefont {T.}~\bibnamefont {Karwal}}, \ and\ \bibinfo {author} {\bibfnamefont {M.}~\bibnamefont {Kamionkowski}},\ }\href {\doibase 10.1103/PhysRevLett.122.221301} {\bibfield  {journal} {\bibinfo  {journal} {Phys. Rev. Lett.}\ }\textbf {\bibinfo {volume} {122}},\ \bibinfo {pages} {221301} (\bibinfo {year} {2019})},\ \Eprint {http://arxiv.org/abs/1811.04083} {arXiv:1811.04083 [astro-ph.CO]} \BibitemShut {NoStop}%
\bibitem [{\citenamefont {Poulin}\ \emph {et~al.}(2018{\natexlab{a}})\citenamefont {Poulin}, \citenamefont {Smith}, \citenamefont {Grin}, \citenamefont {Karwal},\ and\ \citenamefont {Kamionkowski}}]{Poulin:2018dzj}%
  \BibitemOpen
  \bibfield  {author} {\bibinfo {author} {\bibfnamefont {V.}~\bibnamefont {Poulin}}, \bibinfo {author} {\bibfnamefont {T.~L.}\ \bibnamefont {Smith}}, \bibinfo {author} {\bibfnamefont {D.}~\bibnamefont {Grin}}, \bibinfo {author} {\bibfnamefont {T.}~\bibnamefont {Karwal}}, \ and\ \bibinfo {author} {\bibfnamefont {M.}~\bibnamefont {Kamionkowski}},\ }\href {\doibase 10.1103/PhysRevD.98.083525} {\bibfield  {journal} {\bibinfo  {journal} {Phys. Rev. D}\ }\textbf {\bibinfo {volume} {98}},\ \bibinfo {pages} {083525} (\bibinfo {year} {2018}{\natexlab{a}})},\ \Eprint {http://arxiv.org/abs/1806.10608} {arXiv:1806.10608 [astro-ph.CO]} \BibitemShut {NoStop}%
\bibitem [{\citenamefont {Agrawal}\ \emph {et~al.}(2019)\citenamefont {Agrawal}, \citenamefont {Cyr-Racine}, \citenamefont {Pinner},\ and\ \citenamefont {Randall}}]{Agrawal:2019lmo}%
  \BibitemOpen
  \bibfield  {author} {\bibinfo {author} {\bibfnamefont {P.}~\bibnamefont {Agrawal}}, \bibinfo {author} {\bibfnamefont {F.-Y.}\ \bibnamefont {Cyr-Racine}}, \bibinfo {author} {\bibfnamefont {D.}~\bibnamefont {Pinner}}, \ and\ \bibinfo {author} {\bibfnamefont {L.}~\bibnamefont {Randall}}\ }(\bibinfo {year} {2019})\ \Eprint {http://arxiv.org/abs/1904.01016} {arXiv:1904.01016 [astro-ph.CO]} \BibitemShut {NoStop}%
\bibitem [{\citenamefont {Kamionkowski}\ and\ \citenamefont {Riess}(2022)}]{Kamionkowski:2022pkx}%
  \BibitemOpen
  \bibfield  {author} {\bibinfo {author} {\bibfnamefont {M.}~\bibnamefont {Kamionkowski}}\ and\ \bibinfo {author} {\bibfnamefont {A.~G.}\ \bibnamefont {Riess}}\ }(\bibinfo {year} {2022})\ \Eprint {http://arxiv.org/abs/2211.04492} {arXiv:2211.04492 [astro-ph.CO]} \BibitemShut {NoStop}%
\bibitem [{\citenamefont {Odintsov}\ \emph {et~al.}(2023)\citenamefont {Odintsov}, \citenamefont {Oikonomou},\ and\ \citenamefont {Sharov}}]{Odintsov:2023cli}%
  \BibitemOpen
  \bibfield  {author} {\bibinfo {author} {\bibfnamefont {S.~D.}\ \bibnamefont {Odintsov}}, \bibinfo {author} {\bibfnamefont {V.~K.}\ \bibnamefont {Oikonomou}}, \ and\ \bibinfo {author} {\bibfnamefont {G.~S.}\ \bibnamefont {Sharov}},\ }\href {\doibase 10.1016/j.physletb.2023.137988} {\bibfield  {journal} {\bibinfo  {journal} {Phys. Lett. B}\ }\textbf {\bibinfo {volume} {843}},\ \bibinfo {pages} {137988} (\bibinfo {year} {2023})},\ \Eprint {http://arxiv.org/abs/2305.17513} {arXiv:2305.17513 [gr-qc]} \BibitemShut {NoStop}%
\bibitem [{\citenamefont {Niedermann}\ and\ \citenamefont {Sloth}(2021)}]{Niedermann:2019olb}%
  \BibitemOpen
  \bibfield  {author} {\bibinfo {author} {\bibfnamefont {F.}~\bibnamefont {Niedermann}}\ and\ \bibinfo {author} {\bibfnamefont {M.~S.}\ \bibnamefont {Sloth}},\ }\href {\doibase 10.1103/PhysRevD.103.L041303} {\bibfield  {journal} {\bibinfo  {journal} {Phys. Rev. D}\ }\textbf {\bibinfo {volume} {103}},\ \bibinfo {pages} {L041303} (\bibinfo {year} {2021})},\ \Eprint {http://arxiv.org/abs/1910.10739} {arXiv:1910.10739 [astro-ph.CO]} \BibitemShut {NoStop}%
\bibitem [{\citenamefont {Cruz}\ \emph {et~al.}(2023)\citenamefont {Cruz}, \citenamefont {Niedermann},\ and\ \citenamefont {Sloth}}]{Cruz:2023lmn}%
  \BibitemOpen
  \bibfield  {author} {\bibinfo {author} {\bibfnamefont {J.~S.}\ \bibnamefont {Cruz}}, \bibinfo {author} {\bibfnamefont {F.}~\bibnamefont {Niedermann}}, \ and\ \bibinfo {author} {\bibfnamefont {M.~S.}\ \bibnamefont {Sloth}}\ }(\bibinfo {year} {2023})\ \Eprint {http://arxiv.org/abs/2305.08895} {arXiv:2305.08895 [astro-ph.CO]} \BibitemShut {NoStop}%
\bibitem [{\citenamefont {Niedermann}\ and\ \citenamefont {Sloth}(2023)}]{Niedermann:2023ssr}%
  \BibitemOpen
  \bibfield  {author} {\bibinfo {author} {\bibfnamefont {F.}~\bibnamefont {Niedermann}}\ and\ \bibinfo {author} {\bibfnamefont {M.~S.}\ \bibnamefont {Sloth}}\ }(\bibinfo {year} {2023})\ \Eprint {http://arxiv.org/abs/2307.03481} {arXiv:2307.03481 [hep-ph]} \BibitemShut {NoStop}%
\bibitem [{\citenamefont {Ye}\ and\ \citenamefont {Piao}(2020{\natexlab{a}})}]{Ye:2020btb}%
  \BibitemOpen
  \bibfield  {author} {\bibinfo {author} {\bibfnamefont {G.}~\bibnamefont {Ye}}\ and\ \bibinfo {author} {\bibfnamefont {Y.-S.}\ \bibnamefont {Piao}},\ }\href {\doibase 10.1103/PhysRevD.101.083507} {\bibfield  {journal} {\bibinfo  {journal} {Phys. Rev. D}\ }\textbf {\bibinfo {volume} {101}},\ \bibinfo {pages} {083507} (\bibinfo {year} {2020}{\natexlab{a}})},\ \Eprint {http://arxiv.org/abs/2001.02451} {arXiv:2001.02451 [astro-ph.CO]} \BibitemShut {NoStop}%
\bibitem [{\citenamefont {Ye}\ and\ \citenamefont {Piao}(2020{\natexlab{b}})}]{Ye:2020oix}%
  \BibitemOpen
  \bibfield  {author} {\bibinfo {author} {\bibfnamefont {G.}~\bibnamefont {Ye}}\ and\ \bibinfo {author} {\bibfnamefont {Y.-S.}\ \bibnamefont {Piao}},\ }\href {\doibase 10.1103/PhysRevD.102.083523} {\bibfield  {journal} {\bibinfo  {journal} {Phys. Rev. D}\ }\textbf {\bibinfo {volume} {102}},\ \bibinfo {pages} {083523} (\bibinfo {year} {2020}{\natexlab{b}})},\ \Eprint {http://arxiv.org/abs/2008.10832} {arXiv:2008.10832 [astro-ph.CO]} \BibitemShut {NoStop}%
\bibitem [{\citenamefont {Ye}\ \emph {et~al.}(2023)\citenamefont {Ye}, \citenamefont {Zhang},\ and\ \citenamefont {Piao}}]{Ye:2021iwa}%
  \BibitemOpen
  \bibfield  {author} {\bibinfo {author} {\bibfnamefont {G.}~\bibnamefont {Ye}}, \bibinfo {author} {\bibfnamefont {J.}~\bibnamefont {Zhang}}, \ and\ \bibinfo {author} {\bibfnamefont {Y.-S.}\ \bibnamefont {Piao}},\ }\href {\doibase 10.1016/j.physletb.2023.137770} {\bibfield  {journal} {\bibinfo  {journal} {Phys. Lett. B}\ }\textbf {\bibinfo {volume} {839}},\ \bibinfo {pages} {137770} (\bibinfo {year} {2023})},\ \Eprint {http://arxiv.org/abs/2107.13391} {arXiv:2107.13391 [astro-ph.CO]} \BibitemShut {NoStop}%
\bibitem [{\citenamefont {Rossi}\ \emph {et~al.}(2019)\citenamefont {Rossi}, \citenamefont {Ballardini}, \citenamefont {Braglia}, \citenamefont {Finelli}, \citenamefont {Paoletti}, \citenamefont {Starobinsky},\ and\ \citenamefont {Umilt\`a}}]{Rossi:2019lgt}%
  \BibitemOpen
  \bibfield  {author} {\bibinfo {author} {\bibfnamefont {M.}~\bibnamefont {Rossi}}, \bibinfo {author} {\bibfnamefont {M.}~\bibnamefont {Ballardini}}, \bibinfo {author} {\bibfnamefont {M.}~\bibnamefont {Braglia}}, \bibinfo {author} {\bibfnamefont {F.}~\bibnamefont {Finelli}}, \bibinfo {author} {\bibfnamefont {D.}~\bibnamefont {Paoletti}}, \bibinfo {author} {\bibfnamefont {A.~A.}\ \bibnamefont {Starobinsky}}, \ and\ \bibinfo {author} {\bibfnamefont {C.}~\bibnamefont {Umilt\`a}},\ }\href {\doibase 10.1103/PhysRevD.100.103524} {\bibfield  {journal} {\bibinfo  {journal} {Phys. Rev. D}\ }\textbf {\bibinfo {volume} {100}},\ \bibinfo {pages} {103524} (\bibinfo {year} {2019})},\ \Eprint {http://arxiv.org/abs/1906.10218} {arXiv:1906.10218 [astro-ph.CO]} \BibitemShut {NoStop}%
\bibitem [{\citenamefont {Braglia}\ \emph {et~al.}(2020)\citenamefont {Braglia}, \citenamefont {Ballardini}, \citenamefont {Emond}, \citenamefont {Finelli}, \citenamefont {Gumrukcuoglu}, \citenamefont {Koyama},\ and\ \citenamefont {Paoletti}}]{Braglia:2020iik}%
  \BibitemOpen
  \bibfield  {author} {\bibinfo {author} {\bibfnamefont {M.}~\bibnamefont {Braglia}}, \bibinfo {author} {\bibfnamefont {M.}~\bibnamefont {Ballardini}}, \bibinfo {author} {\bibfnamefont {W.~T.}\ \bibnamefont {Emond}}, \bibinfo {author} {\bibfnamefont {F.}~\bibnamefont {Finelli}}, \bibinfo {author} {\bibfnamefont {A.~E.}\ \bibnamefont {Gumrukcuoglu}}, \bibinfo {author} {\bibfnamefont {K.}~\bibnamefont {Koyama}}, \ and\ \bibinfo {author} {\bibfnamefont {D.}~\bibnamefont {Paoletti}},\ }\href {\doibase 10.1103/PhysRevD.102.023529} {\bibfield  {journal} {\bibinfo  {journal} {Phys. Rev. D}\ }\textbf {\bibinfo {volume} {102}},\ \bibinfo {pages} {023529} (\bibinfo {year} {2020})},\ \Eprint {http://arxiv.org/abs/2004.11161} {arXiv:2004.11161 [astro-ph.CO]} \BibitemShut {NoStop}%
\bibitem [{\citenamefont {Adi}\ and\ \citenamefont {Kovetz}(2021)}]{Adi:2020qqf}%
  \BibitemOpen
  \bibfield  {author} {\bibinfo {author} {\bibfnamefont {T.}~\bibnamefont {Adi}}\ and\ \bibinfo {author} {\bibfnamefont {E.~D.}\ \bibnamefont {Kovetz}},\ }\href {\doibase 10.1103/PhysRevD.103.023530} {\bibfield  {journal} {\bibinfo  {journal} {Phys. Rev. D}\ }\textbf {\bibinfo {volume} {103}},\ \bibinfo {pages} {023530} (\bibinfo {year} {2021})},\ \Eprint {http://arxiv.org/abs/2011.13853} {arXiv:2011.13853 [astro-ph.CO]} \BibitemShut {NoStop}%
\bibitem [{\citenamefont {Braglia}\ \emph {et~al.}(2021)\citenamefont {Braglia}, \citenamefont {Ballardini}, \citenamefont {Finelli},\ and\ \citenamefont {Koyama}}]{Braglia:2020auw}%
  \BibitemOpen
  \bibfield  {author} {\bibinfo {author} {\bibfnamefont {M.}~\bibnamefont {Braglia}}, \bibinfo {author} {\bibfnamefont {M.}~\bibnamefont {Ballardini}}, \bibinfo {author} {\bibfnamefont {F.}~\bibnamefont {Finelli}}, \ and\ \bibinfo {author} {\bibfnamefont {K.}~\bibnamefont {Koyama}},\ }\href {\doibase 10.1103/PhysRevD.103.043528} {\bibfield  {journal} {\bibinfo  {journal} {Phys. Rev. D}\ }\textbf {\bibinfo {volume} {103}},\ \bibinfo {pages} {043528} (\bibinfo {year} {2021})},\ \Eprint {http://arxiv.org/abs/2011.12934} {arXiv:2011.12934 [astro-ph.CO]} \BibitemShut {NoStop}%
\bibitem [{\citenamefont {Ballardini}\ \emph {et~al.}(2020)\citenamefont {Ballardini}, \citenamefont {Braglia}, \citenamefont {Finelli}, \citenamefont {Paoletti}, \citenamefont {Starobinsky},\ and\ \citenamefont {Umilt\`a}}]{Ballardini:2020iws}%
  \BibitemOpen
  \bibfield  {author} {\bibinfo {author} {\bibfnamefont {M.}~\bibnamefont {Ballardini}}, \bibinfo {author} {\bibfnamefont {M.}~\bibnamefont {Braglia}}, \bibinfo {author} {\bibfnamefont {F.}~\bibnamefont {Finelli}}, \bibinfo {author} {\bibfnamefont {D.}~\bibnamefont {Paoletti}}, \bibinfo {author} {\bibfnamefont {A.~A.}\ \bibnamefont {Starobinsky}}, \ and\ \bibinfo {author} {\bibfnamefont {C.}~\bibnamefont {Umilt\`a}},\ }\href {\doibase 10.1088/1475-7516/2020/10/044} {\bibfield  {journal} {\bibinfo  {journal} {JCAP}\ }\textbf {\bibinfo {volume} {10}},\ \bibinfo {pages} {044} (\bibinfo {year} {2020})},\ \Eprint {http://arxiv.org/abs/2004.14349} {arXiv:2004.14349 [astro-ph.CO]} \BibitemShut {NoStop}%
\bibitem [{\citenamefont {Franco~Abell\'an}\ \emph {et~al.}(2023)\citenamefont {Franco~Abell\'an}, \citenamefont {Braglia}, \citenamefont {Ballardini}, \citenamefont {Finelli},\ and\ \citenamefont {Poulin}}]{FrancoAbellan:2023gec}%
  \BibitemOpen
  \bibfield  {author} {\bibinfo {author} {\bibfnamefont {G.}~\bibnamefont {Franco~Abell\'an}}, \bibinfo {author} {\bibfnamefont {M.}~\bibnamefont {Braglia}}, \bibinfo {author} {\bibfnamefont {M.}~\bibnamefont {Ballardini}}, \bibinfo {author} {\bibfnamefont {F.}~\bibnamefont {Finelli}}, \ and\ \bibinfo {author} {\bibfnamefont {V.}~\bibnamefont {Poulin}}\ }(\bibinfo {year} {2023})\ \Eprint {http://arxiv.org/abs/2308.12345} {arXiv:2308.12345 [astro-ph.CO]} \BibitemShut {NoStop}%
\bibitem [{\citenamefont {Petronikolou}\ and\ \citenamefont {Saridakis}(2023)}]{Petronikolou:2023cwu}%
  \BibitemOpen
  \bibfield  {author} {\bibinfo {author} {\bibfnamefont {M.}~\bibnamefont {Petronikolou}}\ and\ \bibinfo {author} {\bibfnamefont {E.~N.}\ \bibnamefont {Saridakis}},\ }\href {\doibase 10.3390/universe9090397} {\bibfield  {journal} {\bibinfo  {journal} {Universe}\ }\textbf {\bibinfo {volume} {9}},\ \bibinfo {pages} {397} (\bibinfo {year} {2023})},\ \Eprint {http://arxiv.org/abs/2308.16044} {arXiv:2308.16044 [gr-qc]} \BibitemShut {NoStop}%
\bibitem [{\citenamefont {Hazra}\ \emph {et~al.}(2022)\citenamefont {Hazra}, \citenamefont {Antony},\ and\ \citenamefont {Shafieloo}}]{Hazra:2022rdl}%
  \BibitemOpen
  \bibfield  {author} {\bibinfo {author} {\bibfnamefont {D.~K.}\ \bibnamefont {Hazra}}, \bibinfo {author} {\bibfnamefont {A.}~\bibnamefont {Antony}}, \ and\ \bibinfo {author} {\bibfnamefont {A.}~\bibnamefont {Shafieloo}},\ }\href {\doibase 10.1088/1475-7516/2022/08/063} {\bibfield  {journal} {\bibinfo  {journal} {JCAP}\ }\textbf {\bibinfo {volume} {08}},\ \bibinfo {pages} {063} (\bibinfo {year} {2022})},\ \Eprint {http://arxiv.org/abs/2201.12000} {arXiv:2201.12000 [astro-ph.CO]} \BibitemShut {NoStop}%
\bibitem [{\citenamefont {Akarsu}\ \emph {et~al.}(2020)\citenamefont {Akarsu}, \citenamefont {Barrow}, \citenamefont {Escamilla},\ and\ \citenamefont {Vazquez}}]{Akarsu:2019hmw}%
  \BibitemOpen
  \bibfield  {author} {\bibinfo {author} {\bibfnamefont {O.}~\bibnamefont {Akarsu}}, \bibinfo {author} {\bibfnamefont {J.~D.}\ \bibnamefont {Barrow}}, \bibinfo {author} {\bibfnamefont {L.~A.}\ \bibnamefont {Escamilla}}, \ and\ \bibinfo {author} {\bibfnamefont {J.~A.}\ \bibnamefont {Vazquez}},\ }\href {\doibase 10.1103/PhysRevD.101.063528} {\bibfield  {journal} {\bibinfo  {journal} {Phys. Rev. D}\ }\textbf {\bibinfo {volume} {101}},\ \bibinfo {pages} {063528} (\bibinfo {year} {2020})},\ \Eprint {http://arxiv.org/abs/1912.08751} {arXiv:1912.08751 [astro-ph.CO]} \BibitemShut {NoStop}%
\bibitem [{\citenamefont {Akarsu}\ \emph {et~al.}(2021)\citenamefont {Akarsu}, \citenamefont {Kumar}, \citenamefont {\"Oz\"ulker},\ and\ \citenamefont {Vazquez}}]{Akarsu:2021fol}%
  \BibitemOpen
  \bibfield  {author} {\bibinfo {author} {\bibfnamefont {O.}~\bibnamefont {Akarsu}}, \bibinfo {author} {\bibfnamefont {S.}~\bibnamefont {Kumar}}, \bibinfo {author} {\bibfnamefont {E.}~\bibnamefont {\"Oz\"ulker}}, \ and\ \bibinfo {author} {\bibfnamefont {J.~A.}\ \bibnamefont {Vazquez}},\ }\href {\doibase 10.1103/PhysRevD.104.123512} {\bibfield  {journal} {\bibinfo  {journal} {Phys. Rev. D}\ }\textbf {\bibinfo {volume} {104}},\ \bibinfo {pages} {123512} (\bibinfo {year} {2021})},\ \Eprint {http://arxiv.org/abs/2108.09239} {arXiv:2108.09239 [astro-ph.CO]} \BibitemShut {NoStop}%
\bibitem [{\citenamefont {Akarsu}\ \emph {et~al.}(2023{\natexlab{a}})\citenamefont {Akarsu}, \citenamefont {Kumar}, \citenamefont {\"Oz\"ulker}, \citenamefont {Vazquez},\ and\ \citenamefont {Yadav}}]{Akarsu:2022typ}%
  \BibitemOpen
  \bibfield  {author} {\bibinfo {author} {\bibfnamefont {O.}~\bibnamefont {Akarsu}}, \bibinfo {author} {\bibfnamefont {S.}~\bibnamefont {Kumar}}, \bibinfo {author} {\bibfnamefont {E.}~\bibnamefont {\"Oz\"ulker}}, \bibinfo {author} {\bibfnamefont {J.~A.}\ \bibnamefont {Vazquez}}, \ and\ \bibinfo {author} {\bibfnamefont {A.}~\bibnamefont {Yadav}},\ }\href {\doibase 10.1103/PhysRevD.108.023513} {\bibfield  {journal} {\bibinfo  {journal} {Phys. Rev. D}\ }\textbf {\bibinfo {volume} {108}},\ \bibinfo {pages} {023513} (\bibinfo {year} {2023}{\natexlab{a}})},\ \Eprint {http://arxiv.org/abs/2211.05742} {arXiv:2211.05742 [astro-ph.CO]} \BibitemShut {NoStop}%
\bibitem [{\citenamefont {Akarsu}\ \emph {et~al.}(2023{\natexlab{b}})\citenamefont {Akarsu}, \citenamefont {Di~Valentino}, \citenamefont {Kumar}, \citenamefont {Nunes}, \citenamefont {Vazquez},\ and\ \citenamefont {Yadav}}]{Akarsu:2023mfb}%
  \BibitemOpen
  \bibfield  {author} {\bibinfo {author} {\bibfnamefont {O.}~\bibnamefont {Akarsu}}, \bibinfo {author} {\bibfnamefont {E.}~\bibnamefont {Di~Valentino}}, \bibinfo {author} {\bibfnamefont {S.}~\bibnamefont {Kumar}}, \bibinfo {author} {\bibfnamefont {R.~C.}\ \bibnamefont {Nunes}}, \bibinfo {author} {\bibfnamefont {J.~A.}\ \bibnamefont {Vazquez}}, \ and\ \bibinfo {author} {\bibfnamefont {A.}~\bibnamefont {Yadav}}\ }(\bibinfo {year} {2023})\ \Eprint {http://arxiv.org/abs/2307.10899} {arXiv:2307.10899 [astro-ph.CO]} \BibitemShut {NoStop}%
\bibitem [{\citenamefont {Di~Valentino}\ \emph {et~al.}(2021{\natexlab{c}})\citenamefont {Di~Valentino}, \citenamefont {Mukherjee},\ and\ \citenamefont {Sen}}]{DiValentino:2020naf}%
  \BibitemOpen
  \bibfield  {author} {\bibinfo {author} {\bibfnamefont {E.}~\bibnamefont {Di~Valentino}}, \bibinfo {author} {\bibfnamefont {A.}~\bibnamefont {Mukherjee}}, \ and\ \bibinfo {author} {\bibfnamefont {A.~A.}\ \bibnamefont {Sen}},\ }\href {\doibase 10.3390/e23040404} {\bibfield  {journal} {\bibinfo  {journal} {Entropy}\ }\textbf {\bibinfo {volume} {23}},\ \bibinfo {pages} {404} (\bibinfo {year} {2021}{\natexlab{c}})},\ \Eprint {http://arxiv.org/abs/2005.12587} {arXiv:2005.12587 [astro-ph.CO]} \BibitemShut {NoStop}%
\bibitem [{\citenamefont {Alestas}\ \emph {et~al.}(2020)\citenamefont {Alestas}, \citenamefont {Kazantzidis},\ and\ \citenamefont {Perivolaropoulos}}]{Alestas:2020mvb}%
  \BibitemOpen
  \bibfield  {author} {\bibinfo {author} {\bibfnamefont {G.}~\bibnamefont {Alestas}}, \bibinfo {author} {\bibfnamefont {L.}~\bibnamefont {Kazantzidis}}, \ and\ \bibinfo {author} {\bibfnamefont {L.}~\bibnamefont {Perivolaropoulos}},\ }\href {\doibase 10.1103/PhysRevD.101.123516} {\bibfield  {journal} {\bibinfo  {journal} {Phys. Rev. D}\ }\textbf {\bibinfo {volume} {101}},\ \bibinfo {pages} {123516} (\bibinfo {year} {2020})},\ \Eprint {http://arxiv.org/abs/2004.08363} {arXiv:2004.08363 [astro-ph.CO]} \BibitemShut {NoStop}%
\bibitem [{\citenamefont {Alestas}\ \emph {et~al.}(2021{\natexlab{a}})\citenamefont {Alestas}, \citenamefont {Kazantzidis},\ and\ \citenamefont {Perivolaropoulos}}]{Alestas:2020zol}%
  \BibitemOpen
  \bibfield  {author} {\bibinfo {author} {\bibfnamefont {G.}~\bibnamefont {Alestas}}, \bibinfo {author} {\bibfnamefont {L.}~\bibnamefont {Kazantzidis}}, \ and\ \bibinfo {author} {\bibfnamefont {L.}~\bibnamefont {Perivolaropoulos}},\ }\href {\doibase 10.1103/PhysRevD.103.083517} {\bibfield  {journal} {\bibinfo  {journal} {Phys. Rev. D}\ }\textbf {\bibinfo {volume} {103}},\ \bibinfo {pages} {083517} (\bibinfo {year} {2021}{\natexlab{a}})},\ \Eprint {http://arxiv.org/abs/2012.13932} {arXiv:2012.13932 [astro-ph.CO]} \BibitemShut {NoStop}%
\bibitem [{\citenamefont {Gangopadhyay}\ \emph {et~al.}(2023{\natexlab{a}})\citenamefont {Gangopadhyay}, \citenamefont {Pacif}, \citenamefont {Sami},\ and\ \citenamefont {Sharma}}]{Gangopadhyay:2022bsh}%
  \BibitemOpen
  \bibfield  {author} {\bibinfo {author} {\bibfnamefont {M.~R.}\ \bibnamefont {Gangopadhyay}}, \bibinfo {author} {\bibfnamefont {S.~K.~J.}\ \bibnamefont {Pacif}}, \bibinfo {author} {\bibfnamefont {M.}~\bibnamefont {Sami}}, \ and\ \bibinfo {author} {\bibfnamefont {M.~K.}\ \bibnamefont {Sharma}},\ }\href {\doibase 10.3390/universe9020083} {\bibfield  {journal} {\bibinfo  {journal} {Universe}\ }\textbf {\bibinfo {volume} {9}},\ \bibinfo {pages} {83} (\bibinfo {year} {2023}{\natexlab{a}})},\ \Eprint {http://arxiv.org/abs/2211.12041} {arXiv:2211.12041 [gr-qc]} \BibitemShut {NoStop}%
\bibitem [{\citenamefont {Basilakos}\ \emph {et~al.}(2023)\citenamefont {Basilakos}, \citenamefont {Lymperis}, \citenamefont {Petronikolou},\ and\ \citenamefont {Saridakis}}]{Basilakos:2023kvk}%
  \BibitemOpen
  \bibfield  {author} {\bibinfo {author} {\bibfnamefont {S.}~\bibnamefont {Basilakos}}, \bibinfo {author} {\bibfnamefont {A.}~\bibnamefont {Lymperis}}, \bibinfo {author} {\bibfnamefont {M.}~\bibnamefont {Petronikolou}}, \ and\ \bibinfo {author} {\bibfnamefont {E.~N.}\ \bibnamefont {Saridakis}}\ }(\bibinfo {year} {2023})\ \Eprint {http://arxiv.org/abs/2308.01200} {arXiv:2308.01200 [gr-qc]} \BibitemShut {NoStop}%
\bibitem [{\citenamefont {Adil}\ \emph {et~al.}(2024)\citenamefont {Adil}, \citenamefont {Akarsu}, \citenamefont {Di~Valentino}, \citenamefont {Nunes}, \citenamefont {\"Oz\"ulker}, \citenamefont {Sen},\ and\ \citenamefont {Specogna}}]{Adil:2023exv}%
  \BibitemOpen
  \bibfield  {author} {\bibinfo {author} {\bibfnamefont {S.~A.}\ \bibnamefont {Adil}}, \bibinfo {author} {\bibfnamefont {O.}~\bibnamefont {Akarsu}}, \bibinfo {author} {\bibfnamefont {E.}~\bibnamefont {Di~Valentino}}, \bibinfo {author} {\bibfnamefont {R.~C.}\ \bibnamefont {Nunes}}, \bibinfo {author} {\bibfnamefont {E.}~\bibnamefont {\"Oz\"ulker}}, \bibinfo {author} {\bibfnamefont {A.~A.}\ \bibnamefont {Sen}}, \ and\ \bibinfo {author} {\bibfnamefont {E.}~\bibnamefont {Specogna}},\ }\href {\doibase 10.1103/PhysRevD.109.023527} {\bibfield  {journal} {\bibinfo  {journal} {Phys. Rev. D}\ }\textbf {\bibinfo {volume} {109}},\ \bibinfo {pages} {023527} (\bibinfo {year} {2024})},\ \Eprint {http://arxiv.org/abs/2306.08046} {arXiv:2306.08046 [astro-ph.CO]} \BibitemShut {NoStop}%
\bibitem [{\citenamefont {Gangopadhyay}\ \emph {et~al.}(2023{\natexlab{b}})\citenamefont {Gangopadhyay}, \citenamefont {Sami},\ and\ \citenamefont {Sharma}}]{Gangopadhyay:2023nli}%
  \BibitemOpen
  \bibfield  {author} {\bibinfo {author} {\bibfnamefont {M.~R.}\ \bibnamefont {Gangopadhyay}}, \bibinfo {author} {\bibfnamefont {M.}~\bibnamefont {Sami}}, \ and\ \bibinfo {author} {\bibfnamefont {M.~K.}\ \bibnamefont {Sharma}},\ }\href {\doibase 10.1103/PhysRevD.108.103526} {\bibfield  {journal} {\bibinfo  {journal} {Phys. Rev. D}\ }\textbf {\bibinfo {volume} {108}},\ \bibinfo {pages} {103526} (\bibinfo {year} {2023}{\natexlab{b}})},\ \Eprint {http://arxiv.org/abs/2303.07301} {arXiv:2303.07301 [astro-ph.CO]} \BibitemShut {NoStop}%
\bibitem [{\citenamefont {Visinelli}\ \emph {et~al.}(2019)\citenamefont {Visinelli}, \citenamefont {Vagnozzi},\ and\ \citenamefont {Danielsson}}]{Visinelli:2019qqu}%
  \BibitemOpen
  \bibfield  {author} {\bibinfo {author} {\bibfnamefont {L.}~\bibnamefont {Visinelli}}, \bibinfo {author} {\bibfnamefont {S.}~\bibnamefont {Vagnozzi}}, \ and\ \bibinfo {author} {\bibfnamefont {U.}~\bibnamefont {Danielsson}},\ }\href {\doibase 10.3390/sym11081035} {\bibfield  {journal} {\bibinfo  {journal} {Symmetry}\ }\textbf {\bibinfo {volume} {11}},\ \bibinfo {pages} {1035} (\bibinfo {year} {2019})},\ \Eprint {http://arxiv.org/abs/1907.07953} {arXiv:1907.07953 [astro-ph.CO]} \BibitemShut {NoStop}%
\bibitem [{\citenamefont {Dutta}\ \emph {et~al.}(2020)\citenamefont {Dutta}, \citenamefont {Ruchika}, \citenamefont {Roy}, \citenamefont {Sen},\ and\ \citenamefont {Sheikh-Jabbari}}]{Dutta:2018vmq}%
  \BibitemOpen
  \bibfield  {author} {\bibinfo {author} {\bibfnamefont {K.}~\bibnamefont {Dutta}}, \bibinfo {author} {\bibnamefont {Ruchika}}, \bibinfo {author} {\bibfnamefont {A.}~\bibnamefont {Roy}}, \bibinfo {author} {\bibfnamefont {A.~A.}\ \bibnamefont {Sen}}, \ and\ \bibinfo {author} {\bibfnamefont {M.~M.}\ \bibnamefont {Sheikh-Jabbari}},\ }\href {\doibase 10.1007/s10714-020-2665-4} {\bibfield  {journal} {\bibinfo  {journal} {Gen. Rel. Grav.}\ }\textbf {\bibinfo {volume} {52}},\ \bibinfo {pages} {15} (\bibinfo {year} {2020})},\ \Eprint {http://arxiv.org/abs/1808.06623} {arXiv:1808.06623 [astro-ph.CO]} \BibitemShut {NoStop}%
\bibitem [{\citenamefont {Sen}\ \emph {et~al.}(2022)\citenamefont {Sen}, \citenamefont {Adil},\ and\ \citenamefont {Sen}}]{Sen:2021wld}%
  \BibitemOpen
  \bibfield  {author} {\bibinfo {author} {\bibfnamefont {A.~A.}\ \bibnamefont {Sen}}, \bibinfo {author} {\bibfnamefont {S.~A.}\ \bibnamefont {Adil}}, \ and\ \bibinfo {author} {\bibfnamefont {S.}~\bibnamefont {Sen}},\ }\href {\doibase 10.1093/mnras/stac2796} {\bibfield  {journal} {\bibinfo  {journal} {Mon. Not. Roy. Astron. Soc.}\ }\textbf {\bibinfo {volume} {518}},\ \bibinfo {pages} {1098} (\bibinfo {year} {2022})},\ \Eprint {http://arxiv.org/abs/2112.10641} {arXiv:2112.10641 [astro-ph.CO]} \BibitemShut {NoStop}%
\bibitem [{\citenamefont {Kumar}\ and\ \citenamefont {Nunes}(2017)}]{Kumar:2017dnp}%
  \BibitemOpen
  \bibfield  {author} {\bibinfo {author} {\bibfnamefont {S.}~\bibnamefont {Kumar}}\ and\ \bibinfo {author} {\bibfnamefont {R.~C.}\ \bibnamefont {Nunes}},\ }\href {\doibase 10.1103/PhysRevD.96.103511} {\bibfield  {journal} {\bibinfo  {journal} {Phys. Rev. D}\ }\textbf {\bibinfo {volume} {96}},\ \bibinfo {pages} {103511} (\bibinfo {year} {2017})},\ \Eprint {http://arxiv.org/abs/1702.02143} {arXiv:1702.02143 [astro-ph.CO]} \BibitemShut {NoStop}%
\bibitem [{\citenamefont {Di~Valentino}\ \emph {et~al.}(2017)\citenamefont {Di~Valentino}, \citenamefont {Melchiorri},\ and\ \citenamefont {Mena}}]{DiValentino:2017iww}%
  \BibitemOpen
  \bibfield  {author} {\bibinfo {author} {\bibfnamefont {E.}~\bibnamefont {Di~Valentino}}, \bibinfo {author} {\bibfnamefont {A.}~\bibnamefont {Melchiorri}}, \ and\ \bibinfo {author} {\bibfnamefont {O.}~\bibnamefont {Mena}},\ }\href {\doibase 10.1103/PhysRevD.96.043503} {\bibfield  {journal} {\bibinfo  {journal} {Phys. Rev. D}\ }\textbf {\bibinfo {volume} {96}},\ \bibinfo {pages} {043503} (\bibinfo {year} {2017})},\ \Eprint {http://arxiv.org/abs/1704.08342} {arXiv:1704.08342 [astro-ph.CO]} \BibitemShut {NoStop}%
\bibitem [{\citenamefont {Yang}\ \emph {et~al.}(2018)\citenamefont {Yang}, \citenamefont {Mukherjee}, \citenamefont {Di~Valentino},\ and\ \citenamefont {Pan}}]{Yang:2018uae}%
  \BibitemOpen
  \bibfield  {author} {\bibinfo {author} {\bibfnamefont {W.}~\bibnamefont {Yang}}, \bibinfo {author} {\bibfnamefont {A.}~\bibnamefont {Mukherjee}}, \bibinfo {author} {\bibfnamefont {E.}~\bibnamefont {Di~Valentino}}, \ and\ \bibinfo {author} {\bibfnamefont {S.}~\bibnamefont {Pan}},\ }\href {\doibase 10.1103/PhysRevD.98.123527} {\bibfield  {journal} {\bibinfo  {journal} {Phys. Rev. D}\ }\textbf {\bibinfo {volume} {98}},\ \bibinfo {pages} {123527} (\bibinfo {year} {2018})},\ \Eprint {http://arxiv.org/abs/1809.06883} {arXiv:1809.06883 [astro-ph.CO]} \BibitemShut {NoStop}%
\bibitem [{\citenamefont {Pan}\ \emph {et~al.}(2019)\citenamefont {Pan}, \citenamefont {Yang}, \citenamefont {Di~Valentino}, \citenamefont {Saridakis},\ and\ \citenamefont {Chakraborty}}]{Pan:2019gop}%
  \BibitemOpen
  \bibfield  {author} {\bibinfo {author} {\bibfnamefont {S.}~\bibnamefont {Pan}}, \bibinfo {author} {\bibfnamefont {W.}~\bibnamefont {Yang}}, \bibinfo {author} {\bibfnamefont {E.}~\bibnamefont {Di~Valentino}}, \bibinfo {author} {\bibfnamefont {E.~N.}\ \bibnamefont {Saridakis}}, \ and\ \bibinfo {author} {\bibfnamefont {S.}~\bibnamefont {Chakraborty}},\ }\href {\doibase 10.1103/PhysRevD.100.103520} {\bibfield  {journal} {\bibinfo  {journal} {Phys. Rev. D}\ }\textbf {\bibinfo {volume} {100}},\ \bibinfo {pages} {103520} (\bibinfo {year} {2019})},\ \Eprint {http://arxiv.org/abs/1907.07540} {arXiv:1907.07540 [astro-ph.CO]} \BibitemShut {NoStop}%
\bibitem [{\citenamefont {Kumar}\ \emph {et~al.}(2019)\citenamefont {Kumar}, \citenamefont {Nunes},\ and\ \citenamefont {Yadav}}]{Kumar:2019wfs}%
  \BibitemOpen
  \bibfield  {author} {\bibinfo {author} {\bibfnamefont {S.}~\bibnamefont {Kumar}}, \bibinfo {author} {\bibfnamefont {R.~C.}\ \bibnamefont {Nunes}}, \ and\ \bibinfo {author} {\bibfnamefont {S.~K.}\ \bibnamefont {Yadav}},\ }\href {\doibase 10.1140/epjc/s10052-019-7087-7} {\bibfield  {journal} {\bibinfo  {journal} {Eur. Phys. J. C}\ }\textbf {\bibinfo {volume} {79}},\ \bibinfo {pages} {576} (\bibinfo {year} {2019})},\ \Eprint {http://arxiv.org/abs/1903.04865} {arXiv:1903.04865 [astro-ph.CO]} \BibitemShut {NoStop}%
\bibitem [{\citenamefont {Di~Valentino}\ \emph {et~al.}(2020{\natexlab{a}})\citenamefont {Di~Valentino}, \citenamefont {Melchiorri}, \citenamefont {Mena},\ and\ \citenamefont {Vagnozzi}}]{DiValentino:2019jae}%
  \BibitemOpen
  \bibfield  {author} {\bibinfo {author} {\bibfnamefont {E.}~\bibnamefont {Di~Valentino}}, \bibinfo {author} {\bibfnamefont {A.}~\bibnamefont {Melchiorri}}, \bibinfo {author} {\bibfnamefont {O.}~\bibnamefont {Mena}}, \ and\ \bibinfo {author} {\bibfnamefont {S.}~\bibnamefont {Vagnozzi}},\ }\href {\doibase 10.1103/PhysRevD.101.063502} {\bibfield  {journal} {\bibinfo  {journal} {Phys. Rev. D}\ }\textbf {\bibinfo {volume} {101}},\ \bibinfo {pages} {063502} (\bibinfo {year} {2020}{\natexlab{a}})},\ \Eprint {http://arxiv.org/abs/1910.09853} {arXiv:1910.09853 [astro-ph.CO]} \BibitemShut {NoStop}%
\bibitem [{\citenamefont {Di~Valentino}\ \emph {et~al.}(2020{\natexlab{b}})\citenamefont {Di~Valentino}, \citenamefont {Melchiorri}, \citenamefont {Mena},\ and\ \citenamefont {Vagnozzi}}]{DiValentino:2019ffd}%
  \BibitemOpen
  \bibfield  {author} {\bibinfo {author} {\bibfnamefont {E.}~\bibnamefont {Di~Valentino}}, \bibinfo {author} {\bibfnamefont {A.}~\bibnamefont {Melchiorri}}, \bibinfo {author} {\bibfnamefont {O.}~\bibnamefont {Mena}}, \ and\ \bibinfo {author} {\bibfnamefont {S.}~\bibnamefont {Vagnozzi}},\ }\href {\doibase 10.1016/j.dark.2020.100666} {\bibfield  {journal} {\bibinfo  {journal} {Phys. Dark Univ.}\ }\textbf {\bibinfo {volume} {30}},\ \bibinfo {pages} {100666} (\bibinfo {year} {2020}{\natexlab{b}})},\ \Eprint {http://arxiv.org/abs/1908.04281} {arXiv:1908.04281 [astro-ph.CO]} \BibitemShut {NoStop}%
\bibitem [{\citenamefont {Lucca}\ and\ \citenamefont {Hooper}(2020)}]{Lucca:2020zjb}%
  \BibitemOpen
  \bibfield  {author} {\bibinfo {author} {\bibfnamefont {M.}~\bibnamefont {Lucca}}\ and\ \bibinfo {author} {\bibfnamefont {D.~C.}\ \bibnamefont {Hooper}},\ }\href {\doibase 10.1103/PhysRevD.102.123502} {\bibfield  {journal} {\bibinfo  {journal} {Phys. Rev. D}\ }\textbf {\bibinfo {volume} {102}},\ \bibinfo {pages} {123502} (\bibinfo {year} {2020})},\ \Eprint {http://arxiv.org/abs/2002.06127} {arXiv:2002.06127 [astro-ph.CO]} \BibitemShut {NoStop}%
\bibitem [{\citenamefont {G\'omez-Valent}\ \emph {et~al.}(2020)\citenamefont {G\'omez-Valent}, \citenamefont {Pettorino},\ and\ \citenamefont {Amendola}}]{Gomez-Valent:2020mqn}%
  \BibitemOpen
  \bibfield  {author} {\bibinfo {author} {\bibfnamefont {A.}~\bibnamefont {G\'omez-Valent}}, \bibinfo {author} {\bibfnamefont {V.}~\bibnamefont {Pettorino}}, \ and\ \bibinfo {author} {\bibfnamefont {L.}~\bibnamefont {Amendola}},\ }\href {\doibase 10.1103/PhysRevD.101.123513} {\bibfield  {journal} {\bibinfo  {journal} {Phys. Rev. D}\ }\textbf {\bibinfo {volume} {101}},\ \bibinfo {pages} {123513} (\bibinfo {year} {2020})},\ \Eprint {http://arxiv.org/abs/2004.00610} {arXiv:2004.00610 [astro-ph.CO]} \BibitemShut {NoStop}%
\bibitem [{\citenamefont {Kumar}(2021)}]{Kumar:2021eev}%
  \BibitemOpen
  \bibfield  {author} {\bibinfo {author} {\bibfnamefont {S.}~\bibnamefont {Kumar}},\ }\href {\doibase 10.1016/j.dark.2021.100862} {\bibfield  {journal} {\bibinfo  {journal} {Phys. Dark Univ.}\ }\textbf {\bibinfo {volume} {33}},\ \bibinfo {pages} {100862} (\bibinfo {year} {2021})},\ \Eprint {http://arxiv.org/abs/2102.12902} {arXiv:2102.12902 [astro-ph.CO]} \BibitemShut {NoStop}%
\bibitem [{\citenamefont {Nunes}\ \emph {et~al.}(2022)\citenamefont {Nunes}, \citenamefont {Vagnozzi}, \citenamefont {Kumar}, \citenamefont {Di~Valentino},\ and\ \citenamefont {Mena}}]{Nunes:2022bhn}%
  \BibitemOpen
  \bibfield  {author} {\bibinfo {author} {\bibfnamefont {R.~C.}\ \bibnamefont {Nunes}}, \bibinfo {author} {\bibfnamefont {S.}~\bibnamefont {Vagnozzi}}, \bibinfo {author} {\bibfnamefont {S.}~\bibnamefont {Kumar}}, \bibinfo {author} {\bibfnamefont {E.}~\bibnamefont {Di~Valentino}}, \ and\ \bibinfo {author} {\bibfnamefont {O.}~\bibnamefont {Mena}},\ }\href {\doibase 10.1103/PhysRevD.105.123506} {\bibfield  {journal} {\bibinfo  {journal} {Phys. Rev. D}\ }\textbf {\bibinfo {volume} {105}},\ \bibinfo {pages} {123506} (\bibinfo {year} {2022})},\ \Eprint {http://arxiv.org/abs/2203.08093} {arXiv:2203.08093 [astro-ph.CO]} \BibitemShut {NoStop}%
\bibitem [{\citenamefont {Bernui}\ \emph {et~al.}(2023)\citenamefont {Bernui}, \citenamefont {Di~Valentino}, \citenamefont {Giar\`e}, \citenamefont {Kumar},\ and\ \citenamefont {Nunes}}]{Bernui:2023byc}%
  \BibitemOpen
  \bibfield  {author} {\bibinfo {author} {\bibfnamefont {A.}~\bibnamefont {Bernui}}, \bibinfo {author} {\bibfnamefont {E.}~\bibnamefont {Di~Valentino}}, \bibinfo {author} {\bibfnamefont {W.}~\bibnamefont {Giar\`e}}, \bibinfo {author} {\bibfnamefont {S.}~\bibnamefont {Kumar}}, \ and\ \bibinfo {author} {\bibfnamefont {R.~C.}\ \bibnamefont {Nunes}},\ }\href {\doibase 10.1103/PhysRevD.107.103531} {\bibfield  {journal} {\bibinfo  {journal} {Phys. Rev. D}\ }\textbf {\bibinfo {volume} {107}},\ \bibinfo {pages} {103531} (\bibinfo {year} {2023})},\ \Eprint {http://arxiv.org/abs/2301.06097} {arXiv:2301.06097 [astro-ph.CO]} \BibitemShut {NoStop}%
\bibitem [{\citenamefont {Aubourg}\ \emph {et~al.}(2015)\citenamefont {Aubourg} \emph {et~al.}}]{Aubourg:2014yra}%
  \BibitemOpen
  \bibfield  {author} {\bibinfo {author} {\bibfnamefont {E.}~\bibnamefont {Aubourg}} \emph {et~al.},\ }\href {\doibase 10.1103/PhysRevD.92.123516} {\bibfield  {journal} {\bibinfo  {journal} {Phys. Rev. D}\ }\textbf {\bibinfo {volume} {92}},\ \bibinfo {pages} {123516} (\bibinfo {year} {2015})},\ \Eprint {http://arxiv.org/abs/1411.1074} {arXiv:1411.1074 [astro-ph.CO]} \BibitemShut {NoStop}%
\bibitem [{\citenamefont {Sahni}\ \emph {et~al.}(2014)\citenamefont {Sahni}, \citenamefont {Shafieloo},\ and\ \citenamefont {Starobinsky}}]{Sahni:2014ooa}%
  \BibitemOpen
  \bibfield  {author} {\bibinfo {author} {\bibfnamefont {V.}~\bibnamefont {Sahni}}, \bibinfo {author} {\bibfnamefont {A.}~\bibnamefont {Shafieloo}}, \ and\ \bibinfo {author} {\bibfnamefont {A.~A.}\ \bibnamefont {Starobinsky}},\ }\href {\doibase 10.1088/2041-8205/793/2/L40} {\bibfield  {journal} {\bibinfo  {journal} {Astrophys. J. Lett.}\ }\textbf {\bibinfo {volume} {793}},\ \bibinfo {pages} {L40} (\bibinfo {year} {2014})},\ \Eprint {http://arxiv.org/abs/1406.2209} {arXiv:1406.2209 [astro-ph.CO]} \BibitemShut {NoStop}%
\bibitem [{\citenamefont {Poulin}\ \emph {et~al.}(2018{\natexlab{b}})\citenamefont {Poulin}, \citenamefont {Boddy}, \citenamefont {Bird},\ and\ \citenamefont {Kamionkowski}}]{Poulin:2018zxs}%
  \BibitemOpen
  \bibfield  {author} {\bibinfo {author} {\bibfnamefont {V.}~\bibnamefont {Poulin}}, \bibinfo {author} {\bibfnamefont {K.~K.}\ \bibnamefont {Boddy}}, \bibinfo {author} {\bibfnamefont {S.}~\bibnamefont {Bird}}, \ and\ \bibinfo {author} {\bibfnamefont {M.}~\bibnamefont {Kamionkowski}},\ }\href {\doibase 10.1103/PhysRevD.97.123504} {\bibfield  {journal} {\bibinfo  {journal} {Phys. Rev. D}\ }\textbf {\bibinfo {volume} {97}},\ \bibinfo {pages} {123504} (\bibinfo {year} {2018}{\natexlab{b}})},\ \Eprint {http://arxiv.org/abs/1803.02474} {arXiv:1803.02474 [astro-ph.CO]} \BibitemShut {NoStop}%
\bibitem [{\citenamefont {Wang}\ \emph {et~al.}(2018)\citenamefont {Wang}, \citenamefont {Pogosian}, \citenamefont {Zhao},\ and\ \citenamefont {Zucca}}]{Wang:2018fng}%
  \BibitemOpen
  \bibfield  {author} {\bibinfo {author} {\bibfnamefont {Y.}~\bibnamefont {Wang}}, \bibinfo {author} {\bibfnamefont {L.}~\bibnamefont {Pogosian}}, \bibinfo {author} {\bibfnamefont {G.-B.}\ \bibnamefont {Zhao}}, \ and\ \bibinfo {author} {\bibfnamefont {A.}~\bibnamefont {Zucca}},\ }\href {\doibase 10.3847/2041-8213/aaf238} {\bibfield  {journal} {\bibinfo  {journal} {Astrophys. J. Lett.}\ }\textbf {\bibinfo {volume} {869}},\ \bibinfo {pages} {L8} (\bibinfo {year} {2018})},\ \Eprint {http://arxiv.org/abs/1807.03772} {arXiv:1807.03772 [astro-ph.CO]} \BibitemShut {NoStop}%
\bibitem [{\citenamefont {Bonilla}\ \emph {et~al.}(2021)\citenamefont {Bonilla}, \citenamefont {Kumar},\ and\ \citenamefont {Nunes}}]{Bonilla:2020wbn}%
  \BibitemOpen
  \bibfield  {author} {\bibinfo {author} {\bibfnamefont {A.}~\bibnamefont {Bonilla}}, \bibinfo {author} {\bibfnamefont {S.}~\bibnamefont {Kumar}}, \ and\ \bibinfo {author} {\bibfnamefont {R.~C.}\ \bibnamefont {Nunes}},\ }\href {\doibase 10.1140/epjc/s10052-021-08925-z} {\bibfield  {journal} {\bibinfo  {journal} {Eur. Phys. J. C}\ }\textbf {\bibinfo {volume} {81}},\ \bibinfo {pages} {127} (\bibinfo {year} {2021})},\ \Eprint {http://arxiv.org/abs/2011.07140} {arXiv:2011.07140 [astro-ph.CO]} \BibitemShut {NoStop}%
\bibitem [{\citenamefont {Escamilla}\ \emph {et~al.}(2023)\citenamefont {Escamilla}, \citenamefont {Akarsu}, \citenamefont {Di~Valentino},\ and\ \citenamefont {Vazquez}}]{Escamilla:2023shf}%
  \BibitemOpen
  \bibfield  {author} {\bibinfo {author} {\bibfnamefont {L.~A.}\ \bibnamefont {Escamilla}}, \bibinfo {author} {\bibfnamefont {O.}~\bibnamefont {Akarsu}}, \bibinfo {author} {\bibfnamefont {E.}~\bibnamefont {Di~Valentino}}, \ and\ \bibinfo {author} {\bibfnamefont {J.~A.}\ \bibnamefont {Vazquez}},\ }\href {\doibase 10.1088/1475-7516/2023/11/051} {\bibfield  {journal} {\bibinfo  {journal} {JCAP}\ }\textbf {\bibinfo {volume} {11}},\ \bibinfo {pages} {051} (\bibinfo {year} {2023})},\ \Eprint {http://arxiv.org/abs/2305.16290} {arXiv:2305.16290 [astro-ph.CO]} \BibitemShut {NoStop}%
\bibitem [{\citenamefont {Escamilla}\ and\ \citenamefont {Vazquez}(2023)}]{Escamilla:2021uoj}%
  \BibitemOpen
  \bibfield  {author} {\bibinfo {author} {\bibfnamefont {L.~A.}\ \bibnamefont {Escamilla}}\ and\ \bibinfo {author} {\bibfnamefont {J.~A.}\ \bibnamefont {Vazquez}},\ }\href {\doibase 10.1140/epjc/s10052-023-11404-2} {\bibfield  {journal} {\bibinfo  {journal} {Eur. Phys. J. C}\ }\textbf {\bibinfo {volume} {83}},\ \bibinfo {pages} {251} (\bibinfo {year} {2023})},\ \Eprint {http://arxiv.org/abs/2111.10457} {arXiv:2111.10457 [astro-ph.CO]} \BibitemShut {NoStop}%
\bibitem [{\citenamefont {Malekjani}\ \emph {et~al.}(2023)\citenamefont {Malekjani}, \citenamefont {Conville}, \citenamefont {Colg\'ain}, \citenamefont {Pourojaghi},\ and\ \citenamefont {Sheikh-Jabbari}}]{Malekjani:2023dky}%
  \BibitemOpen
  \bibfield  {author} {\bibinfo {author} {\bibfnamefont {M.}~\bibnamefont {Malekjani}}, \bibinfo {author} {\bibfnamefont {R.~M.}\ \bibnamefont {Conville}}, \bibinfo {author} {\bibfnamefont {E.~O.}\ \bibnamefont {Colg\'ain}}, \bibinfo {author} {\bibfnamefont {S.}~\bibnamefont {Pourojaghi}}, \ and\ \bibinfo {author} {\bibfnamefont {M.~M.}\ \bibnamefont {Sheikh-Jabbari}},\ }\href@noop {} {\  (\bibinfo {year} {2023})},\ \Eprint {http://arxiv.org/abs/2301.12725} {arXiv:2301.12725 [astro-ph.CO]} \BibitemShut {NoStop}%
\bibitem [{\citenamefont {Akarsu}\ \emph {et~al.}(2023{\natexlab{c}})\citenamefont {Akarsu}, \citenamefont {Colgain}, \citenamefont {\"Ozulker}, \citenamefont {Thakur},\ and\ \citenamefont {Yin}}]{Akarsu:2022lhx}%
  \BibitemOpen
  \bibfield  {author} {\bibinfo {author} {\bibfnamefont {O.}~\bibnamefont {Akarsu}}, \bibinfo {author} {\bibfnamefont {E.~O.}\ \bibnamefont {Colgain}}, \bibinfo {author} {\bibfnamefont {E.}~\bibnamefont {\"Ozulker}}, \bibinfo {author} {\bibfnamefont {S.}~\bibnamefont {Thakur}}, \ and\ \bibinfo {author} {\bibfnamefont {L.}~\bibnamefont {Yin}},\ }\href {\doibase 10.1103/PhysRevD.107.123526} {\bibfield  {journal} {\bibinfo  {journal} {Phys. Rev. D}\ }\textbf {\bibinfo {volume} {107}},\ \bibinfo {pages} {123526} (\bibinfo {year} {2023}{\natexlab{c}})},\ \Eprint {http://arxiv.org/abs/2207.10609} {arXiv:2207.10609 [astro-ph.CO]} \BibitemShut {NoStop}%
\bibitem [{\citenamefont {Calder\'on}\ \emph {et~al.}(2021)\citenamefont {Calder\'on}, \citenamefont {Gannouji}, \citenamefont {L'Huillier},\ and\ \citenamefont {Polarski}}]{Calderon:2020hoc}%
  \BibitemOpen
  \bibfield  {author} {\bibinfo {author} {\bibfnamefont {R.}~\bibnamefont {Calder\'on}}, \bibinfo {author} {\bibfnamefont {R.}~\bibnamefont {Gannouji}}, \bibinfo {author} {\bibfnamefont {B.}~\bibnamefont {L'Huillier}}, \ and\ \bibinfo {author} {\bibfnamefont {D.}~\bibnamefont {Polarski}},\ }\href {\doibase 10.1103/PhysRevD.103.023526} {\bibfield  {journal} {\bibinfo  {journal} {Phys. Rev. D}\ }\textbf {\bibinfo {volume} {103}},\ \bibinfo {pages} {023526} (\bibinfo {year} {2021})},\ \Eprint {http://arxiv.org/abs/2008.10237} {arXiv:2008.10237 [astro-ph.CO]} \BibitemShut {NoStop}%
\bibitem [{\citenamefont {Marra}\ and\ \citenamefont {Perivolaropoulos}(2021)}]{Marra:2021fvf}%
  \BibitemOpen
  \bibfield  {author} {\bibinfo {author} {\bibfnamefont {V.}~\bibnamefont {Marra}}\ and\ \bibinfo {author} {\bibfnamefont {L.}~\bibnamefont {Perivolaropoulos}},\ }\href {\doibase 10.1103/PhysRevD.104.L021303} {\bibfield  {journal} {\bibinfo  {journal} {Phys. Rev. D}\ }\textbf {\bibinfo {volume} {104}},\ \bibinfo {pages} {L021303} (\bibinfo {year} {2021})},\ \Eprint {http://arxiv.org/abs/2102.06012} {arXiv:2102.06012 [astro-ph.CO]} \BibitemShut {NoStop}%
\bibitem [{\citenamefont {Alestas}\ \emph {et~al.}(2021{\natexlab{b}})\citenamefont {Alestas}, \citenamefont {Antoniou},\ and\ \citenamefont {Perivolaropoulos}}]{Alestas:2021nmi}%
  \BibitemOpen
  \bibfield  {author} {\bibinfo {author} {\bibfnamefont {G.}~\bibnamefont {Alestas}}, \bibinfo {author} {\bibfnamefont {I.}~\bibnamefont {Antoniou}}, \ and\ \bibinfo {author} {\bibfnamefont {L.}~\bibnamefont {Perivolaropoulos}},\ }\href {\doibase 10.3390/universe7100366} {\bibfield  {journal} {\bibinfo  {journal} {Universe}\ }\textbf {\bibinfo {volume} {7}},\ \bibinfo {pages} {366} (\bibinfo {year} {2021}{\natexlab{b}})},\ \Eprint {http://arxiv.org/abs/2104.14481} {arXiv:2104.14481 [astro-ph.CO]} \BibitemShut {NoStop}%
\bibitem [{\citenamefont {Alestas}\ \emph {et~al.}(2022)\citenamefont {Alestas}, \citenamefont {Camarena}, \citenamefont {Di~Valentino}, \citenamefont {Kazantzidis}, \citenamefont {Marra}, \citenamefont {Nesseris},\ and\ \citenamefont {Perivolaropoulos}}]{Alestas:2021luu}%
  \BibitemOpen
  \bibfield  {author} {\bibinfo {author} {\bibfnamefont {G.}~\bibnamefont {Alestas}}, \bibinfo {author} {\bibfnamefont {D.}~\bibnamefont {Camarena}}, \bibinfo {author} {\bibfnamefont {E.}~\bibnamefont {Di~Valentino}}, \bibinfo {author} {\bibfnamefont {L.}~\bibnamefont {Kazantzidis}}, \bibinfo {author} {\bibfnamefont {V.}~\bibnamefont {Marra}}, \bibinfo {author} {\bibfnamefont {S.}~\bibnamefont {Nesseris}}, \ and\ \bibinfo {author} {\bibfnamefont {L.}~\bibnamefont {Perivolaropoulos}},\ }\href {\doibase 10.1103/PhysRevD.105.063538} {\bibfield  {journal} {\bibinfo  {journal} {Phys. Rev. D}\ }\textbf {\bibinfo {volume} {105}},\ \bibinfo {pages} {063538} (\bibinfo {year} {2022})},\ \Eprint {http://arxiv.org/abs/2110.04336} {arXiv:2110.04336 [astro-ph.CO]} \BibitemShut {NoStop}%
\bibitem [{\citenamefont {Perivolaropoulos}\ and\ \citenamefont {Skara}(2021)}]{Perivolaropoulos:2021bds}%
  \BibitemOpen
  \bibfield  {author} {\bibinfo {author} {\bibfnamefont {L.}~\bibnamefont {Perivolaropoulos}}\ and\ \bibinfo {author} {\bibfnamefont {F.}~\bibnamefont {Skara}},\ }\href {\doibase 10.1103/PhysRevD.104.123511} {\bibfield  {journal} {\bibinfo  {journal} {Phys. Rev. D}\ }\textbf {\bibinfo {volume} {104}},\ \bibinfo {pages} {123511} (\bibinfo {year} {2021})},\ \Eprint {http://arxiv.org/abs/2109.04406} {arXiv:2109.04406 [astro-ph.CO]} \BibitemShut {NoStop}%
\bibitem [{\citenamefont {Pan}\ \emph {et~al.}(2023)\citenamefont {Pan}, \citenamefont {Seto}, \citenamefont {Takahashi},\ and\ \citenamefont {Toda}}]{Pan:2023frx}%
  \BibitemOpen
  \bibfield  {author} {\bibinfo {author} {\bibfnamefont {S.}~\bibnamefont {Pan}}, \bibinfo {author} {\bibfnamefont {O.}~\bibnamefont {Seto}}, \bibinfo {author} {\bibfnamefont {T.}~\bibnamefont {Takahashi}}, \ and\ \bibinfo {author} {\bibfnamefont {Y.}~\bibnamefont {Toda}},\ }\href@noop {} {\  (\bibinfo {year} {2023})},\ \Eprint {http://arxiv.org/abs/2312.15435} {arXiv:2312.15435 [astro-ph.CO]} \BibitemShut {NoStop}%
\bibitem [{\citenamefont {Naidoo}\ \emph {et~al.}(2022)\citenamefont {Naidoo}, \citenamefont {Jaber}, \citenamefont {Hellwing},\ and\ \citenamefont {Bilicki}}]{Naidoo:2022rda}%
  \BibitemOpen
  \bibfield  {author} {\bibinfo {author} {\bibfnamefont {K.}~\bibnamefont {Naidoo}}, \bibinfo {author} {\bibfnamefont {M.}~\bibnamefont {Jaber}}, \bibinfo {author} {\bibfnamefont {W.~A.}\ \bibnamefont {Hellwing}}, \ and\ \bibinfo {author} {\bibfnamefont {M.}~\bibnamefont {Bilicki}},\ }\href@noop {} {\  (\bibinfo {year} {2022})},\ \Eprint {http://arxiv.org/abs/2209.08102} {arXiv:2209.08102 [astro-ph.CO]} \BibitemShut {NoStop}%
\bibitem [{\citenamefont {Boylan-Kolchin}(2023)}]{Boylan-Kolchin:2022kae}%
  \BibitemOpen
  \bibfield  {author} {\bibinfo {author} {\bibfnamefont {M.}~\bibnamefont {Boylan-Kolchin}},\ }\href {\doibase 10.1038/s41550-023-01937-7} {\bibfield  {journal} {\bibinfo  {journal} {Nature Astron.}\ }\textbf {\bibinfo {volume} {7}},\ \bibinfo {pages} {731} (\bibinfo {year} {2023})},\ \Eprint {http://arxiv.org/abs/2208.01611} {arXiv:2208.01611 [astro-ph.CO]} \BibitemShut {NoStop}%
\bibitem [{\citenamefont {{Labb{\'e}}}\ \emph {et~al.}(2023)\citenamefont {{Labb{\'e}}}, \citenamefont {{van Dokkum}}, \citenamefont {{Nelson}}, \citenamefont {{Bezanson}}, \citenamefont {{Suess}}, \citenamefont {{Leja}}, \citenamefont {{Brammer}}, \citenamefont {{Whitaker}}, \citenamefont {{Mathews}}, \citenamefont {{Stefanon}},\ and\ \citenamefont {{Wang}}}]{2023Natur.616..266L}%
  \BibitemOpen
  \bibfield  {author} {\bibinfo {author} {\bibfnamefont {I.}~\bibnamefont {{Labb{\'e}}}}, \bibinfo {author} {\bibfnamefont {P.}~\bibnamefont {{van Dokkum}}}, \bibinfo {author} {\bibfnamefont {E.}~\bibnamefont {{Nelson}}}, \bibinfo {author} {\bibfnamefont {R.}~\bibnamefont {{Bezanson}}}, \bibinfo {author} {\bibfnamefont {K.~A.}\ \bibnamefont {{Suess}}}, \bibinfo {author} {\bibfnamefont {J.}~\bibnamefont {{Leja}}}, \bibinfo {author} {\bibfnamefont {G.}~\bibnamefont {{Brammer}}}, \bibinfo {author} {\bibfnamefont {K.}~\bibnamefont {{Whitaker}}}, \bibinfo {author} {\bibfnamefont {E.}~\bibnamefont {{Mathews}}}, \bibinfo {author} {\bibfnamefont {M.}~\bibnamefont {{Stefanon}}}, \ and\ \bibinfo {author} {\bibfnamefont {B.}~\bibnamefont {{Wang}}},\ }\href {\doibase 10.1038/s41586-023-05786-2} {\bibfield  {journal} {\bibinfo  {journal} {\nat}\ }\textbf {\bibinfo {volume} {616}},\ \bibinfo {pages} {266} (\bibinfo {year} {2023})},\ \Eprint {http://arxiv.org/abs/2207.12446} {arXiv:2207.12446 [astro-ph.GA]} \BibitemShut
  {NoStop}%
\bibitem [{\citenamefont {Menci}\ \emph {et~al.}(2022)\citenamefont {Menci}, \citenamefont {Castellano}, \citenamefont {Santini}, \citenamefont {Merlin}, \citenamefont {Fontana},\ and\ \citenamefont {Shankar}}]{Menci:2022wia}%
  \BibitemOpen
  \bibfield  {author} {\bibinfo {author} {\bibfnamefont {N.}~\bibnamefont {Menci}}, \bibinfo {author} {\bibfnamefont {M.}~\bibnamefont {Castellano}}, \bibinfo {author} {\bibfnamefont {P.}~\bibnamefont {Santini}}, \bibinfo {author} {\bibfnamefont {E.}~\bibnamefont {Merlin}}, \bibinfo {author} {\bibfnamefont {A.}~\bibnamefont {Fontana}}, \ and\ \bibinfo {author} {\bibfnamefont {F.}~\bibnamefont {Shankar}},\ }\href {\doibase 10.3847/2041-8213/ac96e9} {\bibfield  {journal} {\bibinfo  {journal} {Astrophys. J. Lett.}\ }\textbf {\bibinfo {volume} {938}},\ \bibinfo {pages} {L5} (\bibinfo {year} {2022})},\ \Eprint {http://arxiv.org/abs/2208.11471} {arXiv:2208.11471 [astro-ph.CO]} \BibitemShut {NoStop}%
\bibitem [{\citenamefont {Biagetti}\ \emph {et~al.}(2023)\citenamefont {Biagetti}, \citenamefont {Franciolini},\ and\ \citenamefont {Riotto}}]{Biagetti:2022ode}%
  \BibitemOpen
  \bibfield  {author} {\bibinfo {author} {\bibfnamefont {M.}~\bibnamefont {Biagetti}}, \bibinfo {author} {\bibfnamefont {G.}~\bibnamefont {Franciolini}}, \ and\ \bibinfo {author} {\bibfnamefont {A.}~\bibnamefont {Riotto}},\ }\href {\doibase 10.3847/1538-4357/acb5ea} {\bibfield  {journal} {\bibinfo  {journal} {Astrophys. J.}\ }\textbf {\bibinfo {volume} {944}},\ \bibinfo {pages} {113} (\bibinfo {year} {2023})},\ \Eprint {http://arxiv.org/abs/2210.04812} {arXiv:2210.04812 [astro-ph.CO]} \BibitemShut {NoStop}%
\bibitem [{\citenamefont {Forconi}\ \emph {et~al.}(2023)\citenamefont {Forconi}, \citenamefont {Ruchika}, \citenamefont {Melchiorri}, \citenamefont {Mena},\ and\ \citenamefont {Menci}}]{Forconi:2023izg}%
  \BibitemOpen
  \bibfield  {author} {\bibinfo {author} {\bibfnamefont {M.}~\bibnamefont {Forconi}}, \bibinfo {author} {\bibnamefont {Ruchika}}, \bibinfo {author} {\bibfnamefont {A.}~\bibnamefont {Melchiorri}}, \bibinfo {author} {\bibfnamefont {O.}~\bibnamefont {Mena}}, \ and\ \bibinfo {author} {\bibfnamefont {N.}~\bibnamefont {Menci}}\ }(\bibinfo {year} {2023})\ \Eprint {http://arxiv.org/abs/2306.07781} {arXiv:2306.07781 [astro-ph.CO]} \BibitemShut {NoStop}%
\bibitem [{\citenamefont {Gupta}(2023)}]{Gupta:2023mgg}%
  \BibitemOpen
  \bibfield  {author} {\bibinfo {author} {\bibfnamefont {R.~P.}\ \bibnamefont {Gupta}},\ }\href {\doibase 10.1093/mnras/stad2032} {\bibfield  {journal} {\bibinfo  {journal} {Mon. Not. Roy. Astron. Soc.}\ }\textbf {\bibinfo {volume} {524}},\ \bibinfo {pages} {3385} (\bibinfo {year} {2023})},\ \Eprint {http://arxiv.org/abs/2309.13100} {arXiv:2309.13100 [astro-ph.CO]} \BibitemShut {NoStop}%
\bibitem [{\citenamefont {Glazebrook}\ \emph {et~al.}(2023)\citenamefont {Glazebrook} \emph {et~al.}}]{Glazebrook:2023vkx}%
  \BibitemOpen
  \bibfield  {author} {\bibinfo {author} {\bibfnamefont {K.}~\bibnamefont {Glazebrook}} \emph {et~al.}\ }(\bibinfo {year} {2023})\ \Eprint {http://arxiv.org/abs/2308.05606} {arXiv:2308.05606 [astro-ph.GA]} \BibitemShut {NoStop}%
\bibitem [{\citenamefont {Adil}\ \emph {et~al.}(2023{\natexlab{b}})\citenamefont {Adil}, \citenamefont {Mukhopadhyay}, \citenamefont {Sen},\ and\ \citenamefont {Vagnozzi}}]{Adil:2023ara}%
  \BibitemOpen
  \bibfield  {author} {\bibinfo {author} {\bibfnamefont {S.~A.}\ \bibnamefont {Adil}}, \bibinfo {author} {\bibfnamefont {U.}~\bibnamefont {Mukhopadhyay}}, \bibinfo {author} {\bibfnamefont {A.~A.}\ \bibnamefont {Sen}}, \ and\ \bibinfo {author} {\bibfnamefont {S.}~\bibnamefont {Vagnozzi}}\ }(\bibinfo {year} {2023})\ \Eprint {http://arxiv.org/abs/2307.12763} {arXiv:2307.12763 [astro-ph.CO]} \BibitemShut {NoStop}%
\bibitem [{\citenamefont {Hirano}\ and\ \citenamefont {Yoshida}(2023)}]{Hirano:2023auh}%
  \BibitemOpen
  \bibfield  {author} {\bibinfo {author} {\bibfnamefont {S.}~\bibnamefont {Hirano}}\ and\ \bibinfo {author} {\bibfnamefont {N.}~\bibnamefont {Yoshida}}\ }(\bibinfo {year} {2023})\ \Eprint {http://arxiv.org/abs/2306.11993} {arXiv:2306.11993 [astro-ph.GA]} \BibitemShut {NoStop}%
\bibitem [{\citenamefont {Parashari}\ and\ \citenamefont {Laha}(2023)}]{Parashari:2023cui}%
  \BibitemOpen
  \bibfield  {author} {\bibinfo {author} {\bibfnamefont {P.}~\bibnamefont {Parashari}}\ and\ \bibinfo {author} {\bibfnamefont {R.}~\bibnamefont {Laha}},\ }\href {\doibase 10.1093/mnrasl/slad107} {\bibfield  {journal} {\bibinfo  {journal} {Mon. Not. Roy. Astron. Soc.}\ }\textbf {\bibinfo {volume} {526}},\ \bibinfo {pages} {L63} (\bibinfo {year} {2023})},\ \Eprint {http://arxiv.org/abs/2305.00999} {arXiv:2305.00999 [astro-ph.CO]} \BibitemShut {NoStop}%
\bibitem [{\citenamefont {Yung}\ \emph {et~al.}(2023)\citenamefont {Yung}, \citenamefont {Somerville}, \citenamefont {Finkelstein}, \citenamefont {Wilkins},\ and\ \citenamefont {Gardner}}]{Yung:2023bng}%
  \BibitemOpen
  \bibfield  {author} {\bibinfo {author} {\bibfnamefont {L.~Y.~A.}\ \bibnamefont {Yung}}, \bibinfo {author} {\bibfnamefont {R.~S.}\ \bibnamefont {Somerville}}, \bibinfo {author} {\bibfnamefont {S.~L.}\ \bibnamefont {Finkelstein}}, \bibinfo {author} {\bibfnamefont {S.~M.}\ \bibnamefont {Wilkins}}, \ and\ \bibinfo {author} {\bibfnamefont {J.~P.}\ \bibnamefont {Gardner}}\ }(\bibinfo {year} {2023})\ \Eprint {http://arxiv.org/abs/2304.04348} {arXiv:2304.04348 [astro-ph.GA]} \BibitemShut {NoStop}%
\bibitem [{\citenamefont {McCaffrey}\ \emph {et~al.}(2023)\citenamefont {McCaffrey}, \citenamefont {Hardin}, \citenamefont {Wise},\ and\ \citenamefont {Regan}}]{McCaffrey:2023qem}%
  \BibitemOpen
  \bibfield  {author} {\bibinfo {author} {\bibfnamefont {J.}~\bibnamefont {McCaffrey}}, \bibinfo {author} {\bibfnamefont {S.}~\bibnamefont {Hardin}}, \bibinfo {author} {\bibfnamefont {J.}~\bibnamefont {Wise}}, \ and\ \bibinfo {author} {\bibfnamefont {J.}~\bibnamefont {Regan}}\ }(\bibinfo {year} {2023})\ \Eprint {http://arxiv.org/abs/2304.13755} {arXiv:2304.13755 [astro-ph.GA]} \BibitemShut {NoStop}%
\bibitem [{\citenamefont {Wang}\ \emph {et~al.}(2023{\natexlab{a}})\citenamefont {Wang}, \citenamefont {Lei}, \citenamefont {Yuan},\ and\ \citenamefont {Fan}}]{Wang:2023xmm}%
  \BibitemOpen
  \bibfield  {author} {\bibinfo {author} {\bibfnamefont {Y.-Y.}\ \bibnamefont {Wang}}, \bibinfo {author} {\bibfnamefont {L.}~\bibnamefont {Lei}}, \bibinfo {author} {\bibfnamefont {G.-W.}\ \bibnamefont {Yuan}}, \ and\ \bibinfo {author} {\bibfnamefont {Y.-Z.}\ \bibnamefont {Fan}},\ }\href {\doibase 10.3847/2041-8213/acf46c} {\bibfield  {journal} {\bibinfo  {journal} {Astrophys. J. Lett.}\ }\textbf {\bibinfo {volume} {954}},\ \bibinfo {pages} {L48} (\bibinfo {year} {2023}{\natexlab{a}})},\ \Eprint {http://arxiv.org/abs/2307.12487} {arXiv:2307.12487 [astro-ph.GA]} \BibitemShut {NoStop}%
\bibitem [{\citenamefont {Wang}\ \emph {et~al.}(2023{\natexlab{b}})\citenamefont {Wang}, \citenamefont {Huang}, \citenamefont {Huang},\ and\ \citenamefont {Liu}}]{Wang:2023gla}%
  \BibitemOpen
  \bibfield  {author} {\bibinfo {author} {\bibfnamefont {J.}~\bibnamefont {Wang}}, \bibinfo {author} {\bibfnamefont {Z.}~\bibnamefont {Huang}}, \bibinfo {author} {\bibfnamefont {L.}~\bibnamefont {Huang}}, \ and\ \bibinfo {author} {\bibfnamefont {J.}~\bibnamefont {Liu}},\ }\href@noop {} {\enquote {\bibinfo {title} {{Quantifying the tension between cosmological models and JWST red candidate massive galaxies}},}\ } (\bibinfo {year} {2023}{\natexlab{b}}),\ \Eprint {http://arxiv.org/abs/2311.02866} {arXiv:2311.02866 [astro-ph.CO]} \BibitemShut {NoStop}%
\bibitem [{\citenamefont {Wang}\ and\ \citenamefont {Liu}(2022)}]{Wang:2022jvx}%
  \BibitemOpen
  \bibfield  {author} {\bibinfo {author} {\bibfnamefont {D.}~\bibnamefont {Wang}}\ and\ \bibinfo {author} {\bibfnamefont {Y.}~\bibnamefont {Liu}}\ }(\bibinfo {year} {2022})\ \Eprint {http://arxiv.org/abs/2301.00347} {arXiv:2301.00347 [astro-ph.CO]} \BibitemShut {NoStop}%
\bibitem [{\citenamefont {Melia}(2023)}]{Melia:2023dsy}%
  \BibitemOpen
  \bibfield  {author} {\bibinfo {author} {\bibfnamefont {F.}~\bibnamefont {Melia}},\ }\href {\doibase 10.1093/mnrasl/slad025} {\bibfield  {journal} {\bibinfo  {journal} {Mon. Not. Roy. Astron. Soc.}\ }\textbf {\bibinfo {volume} {521}},\ \bibinfo {pages} {L85} (\bibinfo {year} {2023})},\ \Eprint {http://arxiv.org/abs/2302.10103} {arXiv:2302.10103 [astro-ph.CO]} \BibitemShut {NoStop}%
\bibitem [{\citenamefont {Paraskevas}\ and\ \citenamefont {Perivolaropoulos}(2023{\natexlab{a}})}]{Paraskevas:2023itu}%
  \BibitemOpen
  \bibfield  {author} {\bibinfo {author} {\bibfnamefont {E.~A.}\ \bibnamefont {Paraskevas}}\ and\ \bibinfo {author} {\bibfnamefont {L.}~\bibnamefont {Perivolaropoulos}}\ }(\bibinfo {year} {2023})\ \Eprint {http://arxiv.org/abs/2308.07046} {arXiv:2308.07046 [astro-ph.CO]} \BibitemShut {NoStop}%
\bibitem [{\citenamefont {Menci}\ \emph {et~al.}(2024)\citenamefont {Menci}, \citenamefont {Adil}, \citenamefont {Mukhopadhyay}, \citenamefont {Sen},\ and\ \citenamefont {Vagnozzi}}]{Menci:2024rbq}%
  \BibitemOpen
  \bibfield  {author} {\bibinfo {author} {\bibfnamefont {N.}~\bibnamefont {Menci}}, \bibinfo {author} {\bibfnamefont {S.~A.}\ \bibnamefont {Adil}}, \bibinfo {author} {\bibfnamefont {U.}~\bibnamefont {Mukhopadhyay}}, \bibinfo {author} {\bibfnamefont {A.~A.}\ \bibnamefont {Sen}}, \ and\ \bibinfo {author} {\bibfnamefont {S.}~\bibnamefont {Vagnozzi}},\ }\href@noop {} {\  (\bibinfo {year} {2024})},\ \Eprint {http://arxiv.org/abs/2401.12659} {arXiv:2401.12659 [astro-ph.CO]} \BibitemShut {NoStop}%
\bibitem [{\citenamefont {Kazantzidis}\ and\ \citenamefont {Perivolaropoulos}(2019)}]{Kazantzidis:2019nuh}%
  \BibitemOpen
  \bibfield  {author} {\bibinfo {author} {\bibfnamefont {L.}~\bibnamefont {Kazantzidis}}\ and\ \bibinfo {author} {\bibfnamefont {L.}~\bibnamefont {Perivolaropoulos}}\ }(\bibinfo {year} {2019})\ \Eprint {http://arxiv.org/abs/1907.03176} {arXiv:1907.03176 [astro-ph.CO]} \BibitemShut {NoStop}%
\bibitem [{\citenamefont {Benisty}(2021)}]{Benisty:2020kdt}%
  \BibitemOpen
  \bibfield  {author} {\bibinfo {author} {\bibfnamefont {D.}~\bibnamefont {Benisty}},\ }\href {\doibase 10.1016/j.dark.2020.100766} {\bibfield  {journal} {\bibinfo  {journal} {Phys. Dark Univ.}\ }\textbf {\bibinfo {volume} {31}},\ \bibinfo {pages} {100766} (\bibinfo {year} {2021})},\ \Eprint {http://arxiv.org/abs/2005.03751} {arXiv:2005.03751 [astro-ph.CO]} \BibitemShut {NoStop}%
\bibitem [{\citenamefont {Nunes}\ and\ \citenamefont {Vagnozzi}(2021)}]{Nunes:2021ipq}%
  \BibitemOpen
  \bibfield  {author} {\bibinfo {author} {\bibfnamefont {R.~C.}\ \bibnamefont {Nunes}}\ and\ \bibinfo {author} {\bibfnamefont {S.}~\bibnamefont {Vagnozzi}},\ }\href {\doibase 10.1093/mnras/stab1613} {\bibfield  {journal} {\bibinfo  {journal} {Mon. Not. Roy. Astron. Soc.}\ }\textbf {\bibinfo {volume} {505}},\ \bibinfo {pages} {5427} (\bibinfo {year} {2021})},\ \Eprint {http://arxiv.org/abs/2106.01208} {arXiv:2106.01208 [astro-ph.CO]} \BibitemShut {NoStop}%
\bibitem [{\citenamefont {Bocquet}\ \emph {et~al.}(2019)\citenamefont {Bocquet} \emph {et~al.}}]{SPT:2018njh}%
  \BibitemOpen
  \bibfield  {author} {\bibinfo {author} {\bibfnamefont {S.}~\bibnamefont {Bocquet}} \emph {et~al.} (\bibinfo {collaboration} {SPT}),\ }\href {\doibase 10.3847/1538-4357/ab1f10} {\bibfield  {journal} {\bibinfo  {journal} {Astrophys. J.}\ }\textbf {\bibinfo {volume} {878}},\ \bibinfo {pages} {55} (\bibinfo {year} {2019})},\ \Eprint {http://arxiv.org/abs/1812.01679} {arXiv:1812.01679 [astro-ph.CO]} \BibitemShut {NoStop}%
\bibitem [{\citenamefont {Maldacena}(1998)}]{Maldacena:1997re}%
  \BibitemOpen
  \bibfield  {author} {\bibinfo {author} {\bibfnamefont {J.~M.}\ \bibnamefont {Maldacena}},\ }\href {\doibase 10.4310/ATMP.1998.v2.n2.a1} {\bibfield  {journal} {\bibinfo  {journal} {Adv. Theor. Math. Phys.}\ }\textbf {\bibinfo {volume} {2}},\ \bibinfo {pages} {231} (\bibinfo {year} {1998})},\ \Eprint {http://arxiv.org/abs/hep-th/9711200} {arXiv:hep-th/9711200} \BibitemShut {NoStop}%
\bibitem [{\citenamefont {Bousso}\ and\ \citenamefont {Polchinski}(2000)}]{Bousso:2000xa}%
  \BibitemOpen
  \bibfield  {author} {\bibinfo {author} {\bibfnamefont {R.}~\bibnamefont {Bousso}}\ and\ \bibinfo {author} {\bibfnamefont {J.}~\bibnamefont {Polchinski}},\ }\href {\doibase 10.1088/1126-6708/2000/06/006} {\bibfield  {journal} {\bibinfo  {journal} {JHEP}\ }\textbf {\bibinfo {volume} {06}},\ \bibinfo {pages} {006} (\bibinfo {year} {2000})},\ \Eprint {http://arxiv.org/abs/hep-th/0004134} {arXiv:hep-th/0004134} \BibitemShut {NoStop}%
\bibitem [{\citenamefont {Anchordoqui}\ \emph {et~al.}(2023)\citenamefont {Anchordoqui}, \citenamefont {Antoniadis},\ and\ \citenamefont {Lust}}]{Anchordoqui:2023woo}%
  \BibitemOpen
  \bibfield  {author} {\bibinfo {author} {\bibfnamefont {L.~A.}\ \bibnamefont {Anchordoqui}}, \bibinfo {author} {\bibfnamefont {I.}~\bibnamefont {Antoniadis}}, \ and\ \bibinfo {author} {\bibfnamefont {D.}~\bibnamefont {Lust}}\ }(\bibinfo {year} {2023})\ \Eprint {http://arxiv.org/abs/2312.12352} {arXiv:2312.12352 [hep-th]} \BibitemShut {NoStop}%
\bibitem [{\citenamefont {Alexandre}\ \emph {et~al.}(2023)\citenamefont {Alexandre}, \citenamefont {Gielen},\ and\ \citenamefont {Magueijo}}]{Alexandre:2023nmh}%
  \BibitemOpen
  \bibfield  {author} {\bibinfo {author} {\bibfnamefont {B.}~\bibnamefont {Alexandre}}, \bibinfo {author} {\bibfnamefont {S.}~\bibnamefont {Gielen}}, \ and\ \bibinfo {author} {\bibfnamefont {J.~a.}\ \bibnamefont {Magueijo}},\ }\href@noop {} {\  (\bibinfo {year} {2023})},\ \Eprint {http://arxiv.org/abs/2306.11502} {arXiv:2306.11502 [hep-th]} \BibitemShut {NoStop}%
\bibitem [{\citenamefont {Akarsu}\ \emph {et~al.}(2024{\natexlab{b}})\citenamefont {Akarsu}, \citenamefont {De~Felice}, \citenamefont {Di~Valentino}, \citenamefont {Kumar}, \citenamefont {Nunes}, \citenamefont {Ozulker}, \citenamefont {Vazquez},\ and\ \citenamefont {Yadav}}]{Akarsu:2024qsi}%
  \BibitemOpen
  \bibfield  {author} {\bibinfo {author} {\bibfnamefont {O.}~\bibnamefont {Akarsu}}, \bibinfo {author} {\bibfnamefont {A.}~\bibnamefont {De~Felice}}, \bibinfo {author} {\bibfnamefont {E.}~\bibnamefont {Di~Valentino}}, \bibinfo {author} {\bibfnamefont {S.}~\bibnamefont {Kumar}}, \bibinfo {author} {\bibfnamefont {R.~C.}\ \bibnamefont {Nunes}}, \bibinfo {author} {\bibfnamefont {E.}~\bibnamefont {Ozulker}}, \bibinfo {author} {\bibfnamefont {J.~A.}\ \bibnamefont {Vazquez}}, \ and\ \bibinfo {author} {\bibfnamefont {A.}~\bibnamefont {Yadav}},\ }\href@noop {} {\  (\bibinfo {year} {2024}{\natexlab{b}})},\ \Eprint {http://arxiv.org/abs/2402.07716} {arXiv:2402.07716 [astro-ph.CO]} \BibitemShut {NoStop}%
\bibitem [{\citenamefont {De~Felice}\ \emph {et~al.}(2020)\citenamefont {De~Felice}, \citenamefont {Doll},\ and\ \citenamefont {Mukohyama}}]{DeFelice:2020eju}%
  \BibitemOpen
  \bibfield  {author} {\bibinfo {author} {\bibfnamefont {A.}~\bibnamefont {De~Felice}}, \bibinfo {author} {\bibfnamefont {A.}~\bibnamefont {Doll}}, \ and\ \bibinfo {author} {\bibfnamefont {S.}~\bibnamefont {Mukohyama}},\ }\href {\doibase 10.1088/1475-7516/2020/09/034} {\bibfield  {journal} {\bibinfo  {journal} {JCAP}\ }\textbf {\bibinfo {volume} {09}},\ \bibinfo {pages} {034} (\bibinfo {year} {2020})},\ \Eprint {http://arxiv.org/abs/2004.12549} {arXiv:2004.12549 [gr-qc]} \BibitemShut {NoStop}%
\bibitem [{\citenamefont {De~Felice}\ \emph {et~al.}(2021)\citenamefont {De~Felice}, \citenamefont {Mukohyama},\ and\ \citenamefont {Pookkillath}}]{DeFelice:2020cpt}%
  \BibitemOpen
  \bibfield  {author} {\bibinfo {author} {\bibfnamefont {A.}~\bibnamefont {De~Felice}}, \bibinfo {author} {\bibfnamefont {S.}~\bibnamefont {Mukohyama}}, \ and\ \bibinfo {author} {\bibfnamefont {M.~C.}\ \bibnamefont {Pookkillath}},\ }\href {\doibase 10.1016/j.physletb.2021.136201} {\bibfield  {journal} {\bibinfo  {journal} {Phys. Lett. B}\ }\textbf {\bibinfo {volume} {816}},\ \bibinfo {pages} {136201} (\bibinfo {year} {2021})},\ \bibinfo {note} {[Erratum: Phys.Lett.B 818, 136364 (2021)]},\ \Eprint {http://arxiv.org/abs/2009.08718} {arXiv:2009.08718 [astro-ph.CO]} \BibitemShut {NoStop}%
\bibitem [{\citenamefont {Garriga}\ \emph {et~al.}(2004{\natexlab{a}})\citenamefont {Garriga}, \citenamefont {Linde},\ and\ \citenamefont {Vilenkin}}]{Garriga:2003hj}%
  \BibitemOpen
  \bibfield  {author} {\bibinfo {author} {\bibfnamefont {J.}~\bibnamefont {Garriga}}, \bibinfo {author} {\bibfnamefont {A.~D.}\ \bibnamefont {Linde}}, \ and\ \bibinfo {author} {\bibfnamefont {A.}~\bibnamefont {Vilenkin}},\ }\href {\doibase 10.1103/PhysRevD.69.063521} {\bibfield  {journal} {\bibinfo  {journal} {Phys. Rev. D}\ }\textbf {\bibinfo {volume} {69}},\ \bibinfo {pages} {063521} (\bibinfo {year} {2004}{\natexlab{a}})},\ \Eprint {http://arxiv.org/abs/hep-th/0310034} {arXiv:hep-th/0310034} \BibitemShut {NoStop}%
\bibitem [{\citenamefont {Garriga}\ \emph {et~al.}(2004{\natexlab{b}})\citenamefont {Garriga}, \citenamefont {Pogosian},\ and\ \citenamefont {Vachaspati}}]{Garriga:2003nm}%
  \BibitemOpen
  \bibfield  {author} {\bibinfo {author} {\bibfnamefont {J.}~\bibnamefont {Garriga}}, \bibinfo {author} {\bibfnamefont {L.}~\bibnamefont {Pogosian}}, \ and\ \bibinfo {author} {\bibfnamefont {T.}~\bibnamefont {Vachaspati}},\ }\href {\doibase 10.1103/PhysRevD.69.063511} {\bibfield  {journal} {\bibinfo  {journal} {Phys. Rev. D}\ }\textbf {\bibinfo {volume} {69}},\ \bibinfo {pages} {063511} (\bibinfo {year} {2004}{\natexlab{b}})},\ \Eprint {http://arxiv.org/abs/astro-ph/0311412} {arXiv:astro-ph/0311412} \BibitemShut {NoStop}%
\bibitem [{\citenamefont {Perivolaropoulos}(2005)}]{Perivolaropoulos:2004yr}%
  \BibitemOpen
  \bibfield  {author} {\bibinfo {author} {\bibfnamefont {L.}~\bibnamefont {Perivolaropoulos}},\ }\href {\doibase 10.1103/PhysRevD.71.063503} {\bibfield  {journal} {\bibinfo  {journal} {Phys. Rev. D}\ }\textbf {\bibinfo {volume} {71}},\ \bibinfo {pages} {063503} (\bibinfo {year} {2005})},\ \Eprint {http://arxiv.org/abs/astro-ph/0412308} {arXiv:astro-ph/0412308} \BibitemShut {NoStop}%
\bibitem [{\citenamefont {Ozulker}(2022)}]{Ozulker:2022slu}%
  \BibitemOpen
  \bibfield  {author} {\bibinfo {author} {\bibfnamefont {E.}~\bibnamefont {Ozulker}},\ }\href {\doibase 10.1103/PhysRevD.106.063509} {\bibfield  {journal} {\bibinfo  {journal} {Phys. Rev. D}\ }\textbf {\bibinfo {volume} {106}},\ \bibinfo {pages} {063509} (\bibinfo {year} {2022})},\ \Eprint {http://arxiv.org/abs/2203.04167} {arXiv:2203.04167 [astro-ph.CO]} \BibitemShut {NoStop}%
\bibitem [{\citenamefont {Barrow}(2004)}]{Barrow:2004xh}%
  \BibitemOpen
  \bibfield  {author} {\bibinfo {author} {\bibfnamefont {J.~D.}\ \bibnamefont {Barrow}},\ }\href {\doibase 10.1088/0264-9381/21/11/L03} {\bibfield  {journal} {\bibinfo  {journal} {Class. Quant. Grav.}\ }\textbf {\bibinfo {volume} {21}},\ \bibinfo {pages} {L79} (\bibinfo {year} {2004})},\ \Eprint {http://arxiv.org/abs/gr-qc/0403084} {arXiv:gr-qc/0403084} \BibitemShut {NoStop}%
\bibitem [{\citenamefont {Nojiri}\ \emph {et~al.}(2005)\citenamefont {Nojiri}, \citenamefont {Odintsov},\ and\ \citenamefont {Tsujikawa}}]{Nojiri:2005sx}%
  \BibitemOpen
  \bibfield  {author} {\bibinfo {author} {\bibfnamefont {S.}~\bibnamefont {Nojiri}}, \bibinfo {author} {\bibfnamefont {S.~D.}\ \bibnamefont {Odintsov}}, \ and\ \bibinfo {author} {\bibfnamefont {S.}~\bibnamefont {Tsujikawa}},\ }\href {\doibase 10.1103/PhysRevD.71.063004} {\bibfield  {journal} {\bibinfo  {journal} {Phys. Rev. D}\ }\textbf {\bibinfo {volume} {71}},\ \bibinfo {pages} {063004} (\bibinfo {year} {2005})},\ \Eprint {http://arxiv.org/abs/hep-th/0501025} {arXiv:hep-th/0501025} \BibitemShut {NoStop}%
\bibitem [{\citenamefont {Fernandez-Jambrina}\ and\ \citenamefont {Lazkoz}(2006)}]{Fernandez-Jambrina:2006tkb}%
  \BibitemOpen
  \bibfield  {author} {\bibinfo {author} {\bibfnamefont {L.}~\bibnamefont {Fernandez-Jambrina}}\ and\ \bibinfo {author} {\bibfnamefont {R.}~\bibnamefont {Lazkoz}},\ }\href {\doibase 10.1103/PhysRevD.74.064030} {\bibfield  {journal} {\bibinfo  {journal} {Phys. Rev. D}\ }\textbf {\bibinfo {volume} {74}},\ \bibinfo {pages} {064030} (\bibinfo {year} {2006})},\ \Eprint {http://arxiv.org/abs/gr-qc/0607073} {arXiv:gr-qc/0607073} \BibitemShut {NoStop}%
\bibitem [{\citenamefont {Balcerzak}\ \emph {et~al.}(2023)\citenamefont {Balcerzak}, \citenamefont {Denkiewicz},\ and\ \citenamefont {Lisaj}}]{Balcerzak:2023ynk}%
  \BibitemOpen
  \bibfield  {author} {\bibinfo {author} {\bibfnamefont {A.}~\bibnamefont {Balcerzak}}, \bibinfo {author} {\bibfnamefont {T.}~\bibnamefont {Denkiewicz}}, \ and\ \bibinfo {author} {\bibfnamefont {M.}~\bibnamefont {Lisaj}},\ }\href {\doibase 10.1140/epjc/s10052-023-12186-3} {\bibfield  {journal} {\bibinfo  {journal} {Eur. Phys. J. C}\ }\textbf {\bibinfo {volume} {83}},\ \bibinfo {pages} {980} (\bibinfo {year} {2023})},\ \Eprint {http://arxiv.org/abs/2308.13293} {arXiv:2308.13293 [gr-qc]} \BibitemShut {NoStop}%
\bibitem [{\citenamefont {de~Haro}\ \emph {et~al.}(2023)\citenamefont {de~Haro}, \citenamefont {Nojiri}, \citenamefont {Odintsov}, \citenamefont {Oikonomou},\ and\ \citenamefont {Pan}}]{deHaro:2023lbq}%
  \BibitemOpen
  \bibfield  {author} {\bibinfo {author} {\bibfnamefont {J.}~\bibnamefont {de~Haro}}, \bibinfo {author} {\bibfnamefont {S.}~\bibnamefont {Nojiri}}, \bibinfo {author} {\bibfnamefont {S.~D.}\ \bibnamefont {Odintsov}}, \bibinfo {author} {\bibfnamefont {V.~K.}\ \bibnamefont {Oikonomou}}, \ and\ \bibinfo {author} {\bibfnamefont {S.}~\bibnamefont {Pan}}\ }(\bibinfo {year} {2023})\ \Eprint {http://arxiv.org/abs/2309.07465} {arXiv:2309.07465 [gr-qc]} \BibitemShut {NoStop}%
\bibitem [{\citenamefont {Trivedi}(2023)}]{Trivedi:2023zlf}%
  \BibitemOpen
  \bibfield  {author} {\bibinfo {author} {\bibfnamefont {O.}~\bibnamefont {Trivedi}}\ }(\bibinfo {year} {2023})\ \Eprint {http://arxiv.org/abs/2309.08954} {arXiv:2309.08954 [gr-qc]} \BibitemShut {NoStop}%
\bibitem [{\citenamefont {Paraskevas}\ and\ \citenamefont {Perivolaropoulos}(2023{\natexlab{b}})}]{Paraskevas:2023aae}%
  \BibitemOpen
  \bibfield  {author} {\bibinfo {author} {\bibfnamefont {E.~A.}\ \bibnamefont {Paraskevas}}\ and\ \bibinfo {author} {\bibfnamefont {L.}~\bibnamefont {Perivolaropoulos}},\ }\href {\doibase 10.3390/universe9070317} {\bibfield  {journal} {\bibinfo  {journal} {Universe}\ }\textbf {\bibinfo {volume} {9}},\ \bibinfo {pages} {317} (\bibinfo {year} {2023}{\natexlab{b}})},\ \Eprint {http://arxiv.org/abs/2307.00298} {arXiv:2307.00298 [astro-ph.CO]} \BibitemShut {NoStop}%
\bibitem [{\citenamefont {Eke}\ \emph {et~al.}(1996)\citenamefont {Eke}, \citenamefont {Cole},\ and\ \citenamefont {Frenk}}]{Eke:1996ds}%
  \BibitemOpen
  \bibfield  {author} {\bibinfo {author} {\bibfnamefont {V.~R.}\ \bibnamefont {Eke}}, \bibinfo {author} {\bibfnamefont {S.}~\bibnamefont {Cole}}, \ and\ \bibinfo {author} {\bibfnamefont {C.~S.}\ \bibnamefont {Frenk}},\ }\href {\doibase 10.1093/mnras/282.1.263} {\bibfield  {journal} {\bibinfo  {journal} {Mon. Not. Roy. Astron. Soc.}\ }\textbf {\bibinfo {volume} {282}},\ \bibinfo {pages} {263} (\bibinfo {year} {1996})},\ \Eprint {http://arxiv.org/abs/astro-ph/9601088} {arXiv:astro-ph/9601088} \BibitemShut {NoStop}%
\bibitem [{\citenamefont {Lokas}\ and\ \citenamefont {Hoffman}(2000)}]{Lokas:2000cn}%
  \BibitemOpen
  \bibfield  {author} {\bibinfo {author} {\bibfnamefont {E.~L.}\ \bibnamefont {Lokas}}\ and\ \bibinfo {author} {\bibfnamefont {Y.}~\bibnamefont {Hoffman}}\ }(\bibinfo {year} {2000})\ \Eprint {http://arxiv.org/abs/astro-ph/0011295} {arXiv:astro-ph/0011295} \BibitemShut {NoStop}%
\bibitem [{\citenamefont {Mota}\ and\ \citenamefont {van~de Bruck}(2004)}]{Mota:2004pa}%
  \BibitemOpen
  \bibfield  {author} {\bibinfo {author} {\bibfnamefont {D.~F.}\ \bibnamefont {Mota}}\ and\ \bibinfo {author} {\bibfnamefont {C.}~\bibnamefont {van~de Bruck}},\ }\href {\doibase 10.1051/0004-6361:20041090} {\bibfield  {journal} {\bibinfo  {journal} {Astron. Astrophys.}\ }\textbf {\bibinfo {volume} {421}},\ \bibinfo {pages} {71} (\bibinfo {year} {2004})},\ \Eprint {http://arxiv.org/abs/astro-ph/0401504} {arXiv:astro-ph/0401504} \BibitemShut {NoStop}%
\bibitem [{\citenamefont {Creminelli}\ \emph {et~al.}(2010)\citenamefont {Creminelli}, \citenamefont {D'Amico}, \citenamefont {Norena}, \citenamefont {Senatore},\ and\ \citenamefont {Vernizzi}}]{Creminelli:2009mu}%
  \BibitemOpen
  \bibfield  {author} {\bibinfo {author} {\bibfnamefont {P.}~\bibnamefont {Creminelli}}, \bibinfo {author} {\bibfnamefont {G.}~\bibnamefont {D'Amico}}, \bibinfo {author} {\bibfnamefont {J.}~\bibnamefont {Norena}}, \bibinfo {author} {\bibfnamefont {L.}~\bibnamefont {Senatore}}, \ and\ \bibinfo {author} {\bibfnamefont {F.}~\bibnamefont {Vernizzi}},\ }\href {\doibase 10.1088/1475-7516/2010/03/027} {\bibfield  {journal} {\bibinfo  {journal} {JCAP}\ }\textbf {\bibinfo {volume} {03}},\ \bibinfo {pages} {027} (\bibinfo {year} {2010})},\ \Eprint {http://arxiv.org/abs/0911.2701} {arXiv:0911.2701 [astro-ph.CO]} \BibitemShut {NoStop}%
\bibitem [{\citenamefont {Pavlidou}\ and\ \citenamefont {Fields}(2005)}]{Pavlidou:2004vq}%
  \BibitemOpen
  \bibfield  {author} {\bibinfo {author} {\bibfnamefont {V.}~\bibnamefont {Pavlidou}}\ and\ \bibinfo {author} {\bibfnamefont {B.~D.}\ \bibnamefont {Fields}},\ }\href {\doibase 10.1103/PhysRevD.71.043510} {\bibfield  {journal} {\bibinfo  {journal} {Phys. Rev. D}\ }\textbf {\bibinfo {volume} {71}},\ \bibinfo {pages} {043510} (\bibinfo {year} {2005})},\ \Eprint {http://arxiv.org/abs/astro-ph/0410338} {arXiv:astro-ph/0410338} \BibitemShut {NoStop}%
\bibitem [{\citenamefont {Horellou}\ and\ \citenamefont {Berge}(2005)}]{Horellou:2005qc}%
  \BibitemOpen
  \bibfield  {author} {\bibinfo {author} {\bibfnamefont {C.}~\bibnamefont {Horellou}}\ and\ \bibinfo {author} {\bibfnamefont {J.}~\bibnamefont {Berge}},\ }\href {\doibase 10.1111/j.1365-2966.2005.09140.x} {\bibfield  {journal} {\bibinfo  {journal} {Mon. Not. Roy. Astron. Soc.}\ }\textbf {\bibinfo {volume} {360}},\ \bibinfo {pages} {1393} (\bibinfo {year} {2005})},\ \Eprint {http://arxiv.org/abs/astro-ph/0504465} {arXiv:astro-ph/0504465} \BibitemShut {NoStop}%
\bibitem [{\citenamefont {Basilakos}(2003)}]{Basilakos:2003bi}%
  \BibitemOpen
  \bibfield  {author} {\bibinfo {author} {\bibfnamefont {S.}~\bibnamefont {Basilakos}},\ }\href {\doibase 10.1086/375154} {\bibfield  {journal} {\bibinfo  {journal} {Astrophys. J.}\ }\textbf {\bibinfo {volume} {590}},\ \bibinfo {pages} {636} (\bibinfo {year} {2003})},\ \Eprint {http://arxiv.org/abs/astro-ph/0303112} {arXiv:astro-ph/0303112} \BibitemShut {NoStop}%
\bibitem [{\citenamefont {Tanoglidis}\ \emph {et~al.}(2015)\citenamefont {Tanoglidis}, \citenamefont {Pavlidou},\ and\ \citenamefont {Tomaras}}]{Tanoglidis:2014lea}%
  \BibitemOpen
  \bibfield  {author} {\bibinfo {author} {\bibfnamefont {D.}~\bibnamefont {Tanoglidis}}, \bibinfo {author} {\bibfnamefont {V.}~\bibnamefont {Pavlidou}}, \ and\ \bibinfo {author} {\bibfnamefont {T.}~\bibnamefont {Tomaras}},\ }\href {\doibase 10.1088/1475-7516/2015/12/060} {\bibfield  {journal} {\bibinfo  {journal} {JCAP}\ }\textbf {\bibinfo {volume} {12}},\ \bibinfo {pages} {060} (\bibinfo {year} {2015})},\ \Eprint {http://arxiv.org/abs/1412.6671} {arXiv:1412.6671 [astro-ph.CO]} \BibitemShut {NoStop}%
\bibitem [{\citenamefont {Tanoglidis}\ \emph {et~al.}(2016)\citenamefont {Tanoglidis}, \citenamefont {Pavlidou},\ and\ \citenamefont {Tomaras}}]{Tanoglidis:2016lrj}%
  \BibitemOpen
  \bibfield  {author} {\bibinfo {author} {\bibfnamefont {D.}~\bibnamefont {Tanoglidis}}, \bibinfo {author} {\bibfnamefont {V.}~\bibnamefont {Pavlidou}}, \ and\ \bibinfo {author} {\bibfnamefont {T.}~\bibnamefont {Tomaras}}\ }(\bibinfo {year} {2016})\ \Eprint {http://arxiv.org/abs/1601.03740} {arXiv:1601.03740 [astro-ph.CO]} \BibitemShut {NoStop}%
\bibitem [{\citenamefont {Pace}\ \emph {et~al.}(2017)\citenamefont {Pace}, \citenamefont {Meyer},\ and\ \citenamefont {Bartelmann}}]{Pace:2017qxv}%
  \BibitemOpen
  \bibfield  {author} {\bibinfo {author} {\bibfnamefont {F.}~\bibnamefont {Pace}}, \bibinfo {author} {\bibfnamefont {S.}~\bibnamefont {Meyer}}, \ and\ \bibinfo {author} {\bibfnamefont {M.}~\bibnamefont {Bartelmann}},\ }\href {\doibase 10.1088/1475-7516/2017/10/040} {\bibfield  {journal} {\bibinfo  {journal} {JCAP}\ }\textbf {\bibinfo {volume} {10}},\ \bibinfo {pages} {040} (\bibinfo {year} {2017})},\ \Eprint {http://arxiv.org/abs/1708.02477} {arXiv:1708.02477 [astro-ph.CO]} \BibitemShut {NoStop}%
\bibitem [{\citenamefont {Pavlidou}\ \emph {et~al.}(2020)\citenamefont {Pavlidou}, \citenamefont {Korkidis}, \citenamefont {Tomaras},\ and\ \citenamefont {Tanoglidis}}]{Pavlidou:2020afx}%
  \BibitemOpen
  \bibfield  {author} {\bibinfo {author} {\bibfnamefont {V.}~\bibnamefont {Pavlidou}}, \bibinfo {author} {\bibfnamefont {G.}~\bibnamefont {Korkidis}}, \bibinfo {author} {\bibfnamefont {T.}~\bibnamefont {Tomaras}}, \ and\ \bibinfo {author} {\bibfnamefont {D.}~\bibnamefont {Tanoglidis}},\ }\href {\doibase 10.1051/0004-6361/201937358} {\bibfield  {journal} {\bibinfo  {journal} {Astron. Astrophys.}\ }\textbf {\bibinfo {volume} {638}},\ \bibinfo {pages} {L8} (\bibinfo {year} {2020})},\ \Eprint {http://arxiv.org/abs/2004.04395} {arXiv:2004.04395 [astro-ph.CO]} \BibitemShut {NoStop}%
\bibitem [{\citenamefont {Press}\ and\ \citenamefont {Schechter}(1974)}]{Press:1973iz}%
  \BibitemOpen
  \bibfield  {author} {\bibinfo {author} {\bibfnamefont {W.~H.}\ \bibnamefont {Press}}\ and\ \bibinfo {author} {\bibfnamefont {P.}~\bibnamefont {Schechter}},\ }\href {\doibase 10.1086/152650} {\bibfield  {journal} {\bibinfo  {journal} {Astrophys. J.}\ }\textbf {\bibinfo {volume} {187}},\ \bibinfo {pages} {425} (\bibinfo {year} {1974})}\BibitemShut {NoStop}%
\bibitem [{\citenamefont {Bond}\ \emph {et~al.}(1991)\citenamefont {Bond}, \citenamefont {Cole}, \citenamefont {Efstathiou},\ and\ \citenamefont {Kaiser}}]{Bond:1990iw}%
  \BibitemOpen
  \bibfield  {author} {\bibinfo {author} {\bibfnamefont {J.~R.}\ \bibnamefont {Bond}}, \bibinfo {author} {\bibfnamefont {S.}~\bibnamefont {Cole}}, \bibinfo {author} {\bibfnamefont {G.}~\bibnamefont {Efstathiou}}, \ and\ \bibinfo {author} {\bibfnamefont {N.}~\bibnamefont {Kaiser}},\ }\href {\doibase 10.1086/170520} {\bibfield  {journal} {\bibinfo  {journal} {Astrophys. J.}\ }\textbf {\bibinfo {volume} {379}},\ \bibinfo {pages} {440} (\bibinfo {year} {1991})}\BibitemShut {NoStop}%
\bibitem [{\citenamefont {Cooray}\ and\ \citenamefont {Sheth}(2002)}]{Cooray:2002dia}%
  \BibitemOpen
  \bibfield  {author} {\bibinfo {author} {\bibfnamefont {A.}~\bibnamefont {Cooray}}\ and\ \bibinfo {author} {\bibfnamefont {R.~K.}\ \bibnamefont {Sheth}},\ }\href {\doibase 10.1016/S0370-1573(02)00276-4} {\bibfield  {journal} {\bibinfo  {journal} {Phys. Rept.}\ }\textbf {\bibinfo {volume} {372}},\ \bibinfo {pages} {1} (\bibinfo {year} {2002})},\ \Eprint {http://arxiv.org/abs/astro-ph/0206508} {arXiv:astro-ph/0206508} \BibitemShut {NoStop}%
\bibitem [{\citenamefont {Asgari}\ \emph {et~al.}(2023)\citenamefont {Asgari}, \citenamefont {Mead},\ and\ \citenamefont {Heymans}}]{Asgari:2023mej}%
  \BibitemOpen
  \bibfield  {author} {\bibinfo {author} {\bibfnamefont {M.}~\bibnamefont {Asgari}}, \bibinfo {author} {\bibfnamefont {A.~J.}\ \bibnamefont {Mead}}, \ and\ \bibinfo {author} {\bibfnamefont {C.}~\bibnamefont {Heymans}}\ }(\bibinfo {year} {2023})\ \Eprint {http://arxiv.org/abs/2303.08752} {arXiv:2303.08752 [astro-ph.CO]} \BibitemShut {NoStop}%
\bibitem [{\citenamefont {Dodelson}\ and\ \citenamefont {Schmidt}(2021{\natexlab{b}})}]{DODELSON2021325}%
  \BibitemOpen
  \bibfield  {author} {\bibinfo {author} {\bibfnamefont {S.}~\bibnamefont {Dodelson}}\ and\ \bibinfo {author} {\bibfnamefont {F.}~\bibnamefont {Schmidt}},\ }in\ \href {\doibase https://doi.org/10.1016/B978-0-12-815948-4.00018-8} {\emph {\bibinfo {booktitle} {Modern Cosmology (Second Edition)}}},\ \bibinfo {editor} {edited by\ \bibinfo {editor} {\bibfnamefont {S.}~\bibnamefont {Dodelson}}\ and\ \bibinfo {editor} {\bibfnamefont {F.}~\bibnamefont {Schmidt}}}\ (\bibinfo  {publisher} {Academic Press},\ \bibinfo {year} {2021})\ \bibinfo {edition} {second edition}\ ed.,\ pp.\ \bibinfo {pages} {325--372}\BibitemShut {NoStop}%
\bibitem [{\citenamefont {Baker}(2001)}]{Baker:2001yc}%
  \BibitemOpen
  \bibfield  {author} {\bibinfo {author} {\bibfnamefont {G.~A.}\ \bibnamefont {Baker}, \bibfnamefont {Jr.}}\ }(\bibinfo {year} {2001})\ \Eprint {http://arxiv.org/abs/astro-ph/0112320} {arXiv:astro-ph/0112320} \BibitemShut {NoStop}%
\bibitem [{\citenamefont {Nesseris}\ and\ \citenamefont {Perivolaropoulos}(2004)}]{Nesseris:2004uj}%
  \BibitemOpen
  \bibfield  {author} {\bibinfo {author} {\bibfnamefont {S.}~\bibnamefont {Nesseris}}\ and\ \bibinfo {author} {\bibfnamefont {L.}~\bibnamefont {Perivolaropoulos}},\ }\href {\doibase 10.1103/PhysRevD.70.123529} {\bibfield  {journal} {\bibinfo  {journal} {Phys. Rev. D}\ }\textbf {\bibinfo {volume} {70}},\ \bibinfo {pages} {123529} (\bibinfo {year} {2004})},\ \Eprint {http://arxiv.org/abs/astro-ph/0410309} {arXiv:astro-ph/0410309} \BibitemShut {NoStop}%
\bibitem [{\citenamefont {Faraoni}\ and\ \citenamefont {Jacques}(2007)}]{Faraoni:2007es}%
  \BibitemOpen
  \bibfield  {author} {\bibinfo {author} {\bibfnamefont {V.}~\bibnamefont {Faraoni}}\ and\ \bibinfo {author} {\bibfnamefont {A.}~\bibnamefont {Jacques}},\ }\href {\doibase 10.1103/PhysRevD.76.063510} {\bibfield  {journal} {\bibinfo  {journal} {Phys. Rev. D}\ }\textbf {\bibinfo {volume} {76}},\ \bibinfo {pages} {063510} (\bibinfo {year} {2007})},\ \Eprint {http://arxiv.org/abs/0707.1350} {arXiv:0707.1350 [gr-qc]} \BibitemShut {NoStop}%
\bibitem [{\citenamefont {Perivolaropoulos}(2016)}]{Perivolaropoulos:2016nhp}%
  \BibitemOpen
  \bibfield  {author} {\bibinfo {author} {\bibfnamefont {L.}~\bibnamefont {Perivolaropoulos}},\ }\href {\doibase 10.1103/PhysRevD.94.124018} {\bibfield  {journal} {\bibinfo  {journal} {Phys. Rev. D}\ }\textbf {\bibinfo {volume} {94}},\ \bibinfo {pages} {124018} (\bibinfo {year} {2016})},\ \Eprint {http://arxiv.org/abs/1609.08528} {arXiv:1609.08528 [gr-qc]} \BibitemShut {NoStop}%
\bibitem [{\citenamefont {Lahav}\ \emph {et~al.}(1991)\citenamefont {Lahav}, \citenamefont {Lilje}, \citenamefont {Primack},\ and\ \citenamefont {Rees}}]{Lahav:1991wc}%
  \BibitemOpen
  \bibfield  {author} {\bibinfo {author} {\bibfnamefont {O.}~\bibnamefont {Lahav}}, \bibinfo {author} {\bibfnamefont {P.~B.}\ \bibnamefont {Lilje}}, \bibinfo {author} {\bibfnamefont {J.~R.}\ \bibnamefont {Primack}}, \ and\ \bibinfo {author} {\bibfnamefont {M.~J.}\ \bibnamefont {Rees}},\ }\href@noop {} {\bibfield  {journal} {\bibinfo  {journal} {Mon. Not. Roy. Astron. Soc.}\ }\textbf {\bibinfo {volume} {251}},\ \bibinfo {pages} {128} (\bibinfo {year} {1991})}\BibitemShut {NoStop}%
\bibitem [{\citenamefont {Iliev}\ and\ \citenamefont {Shapiro}(2001)}]{Iliev:2001he}%
  \BibitemOpen
  \bibfield  {author} {\bibinfo {author} {\bibfnamefont {I.~T.}\ \bibnamefont {Iliev}}\ and\ \bibinfo {author} {\bibfnamefont {P.~R.}\ \bibnamefont {Shapiro}},\ }\href {\doibase 10.1046/j.1365-8711.2001.04422.x} {\bibfield  {journal} {\bibinfo  {journal} {Mon. Not. Roy. Astron. Soc.}\ }\textbf {\bibinfo {volume} {325}},\ \bibinfo {pages} {468} (\bibinfo {year} {2001})},\ \Eprint {http://arxiv.org/abs/astro-ph/0101067} {arXiv:astro-ph/0101067} \BibitemShut {NoStop}%
\bibitem [{\citenamefont {Maor}\ and\ \citenamefont {Lahav}(2005)}]{Maor:2005hq}%
  \BibitemOpen
  \bibfield  {author} {\bibinfo {author} {\bibfnamefont {I.}~\bibnamefont {Maor}}\ and\ \bibinfo {author} {\bibfnamefont {O.}~\bibnamefont {Lahav}},\ }\href {\doibase 10.1088/1475-7516/2005/07/003} {\bibfield  {journal} {\bibinfo  {journal} {JCAP}\ }\textbf {\bibinfo {volume} {07}},\ \bibinfo {pages} {003} (\bibinfo {year} {2005})},\ \Eprint {http://arxiv.org/abs/astro-ph/0505308} {arXiv:astro-ph/0505308} \BibitemShut {NoStop}%
\bibitem [{\citenamefont {Saha}\ \emph {et~al.}(2023)\citenamefont {Saha}, \citenamefont {Dey},\ and\ \citenamefont {Bhattacharya}}]{Saha:2023zos}%
  \BibitemOpen
  \bibfield  {author} {\bibinfo {author} {\bibfnamefont {P.}~\bibnamefont {Saha}}, \bibinfo {author} {\bibfnamefont {D.}~\bibnamefont {Dey}}, \ and\ \bibinfo {author} {\bibfnamefont {K.}~\bibnamefont {Bhattacharya}}\ }(\bibinfo {year} {2023})\ \Eprint {http://arxiv.org/abs/2306.00805} {arXiv:2306.00805 [gr-qc]} \BibitemShut {NoStop}%
\bibitem [{\citenamefont {Battye}\ and\ \citenamefont {Weller}(2003)}]{Battye:2003bm}%
  \BibitemOpen
  \bibfield  {author} {\bibinfo {author} {\bibfnamefont {R.~A.}\ \bibnamefont {Battye}}\ and\ \bibinfo {author} {\bibfnamefont {J.}~\bibnamefont {Weller}},\ }\href {\doibase 10.1103/PhysRevD.68.083506} {\bibfield  {journal} {\bibinfo  {journal} {Phys. Rev. D}\ }\textbf {\bibinfo {volume} {68}},\ \bibinfo {pages} {083506} (\bibinfo {year} {2003})},\ \Eprint {http://arxiv.org/abs/astro-ph/0305568} {arXiv:astro-ph/0305568} \BibitemShut {NoStop}%
\bibitem [{\citenamefont {Wang}\ and\ \citenamefont {Steinhardt}(1998)}]{Wang:1998gt}%
  \BibitemOpen
  \bibfield  {author} {\bibinfo {author} {\bibfnamefont {L.-M.}\ \bibnamefont {Wang}}\ and\ \bibinfo {author} {\bibfnamefont {P.~J.}\ \bibnamefont {Steinhardt}},\ }\href {\doibase 10.1086/306436} {\bibfield  {journal} {\bibinfo  {journal} {Astrophys. J.}\ }\textbf {\bibinfo {volume} {508}},\ \bibinfo {pages} {483} (\bibinfo {year} {1998})},\ \Eprint {http://arxiv.org/abs/astro-ph/9804015} {arXiv:astro-ph/9804015} \BibitemShut {NoStop}%
\bibitem [{\citenamefont {Weinberg}\ and\ \citenamefont {Kamionkowski}(2003)}]{Weinberg:2002rd}%
  \BibitemOpen
  \bibfield  {author} {\bibinfo {author} {\bibfnamefont {N.~N.}\ \bibnamefont {Weinberg}}\ and\ \bibinfo {author} {\bibfnamefont {M.}~\bibnamefont {Kamionkowski}},\ }\href {\doibase 10.1046/j.1365-8711.2003.06421.x} {\bibfield  {journal} {\bibinfo  {journal} {Mon. Not. Roy. Astron. Soc.}\ }\textbf {\bibinfo {volume} {341}},\ \bibinfo {pages} {251} (\bibinfo {year} {2003})},\ \Eprint {http://arxiv.org/abs/astro-ph/0210134} {arXiv:astro-ph/0210134} \BibitemShut {NoStop}%
\bibitem [{\citenamefont {Einstein}\ and\ \citenamefont {Straus}(1945)}]{Einstein:1945id}%
  \BibitemOpen
  \bibfield  {author} {\bibinfo {author} {\bibfnamefont {A.}~\bibnamefont {Einstein}}\ and\ \bibinfo {author} {\bibfnamefont {E.~G.}\ \bibnamefont {Straus}},\ }\href {\doibase 10.1103/RevModPhys.17.120} {\bibfield  {journal} {\bibinfo  {journal} {Rev. Mod. Phys.}\ }\textbf {\bibinfo {volume} {17}},\ \bibinfo {pages} {120} (\bibinfo {year} {1945})}\BibitemShut {NoStop}%
\bibitem [{\citenamefont {Nandra}\ \emph {et~al.}(2012{\natexlab{a}})\citenamefont {Nandra}, \citenamefont {Lasenby},\ and\ \citenamefont {Hobson}}]{Nandra:2011ug}%
  \BibitemOpen
  \bibfield  {author} {\bibinfo {author} {\bibfnamefont {R.}~\bibnamefont {Nandra}}, \bibinfo {author} {\bibfnamefont {A.~N.}\ \bibnamefont {Lasenby}}, \ and\ \bibinfo {author} {\bibfnamefont {M.~P.}\ \bibnamefont {Hobson}},\ }\href {\doibase 10.1111/j.1365-2966.2012.20618.x} {\bibfield  {journal} {\bibinfo  {journal} {Mon. Not. Roy. Astron. Soc.}\ }\textbf {\bibinfo {volume} {422}},\ \bibinfo {pages} {2931} (\bibinfo {year} {2012}{\natexlab{a}})},\ \Eprint {http://arxiv.org/abs/1104.4447} {arXiv:1104.4447 [gr-qc]} \BibitemShut {NoStop}%
\bibitem [{\citenamefont {Nandra}\ \emph {et~al.}(2012{\natexlab{b}})\citenamefont {Nandra}, \citenamefont {Lasenby},\ and\ \citenamefont {Hobson}}]{Nandra:2011ui}%
  \BibitemOpen
  \bibfield  {author} {\bibinfo {author} {\bibfnamefont {R.}~\bibnamefont {Nandra}}, \bibinfo {author} {\bibfnamefont {A.~N.}\ \bibnamefont {Lasenby}}, \ and\ \bibinfo {author} {\bibfnamefont {M.~P.}\ \bibnamefont {Hobson}},\ }\href {\doibase 10.1111/j.1365-2966.2012.20617.x} {\bibfield  {journal} {\bibinfo  {journal} {Mon. Not. Roy. Astron. Soc.}\ }\textbf {\bibinfo {volume} {422}},\ \bibinfo {pages} {2945} (\bibinfo {year} {2012}{\natexlab{b}})},\ \Eprint {http://arxiv.org/abs/1104.4458} {arXiv:1104.4458 [gr-qc]} \BibitemShut {NoStop}%
\bibitem [{\citenamefont {Bouhmadi-L\'opez}\ \emph {et~al.}(2015)\citenamefont {Bouhmadi-L\'opez}, \citenamefont {Chen},\ and\ \citenamefont {Chen}}]{Bouhmadi-Lopez:2014jfa}%
  \BibitemOpen
  \bibfield  {author} {\bibinfo {author} {\bibfnamefont {M.}~\bibnamefont {Bouhmadi-L\'opez}}, \bibinfo {author} {\bibfnamefont {C.-Y.}\ \bibnamefont {Chen}}, \ and\ \bibinfo {author} {\bibfnamefont {P.}~\bibnamefont {Chen}},\ }\href {\doibase 10.1140/epjc/s10052-015-3257-4} {\bibfield  {journal} {\bibinfo  {journal} {Eur. Phys. J. C}\ }\textbf {\bibinfo {volume} {75}},\ \bibinfo {pages} {90} (\bibinfo {year} {2015})},\ \Eprint {http://arxiv.org/abs/1406.6157} {arXiv:1406.6157 [gr-qc]} \BibitemShut {NoStop}%
\bibitem [{\citenamefont {Antoniou}\ and\ \citenamefont {Perivolaropoulos}(2016)}]{Antoniou:2016obw}%
  \BibitemOpen
  \bibfield  {author} {\bibinfo {author} {\bibfnamefont {I.}~\bibnamefont {Antoniou}}\ and\ \bibinfo {author} {\bibfnamefont {L.}~\bibnamefont {Perivolaropoulos}},\ }\href {\doibase 10.1103/PhysRevD.93.123520} {\bibfield  {journal} {\bibinfo  {journal} {Phys. Rev. D}\ }\textbf {\bibinfo {volume} {93}},\ \bibinfo {pages} {123520} (\bibinfo {year} {2016})},\ \Eprint {http://arxiv.org/abs/1603.02569} {arXiv:1603.02569 [gr-qc]} \BibitemShut {NoStop}%
\bibitem [{\citenamefont {Kravtsov}\ and\ \citenamefont {Borgani}(2012)}]{Kravtsov:2012zs}%
  \BibitemOpen
  \bibfield  {author} {\bibinfo {author} {\bibfnamefont {A.}~\bibnamefont {Kravtsov}}\ and\ \bibinfo {author} {\bibfnamefont {S.}~\bibnamefont {Borgani}},\ }\href {\doibase 10.1146/annurev-astro-081811-125502} {\bibfield  {journal} {\bibinfo  {journal} {Ann. Rev. Astron. Astrophys.}\ }\textbf {\bibinfo {volume} {50}},\ \bibinfo {pages} {353} (\bibinfo {year} {2012})},\ \Eprint {http://arxiv.org/abs/1205.5556} {arXiv:1205.5556 [astro-ph.CO]} \BibitemShut {NoStop}%
\bibitem [{\citenamefont {Gao}\ \emph {et~al.}(2020)\citenamefont {Gao}, \citenamefont {Zou}, \citenamefont {Zhou},\ and\ \citenamefont {Kong}}]{Gao:2019tfj}%
  \BibitemOpen
  \bibfield  {author} {\bibinfo {author} {\bibfnamefont {J.}~\bibnamefont {Gao}}, \bibinfo {author} {\bibfnamefont {H.}~\bibnamefont {Zou}}, \bibinfo {author} {\bibfnamefont {X.}~\bibnamefont {Zhou}}, \ and\ \bibinfo {author} {\bibfnamefont {X.}~\bibnamefont {Kong}},\ }\href {\doibase 10.1088/1538-3873/ab6151} {\bibfield  {journal} {\bibinfo  {journal} {Publ. Astron. Soc. Pac.}\ }\textbf {\bibinfo {volume} {132}},\ \bibinfo {pages} {024101} (\bibinfo {year} {2020})},\ \Eprint {http://arxiv.org/abs/1912.10909} {arXiv:1912.10909 [astro-ph.GA]} \BibitemShut {NoStop}%
\bibitem [{\citenamefont {Harvey}\ \emph {et~al.}(2020)\citenamefont {Harvey}, \citenamefont {Robertson}, \citenamefont {Tam}, \citenamefont {Jauzac}, \citenamefont {Massey}, \citenamefont {Rhodes},\ and\ \citenamefont {McCarthy}}]{Harvey:2020gsy}%
  \BibitemOpen
  \bibfield  {author} {\bibinfo {author} {\bibfnamefont {D.}~\bibnamefont {Harvey}}, \bibinfo {author} {\bibfnamefont {A.}~\bibnamefont {Robertson}}, \bibinfo {author} {\bibfnamefont {S.-I.}\ \bibnamefont {Tam}}, \bibinfo {author} {\bibfnamefont {M.}~\bibnamefont {Jauzac}}, \bibinfo {author} {\bibfnamefont {R.}~\bibnamefont {Massey}}, \bibinfo {author} {\bibfnamefont {J.}~\bibnamefont {Rhodes}}, \ and\ \bibinfo {author} {\bibfnamefont {I.~G.}\ \bibnamefont {McCarthy}},\ }\href {\doibase 10.1093/mnras/staa3193} {\bibfield  {journal} {\bibinfo  {journal} {Mon. Not. Roy. Astron. Soc.}\ }\textbf {\bibinfo {volume} {500}},\ \bibinfo {pages} {2627} (\bibinfo {year} {2020})},\ \Eprint {http://arxiv.org/abs/2011.01945} {arXiv:2011.01945 [astro-ph.CO]} \BibitemShut {NoStop}%
\bibitem [{\citenamefont {Fernandez}\ \emph {et~al.}(2022)\citenamefont {Fernandez}, \citenamefont {Cueli}, \citenamefont {Gonz\'alez-Nuevo}, \citenamefont {Bonavera}, \citenamefont {Crespo}, \citenamefont {Casas},\ and\ \citenamefont {Lapi}}]{Fernandez:2021jlm}%
  \BibitemOpen
  \bibfield  {author} {\bibinfo {author} {\bibfnamefont {L.}~\bibnamefont {Fernandez}}, \bibinfo {author} {\bibfnamefont {M.~M.}\ \bibnamefont {Cueli}}, \bibinfo {author} {\bibfnamefont {J.}~\bibnamefont {Gonz\'alez-Nuevo}}, \bibinfo {author} {\bibfnamefont {L.}~\bibnamefont {Bonavera}}, \bibinfo {author} {\bibfnamefont {D.}~\bibnamefont {Crespo}}, \bibinfo {author} {\bibfnamefont {J.~M.}\ \bibnamefont {Casas}}, \ and\ \bibinfo {author} {\bibfnamefont {A.}~\bibnamefont {Lapi}},\ }\href {\doibase 10.1051/0004-6361/202141905} {\bibfield  {journal} {\bibinfo  {journal} {Astron. Astrophys.}\ }\textbf {\bibinfo {volume} {658}},\ \bibinfo {pages} {A19} (\bibinfo {year} {2022})},\ \Eprint {http://arxiv.org/abs/2111.05422} {arXiv:2111.05422 [astro-ph.CO]} \BibitemShut {NoStop}%
\bibitem [{\citenamefont {Sankhyayan}\ \emph {et~al.}(2023)\citenamefont {Sankhyayan}, \citenamefont {Bagchi}, \citenamefont {Tempel}, \citenamefont {More}, \citenamefont {Einasto}, \citenamefont {Dabhade}, \citenamefont {Raychaudhury}, \citenamefont {Athreya},\ and\ \citenamefont {Hein\"am\"aki}}]{Sankhyayan:2023tii}%
  \BibitemOpen
  \bibfield  {author} {\bibinfo {author} {\bibfnamefont {S.}~\bibnamefont {Sankhyayan}}, \bibinfo {author} {\bibfnamefont {J.}~\bibnamefont {Bagchi}}, \bibinfo {author} {\bibfnamefont {E.}~\bibnamefont {Tempel}}, \bibinfo {author} {\bibfnamefont {S.}~\bibnamefont {More}}, \bibinfo {author} {\bibfnamefont {M.}~\bibnamefont {Einasto}}, \bibinfo {author} {\bibfnamefont {P.}~\bibnamefont {Dabhade}}, \bibinfo {author} {\bibfnamefont {S.}~\bibnamefont {Raychaudhury}}, \bibinfo {author} {\bibfnamefont {R.}~\bibnamefont {Athreya}}, \ and\ \bibinfo {author} {\bibfnamefont {P.}~\bibnamefont {Hein\"am\"aki}},\ }\href {\doibase 10.3847/1538-4357/acfaeb} {\bibfield  {journal} {\bibinfo  {journal} {Astrophys. J.}\ }\textbf {\bibinfo {volume} {958}},\ \bibinfo {pages} {62} (\bibinfo {year} {2023})},\ \Eprint {http://arxiv.org/abs/2309.06251} {arXiv:2309.06251 [astro-ph.CO]} \BibitemShut {NoStop}%
\bibitem [{\citenamefont {Arfken}\ \emph {et~al.}(2012)\citenamefont {Arfken}, \citenamefont {Weber},\ and\ \citenamefont {Harris}}]{arfken2012mathematical}%
  \BibitemOpen
  \bibfield  {author} {\bibinfo {author} {\bibfnamefont {G.}~\bibnamefont {Arfken}}, \bibinfo {author} {\bibfnamefont {H.}~\bibnamefont {Weber}}, \ and\ \bibinfo {author} {\bibfnamefont {F.}~\bibnamefont {Harris}},\ }\href {https://books.google.gr/books?id=qLFo\_Z-PoGIC} {\emph {\bibinfo {title} {Mathematical Methods for Physicists: A Comprehensive Guide}}}\ (\bibinfo  {publisher} {Elsevier Science},\ \bibinfo {year} {2012})\BibitemShut {NoStop}%
\bibitem [{\citenamefont {Knox}\ and\ \citenamefont {Millea}(2020)}]{Knox:2019rjx}%
  \BibitemOpen
  \bibfield  {author} {\bibinfo {author} {\bibfnamefont {L.}~\bibnamefont {Knox}}\ and\ \bibinfo {author} {\bibfnamefont {M.}~\bibnamefont {Millea}},\ }\href {\doibase 10.1103/PhysRevD.101.043533} {\bibfield  {journal} {\bibinfo  {journal} {Phys. Rev. D}\ }\textbf {\bibinfo {volume} {101}},\ \bibinfo {pages} {043533} (\bibinfo {year} {2020})},\ \Eprint {http://arxiv.org/abs/1908.03663} {arXiv:1908.03663 [astro-ph.CO]} \BibitemShut {NoStop}%
\bibitem [{\citenamefont {Shah}\ \emph {et~al.}(2021)\citenamefont {Shah}, \citenamefont {Lemos},\ and\ \citenamefont {Lahav}}]{Shah:2021onj}%
  \BibitemOpen
  \bibfield  {author} {\bibinfo {author} {\bibfnamefont {P.}~\bibnamefont {Shah}}, \bibinfo {author} {\bibfnamefont {P.}~\bibnamefont {Lemos}}, \ and\ \bibinfo {author} {\bibfnamefont {O.}~\bibnamefont {Lahav}},\ }\href {\doibase 10.1007/s00159-021-00137-4} {\bibfield  {journal} {\bibinfo  {journal} {Astron. Astrophys. Rev.}\ }\textbf {\bibinfo {volume} {29}},\ \bibinfo {pages} {9} (\bibinfo {year} {2021})},\ \Eprint {http://arxiv.org/abs/2109.01161} {arXiv:2109.01161 [astro-ph.CO]} \BibitemShut {NoStop}%
\bibitem [{\citenamefont {Vagnozzi}(2020)}]{Vagnozzi:2019ezj}%
  \BibitemOpen
  \bibfield  {author} {\bibinfo {author} {\bibfnamefont {S.}~\bibnamefont {Vagnozzi}},\ }\href {\doibase 10.1103/PhysRevD.102.023518} {\bibfield  {journal} {\bibinfo  {journal} {Phys. Rev. D}\ }\textbf {\bibinfo {volume} {102}},\ \bibinfo {pages} {023518} (\bibinfo {year} {2020})},\ \Eprint {http://arxiv.org/abs/1907.07569} {arXiv:1907.07569 [astro-ph.CO]} \BibitemShut {NoStop}%
\bibitem [{\citenamefont {Komatsu}\ \emph {et~al.}(2009)\citenamefont {Komatsu} \emph {et~al.}}]{WMAP:2008lyn}%
  \BibitemOpen
  \bibfield  {author} {\bibinfo {author} {\bibfnamefont {E.}~\bibnamefont {Komatsu}} \emph {et~al.} (\bibinfo {collaboration} {WMAP}),\ }\href {\doibase 10.1088/0067-0049/180/2/330} {\bibfield  {journal} {\bibinfo  {journal} {Astrophys. J. Suppl.}\ }\textbf {\bibinfo {volume} {180}},\ \bibinfo {pages} {330} (\bibinfo {year} {2009})},\ \Eprint {http://arxiv.org/abs/0803.0547} {arXiv:0803.0547 [astro-ph]} \BibitemShut {NoStop}%
\bibitem [{\citenamefont {Komatsu}\ \emph {et~al.}(2011)\citenamefont {Komatsu}, \citenamefont {Smith}, \citenamefont {Dunkley}, \citenamefont {Bennett}, \citenamefont {Gold}, \citenamefont {Hinshaw}, \citenamefont {Jarosik}, \citenamefont {Larson}, \citenamefont {Nolta}, \citenamefont {Page}, \citenamefont {Spergel}, \citenamefont {Halpern}, \citenamefont {Hill}, \citenamefont {Kogut}, \citenamefont {Limon}, \citenamefont {Meyer}, \citenamefont {Odegard}, \citenamefont {Tucker}, \citenamefont {Weiland}, \citenamefont {Wollack},\ and\ \citenamefont {Wright}}]{Komatsu_2011}%
  \BibitemOpen
  \bibfield  {author} {\bibinfo {author} {\bibfnamefont {E.}~\bibnamefont {Komatsu}}, \bibinfo {author} {\bibfnamefont {K.~M.}\ \bibnamefont {Smith}}, \bibinfo {author} {\bibfnamefont {J.}~\bibnamefont {Dunkley}}, \bibinfo {author} {\bibfnamefont {C.~L.}\ \bibnamefont {Bennett}}, \bibinfo {author} {\bibfnamefont {B.}~\bibnamefont {Gold}}, \bibinfo {author} {\bibfnamefont {G.}~\bibnamefont {Hinshaw}}, \bibinfo {author} {\bibfnamefont {N.}~\bibnamefont {Jarosik}}, \bibinfo {author} {\bibfnamefont {D.}~\bibnamefont {Larson}}, \bibinfo {author} {\bibfnamefont {M.~R.}\ \bibnamefont {Nolta}}, \bibinfo {author} {\bibfnamefont {L.}~\bibnamefont {Page}}, \bibinfo {author} {\bibfnamefont {D.~N.}\ \bibnamefont {Spergel}}, \bibinfo {author} {\bibfnamefont {M.}~\bibnamefont {Halpern}}, \bibinfo {author} {\bibfnamefont {R.~S.}\ \bibnamefont {Hill}}, \bibinfo {author} {\bibfnamefont {A.}~\bibnamefont {Kogut}}, \bibinfo {author} {\bibfnamefont {M.}~\bibnamefont {Limon}}, \bibinfo {author} {\bibfnamefont {S.~S.}\ \bibnamefont
  {Meyer}}, \bibinfo {author} {\bibfnamefont {N.}~\bibnamefont {Odegard}}, \bibinfo {author} {\bibfnamefont {G.~S.}\ \bibnamefont {Tucker}}, \bibinfo {author} {\bibfnamefont {J.~L.}\ \bibnamefont {Weiland}}, \bibinfo {author} {\bibfnamefont {E.}~\bibnamefont {Wollack}}, \ and\ \bibinfo {author} {\bibfnamefont {E.~L.}\ \bibnamefont {Wright}},\ }\href {\doibase 10.1088/0067-0049/192/2/18} {\bibfield  {journal} {\bibinfo  {journal} {The Astrophysical Journal Supplement Series}\ }\textbf {\bibinfo {volume} {192}},\ \bibinfo {pages} {18} (\bibinfo {year} {2011})}\BibitemShut {NoStop}%
\bibitem [{\citenamefont {Wang}\ and\ \citenamefont {Mukherjee}(2007)}]{Wang:2007mza}%
  \BibitemOpen
  \bibfield  {author} {\bibinfo {author} {\bibfnamefont {Y.}~\bibnamefont {Wang}}\ and\ \bibinfo {author} {\bibfnamefont {P.}~\bibnamefont {Mukherjee}},\ }\href {\doibase 10.1103/PhysRevD.76.103533} {\bibfield  {journal} {\bibinfo  {journal} {Phys. Rev. D}\ }\textbf {\bibinfo {volume} {76}},\ \bibinfo {pages} {103533} (\bibinfo {year} {2007})},\ \Eprint {http://arxiv.org/abs/astro-ph/0703780} {arXiv:astro-ph/0703780} \BibitemShut {NoStop}%
\bibitem [{\citenamefont {Hu}\ and\ \citenamefont {Sugiyama}(1996)}]{Hu:1995en}%
  \BibitemOpen
  \bibfield  {author} {\bibinfo {author} {\bibfnamefont {W.}~\bibnamefont {Hu}}\ and\ \bibinfo {author} {\bibfnamefont {N.}~\bibnamefont {Sugiyama}},\ }\href {\doibase 10.1086/177989} {\bibfield  {journal} {\bibinfo  {journal} {Astrophys. J.}\ }\textbf {\bibinfo {volume} {471}},\ \bibinfo {pages} {542} (\bibinfo {year} {1996})},\ \Eprint {http://arxiv.org/abs/astro-ph/9510117} {arXiv:astro-ph/9510117} \BibitemShut {NoStop}%
\bibitem [{\citenamefont {Wang}\ and\ \citenamefont {Wang}(2013)}]{Wang:2013mha}%
  \BibitemOpen
  \bibfield  {author} {\bibinfo {author} {\bibfnamefont {Y.}~\bibnamefont {Wang}}\ and\ \bibinfo {author} {\bibfnamefont {S.}~\bibnamefont {Wang}},\ }\href {\doibase 10.1103/PhysRevD.88.043522} {\bibfield  {journal} {\bibinfo  {journal} {Phys. Rev. D}\ }\textbf {\bibinfo {volume} {88}},\ \bibinfo {pages} {043522} (\bibinfo {year} {2013})},\ \bibinfo {note} {[Erratum: Phys.Rev.D 88, 069903 (2013)]},\ \Eprint {http://arxiv.org/abs/1304.4514} {arXiv:1304.4514 [astro-ph.CO]} \BibitemShut {NoStop}%
\bibitem [{\citenamefont {Huang}\ \emph {et~al.}(2015)\citenamefont {Huang}, \citenamefont {Wang},\ and\ \citenamefont {Wang}}]{Huang:2015vpa}%
  \BibitemOpen
  \bibfield  {author} {\bibinfo {author} {\bibfnamefont {Q.-G.}\ \bibnamefont {Huang}}, \bibinfo {author} {\bibfnamefont {K.}~\bibnamefont {Wang}}, \ and\ \bibinfo {author} {\bibfnamefont {S.}~\bibnamefont {Wang}},\ }\href {\doibase 10.1088/1475-7516/2015/12/022} {\bibfield  {journal} {\bibinfo  {journal} {JCAP}\ }\textbf {\bibinfo {volume} {12}},\ \bibinfo {pages} {022} (\bibinfo {year} {2015})},\ \Eprint {http://arxiv.org/abs/1509.00969} {arXiv:1509.00969 [astro-ph.CO]} \BibitemShut {NoStop}%
\bibitem [{\citenamefont {Chen}\ \emph {et~al.}(2019)\citenamefont {Chen}, \citenamefont {Huang},\ and\ \citenamefont {Wang}}]{Chen:2018dbv}%
  \BibitemOpen
  \bibfield  {author} {\bibinfo {author} {\bibfnamefont {L.}~\bibnamefont {Chen}}, \bibinfo {author} {\bibfnamefont {Q.-G.}\ \bibnamefont {Huang}}, \ and\ \bibinfo {author} {\bibfnamefont {K.}~\bibnamefont {Wang}},\ }\href {\doibase 10.1088/1475-7516/2019/02/028} {\bibfield  {journal} {\bibinfo  {journal} {JCAP}\ }\textbf {\bibinfo {volume} {02}},\ \bibinfo {pages} {028} (\bibinfo {year} {2019})},\ \Eprint {http://arxiv.org/abs/1808.05724} {arXiv:1808.05724 [astro-ph.CO]} \BibitemShut {NoStop}%
\bibitem [{\citenamefont {Elgaroy}\ and\ \citenamefont {Multamaki}(2007)}]{Elgaroy:2007bv}%
  \BibitemOpen
  \bibfield  {author} {\bibinfo {author} {\bibfnamefont {O.}~\bibnamefont {Elgaroy}}\ and\ \bibinfo {author} {\bibfnamefont {T.}~\bibnamefont {Multamaki}},\ }\href {\doibase 10.1051/0004-6361:20077292} {\bibfield  {journal} {\bibinfo  {journal} {Astron. Astrophys.}\ }\textbf {\bibinfo {volume} {471}},\ \bibinfo {pages} {65} (\bibinfo {year} {2007})},\ \Eprint {http://arxiv.org/abs/astro-ph/0702343} {arXiv:astro-ph/0702343} \BibitemShut {NoStop}%
\end{thebibliography}%

\end{document}